\documentclass[a4paper,11pt]{article}
\pdfoutput=1 

\usepackage{jheppub} 

\usepackage[T1]{fontenc} 
\usepackage[dvipsnames]{xcolor}
\usepackage{amsmath}

\newcommand{\slashed}[1]{\displaystyle{\not} #1}    
\newcommand{\xdownarrow}[1]{    
  {\left\downarrow\vbox to #1{}\right.\kern-\nulldelimiterspace}
}

\title{\boldmath Off-shell vs on-shell modelling of  top
  quarks in  photon associated production}

\author[a]{G. Bevilacqua,}
\author[b]{H. B. Hartanto,}
\author[c]{M. Kraus,}
\author[d]{T. Weber }
\author[d]{and M. Worek }

\affiliation[a]{MTA-DE Particle Physics Research Group, University of
Debrecen, H-4010 Debrecen, PBox 105, Hungary} 
\affiliation[b]{Institute for
Particle Physics Phenomenology, Department of Physics, 
Durham University, Durham, DH1 3LE, UK} 
\affiliation[c]{Physics Department, Florida State University,
Tallahassee, FL 32306-4350, USA}
\affiliation[d]{ Institute for Theoretical Particle Physics
and Cosmology, RWTH Aachen University, D-52056 Aachen, Germany}
 
\emailAdd{\\
giuseppe.bevilacqua@science.unideb.hu,
  heribertus.b.hartanto@durham.ac.uk,\\
  mkraus@hep.fsu.edu, \\tweber@physik.rwth-aachen.de,\\
  worek@physik.rwth-aachen.de}

\abstract{
  We present a comparative study of various approaches for modelling
of the $e^+ \nu_e \mu^- \bar{\nu}_\mu b\bar{b} \gamma$ final state in
$t\bar{t}\gamma$ production at the LHC. Working at the NLO in QCD we
compare the fully realistic description of the top quark decay chain
with the one provided by the narrow-width-approximation. The former
approach comprises all double, single and non-resonant diagrams,
interferences, and off-shell effects of the top quarks. The latter
incorporates only double resonant  contributions and restricts the
unstable top quarks to on-shell states. We confirm that for the
integrated cross sections the finite top quark width effects are small
and of the order of ${\cal O}(\Gamma_t/m_t)$. We show, however, that they
are strongly enhanced for more exclusive observables. In
addition, we investigate fractions of events where the photon is
radiated either in the production or in the decay stage. We find that
large fraction of isolated photons comes from radiative decays of top
quarks. Based on our findings, selection criteria might be developed
to reduce such contributions, that constitute a background
for the measurement of the anomalous couplings in the $t\bar{t}\gamma$
vertex.}

\dedicated{\rm TTK-19-52,  P3H-19-051, IPPP/19/98}


\begin{document} 
\maketitle
\flushbottom

%
\section{Introduction}
%

Higher order predictions for top quark pair production allow us to
deepen our understanding of the Standard Model (SM). By carefully
studying with very high accuracy the properties of the heaviest
particle discovered so far, physicists might shed some light on
physics beyond the SM, so sought-after at the LHC. Besides
$t\bar{t}(j)$ production, however, more exclusive $t\bar{t}V,\,
V=\gamma, Z,H,W^\pm$ final states are produced and thoroughly analysed
at the LHC. Even though $t\bar{t}V$ cross sections are orders of
magnitude smaller than those of $\sigma_{pp\to t\bar{t}}$ and
$\sigma_{pp \to t\bar{t}j}$, they add greatly to the already rich top
quark research plans, which are carried out by the ATLAS and CMS
collaborations. In particular, among all associated processes the
associated production of top quark pairs with a photon has the highest
production rate at the LHC. First evidence for $t\bar{t}\gamma$ was
reported by the CDF collaboration \cite{Aaltonen:2011sp}, whereas the
observation of the process with a significance of $5.3\sigma$ was
established at $\sqrt{s}=7$ TeV by the ATLAS collaboration
\cite{Aad:2015uwa}. Both ATLAS and CMS measured $t\bar{t}\gamma$ cross
section at $\sqrt{s}=8$ TeV \cite{Aaboud:2017era, Sirunyan:2017iyh}. But
only recently first measurements of the differential cross sections at
$13$ TeV have been performed by the ATLAS collaboration
\cite{Aaboud:2018hip,ATLAS:2019gkg}.

The $t\bar{t}\gamma$ process
probes the $t\gamma$ electroweak coupling, thus, provides a direct way
to measure the top quark electric charge \cite{Baur:2001si}. The
latter is known to be $Q_t=+2/3$, i.e. consistent with the SM, albeit
has only been measured indirectly in $t\bar{t}$ production
\cite{Aad:2013uza,Aaltonen:2013sgl}. More exotic physics scenarios,
that propose a heavy quark of electric charge $Q_t= -4/3$, instead of
the SM top quark, have been excluded with a significance of more than
$8\sigma$. Not only the strength but also the structure of
$t\bar{t}\gamma$ vertex can be examined with the help of the $pp\to
t\bar{t}\gamma$ production process. Based on the fundamental principle
of gauge symmetry, the $t\bar{t}\gamma$ vertex, which includes the SM
coupling given by the top-quark electric charge $Q_t$, and contributions
from dimension-six effective operators, can be parametrised using only
$\gamma^\mu$ and $\sigma^{\mu\nu}q_\nu$, where $q=(p_{\bar t}-p_t)$ is
the outgoing photon momentum \cite{AguilarSaavedra:2008zc}. Because
the $\gamma^\mu$ term does not receive corrections from dimension-six
gauge invariant operators the electroweak top anomalous interactions
can be described in terms of only two anomalous couplings (two Wilson
coefficients) that are the coefficients of the effective
$\sigma^{\mu\nu}q_\nu$ interactions. The latter should be constrained
at the LHC already at $13$ TeV with an integrated luminosity of about
$300$ ${\rm fb}^{-1}$
\cite{Baur:2004uw,Bouzas:2012av,Schulze:2016qas,Bylund:2016phk}.

Furthermore, production of top quark pairs in association with a
photon can be employed to obtain predictions for integrated and
differential cross section ratios  \cite{Bevilacqua:2018dny}, that are
defined according to 
\begin{equation}
{\cal R} =\frac{\sigma_{t\bar{t}\gamma}}{\sigma_{t\bar{t}}} \,,
  \quad  \quad  \quad  \quad {\rm and} \quad \quad \quad  \quad 
  {\cal R}_X= \left(
\frac{d\sigma_{t\bar{t}\gamma}}{dX}
  \right)
  \left(\frac{
d\sigma_{t\bar{t}}}{dX}
    \right)^{-1}\,,
\end{equation}  
where $X$ stands for the kinematic observable under consideration 
(e.g. $M_{b\bar{b}},\, M_{\ell\ell},\, \Delta
\phi_{\ell\ell}$). The cross section ratios have many advantages
compared to absolute cross sections. They are, for example, more
stable against radiative corrections and have reduced scale
dependence. Considering that $t\bar{t}\gamma$ and $t\bar{t}$ are very
similar processes from the QCD point of view many theoretical and
experimental uncertainties cancel in ratio. Consequently, ${\cal R}$
and ${\cal R}_X$ have enhanced predictive power and are interesting
not only to study the $t\bar{t}\gamma$ process with the highest 
precision that until now has only been reserved for  the top quark
pair production at NNLO in QCD, but also to
probe new physics. The latter might reveal itself once
sufficiently precise theoretical predictions are compared with 
appropriately precise experimental data.

Finally, the $t\bar{t}$ charge asymmetry, $A_C$, the di-leptonic
charge asymmetry, $A_{C}^{\ell\ell}$, as well as, the laboratory frame
single asymmetry, $A_{C}^{t\ell}$, and the single lepton asymmetry
defined in the $t\bar{t}$ rest frame, $A_{C}^\ell$ , can be
investigated in $t\bar{t}\gamma$ production at the LHC and the high
luminosity LHC. They provide complementary information to the measured
asymmetries in $t\bar{t}$ production
\cite{Aguilar-Saavedra:2014vta,Aguilar-Saavedra:2018gfv,
Bergner:2018lgm}.

To be able to provide reliable and very precise theoretical
predictions for the $pp\to t\bar{t}\gamma$ production process higher
order effects in $\alpha_s$ must be incorporated. NLO QCD corrections
to the inclusive $t\bar{t}\gamma$ production with on-shell top quarks
have been calculated for the first time in Ref.~\cite{PengFei:2009ph}
and afterwards recomputed in
Refs.~\cite{PengFei:2011qg,Maltoni:2015ena}.  In
Ref.~\cite{Duan:2016qlc} results with NLO electroweak corrections have
been provided. In all these cases, however, top quarks were treated as
stable particles. Such predictions may give us a general idea of the
size of the NLO corrections. Because they lack top quark decays,
however, they are neither capable to ensure a reliable description of
the fiducial phase space regions nor to give us a glimpse into the top
quark radiation pattern. For more realistic studies top quark decays
are needed. First attempt in this direction has been carried out in
Ref.~\cite{Kardos:2014zba} where NLO QCD theoretical predictions for
stable top quarks and a hard photon have been matched with parton
shower (PS) Monte Carlo (MC) programs using the \textsc{PowHel}
framework. The \textsc{PowHel} approach relies on the
\textsc{Powheg-Box} system \cite{Frixione:2007vw, Alioli:2010xd} and
allows for the matching between fixed order computation at NLO in QCD
and parton shower simulation, followed by hadronisation and hadronic
decays. The former is provided by the \textsc{Helac-Nlo} MC program
\cite{Bevilacqua:2011xh} and the latter by the general purpose MC
program like \textsc{Pythia} \cite{Sjostrand:2014zea} or
\textsc{Herwig} \cite{Bahr:2008pv}. In Ref.~\cite{Kardos:2014zba} top
quark decays have been treated in the PS approximation omitting
$t\bar{t}$ spin correlations and photon emission in parton shower
evolution. First fully realistic theoretical predictions for
$t\bar{t}\gamma$ have been presented in Ref.~\cite{Melnikov:2011ta}
where top quark decays in the narrow width approximation (NWA) have
been included, maintaining spin correlations of final state
particles. In addition, photon radiation off top quark decay products
has been incorporated. They brought along a significant contribution
to the cross section. Finally, in Ref.~\cite{Bevilacqua:2018woc} a
complete description of top quark pair production in association with
a hard photon in the di-lepton top quark decay channel has been
presented.  It is based on matrix elements for $e^+\nu_e\mu^-\nu_\mu
b\bar{b}\gamma$ production and included all resonant and non-resonant
diagrams, interferences, and off-shell effects of the top quarks and
the $W$ gauge bosons. This calculation constituted the first full
computation for the $t\bar{t}\gamma$ production process at NLO in QCD.

Having different theoretical approaches available for the modelling of
top quark decays, it is only natural to investigate whether the full
result is always mandatory for the description of various
observables. In other words when it might be safe to replace the full
result by the one from the NWA. The goal of this paper is, therefore, to
compare these two approaches that we shall refer to as \textit{NWA}
and \textit{full off-shell}. Furthermore, we shall approximate the
NLO+PS results featured in Ref.~\cite{Kardos:2014zba} by the ones in
NWA that keeps the isolated photon emission only in the
production stage and allows only LO top quark decays.
For brevity, we will refer to this last approach as NWA${}_{\rm
LOdecay}$. Another motivation for the paper is more
theoretical.  We shall carry out the systematic comparison within one
framework. To this end we have implemented the full NWA for top quark
related processes into \textsc{Helac-Nlo}.  
This required substantial modifications of
both parts of the program: \textsc{Helac-1Loop}
\cite{vanHameren:2009dr} and \textsc{Helac-Dipoles}
\cite{Czakon:2009ss} and an inclusion of an additional polarised
dipole into our implementation of the Catani-Seymour subtraction
scheme. Having a fully flexible MC program with both options allows us to
investigate the fraction of events with photon radiation in the
production and in the decays and compare double-, single- and
non-resonant contributions of the full off-shell result to the NWA.

The rest of the paper is organised as follows. In
Section~\ref{Implementation} details of the NWA implementation into
the \textsc{Helac-Nlo} system are shortly outlined. In Section~\ref{setup}
the SM input parameters and the cuts on final states are listed. Stability
of the full off-shell result with respect to the transverse momentum
cut on the bottom jet is examined in
Section~\ref{stability}. Numerical results for the integrated cross
sections are presented in Section~\ref{integrated}, while 
various differential cross sections are studied  in
Section~\ref{differential}. Finally, in Section~\ref{conclusions} we
give our conclusions.

%
\section{The Narrow Width Approximation}
\label{Implementation}
%

The NWA offers a conceptually easy and powerful method
for computing cross sections for processes comprising the unstable
resonances when  the width $(\Gamma)$ of the unstable particle is much
smaller than its mass $(M)$, see
e.g. \cite{Kauer:2001sp,Uhlemann:2008pm}. Therefore, for the 
top quark the NWA allows to make predictions for realistic
final states described in terms of light and $b$-jets, charged leptons
as well as missing transverse momentum and  gives the opportunity
for direct comparisons with cross section measurements in the fiducial
phase space regions.  Prominent examples include the calculations for
the $pp\to t\bar{t}(j)$ process at NLO QCD and theoretical predictions
for $t\bar{t}$ at NNLO in QCD
\cite{Bernreuther:2004jv,Melnikov:2009dn,Melnikov:2011qx,Campbell:2012uf,
Behring:2019iiv}. The NWA is well established and allows factorisation
of the cross section into production times decays due to the following
relation
\begin{equation}
\label{Eq:NWA}
\frac{1}{\left(p_t^2 - m_t^2\right)^2 + m_t^2\Gamma_t^2}
\stackrel{\Gamma_t/m_t \to 0}{\longrightarrow}
\frac{\pi}{m_t\Gamma_t} \, \delta\left(p_t^2-m_t^2\right) +
\mathcal{O}\left(\frac{\Gamma_t}{m_t}\right) \,,
\end{equation}
where $\Gamma_t$ and $m_t$ are the width and mass of the top
quark. All effects related to the off-shellness of the top quarks as
well as non-resonant contributions are systematically neglected in the
computation of scattering amplitudes. The neglected contributions are
suppressed by the $\Gamma_t/m_t$ ratio for sufficiently inclusive
observables \cite{Fadin:1993kt}, although they can be enhanced for
various differential cross sections.  In specific phase space regions,
like in the high $p_T$ region of various dimensionful observables or
in the vicinity of kinematical thresholds and edges, they contribute
up to $20\%-50\%$ \cite{AlcarazMaestre:2012vp,Denner:2012mx}. Although
there is no doubt that the full off-shell calculations should be used
if available, as they provide the most realistic description of the
processes under consideration, they are computationally demanding and
have practical limitations already at the NLO in perturbation
theory. For example, obtaining NLO QCD predictions with full off-shell
effects for the $t\bar{t}t\bar{t}$ production process is currently
very difficult to imagine.
%
\begin{figure}[t]
\begin{center}
  \includegraphics[width=0.65\textwidth]{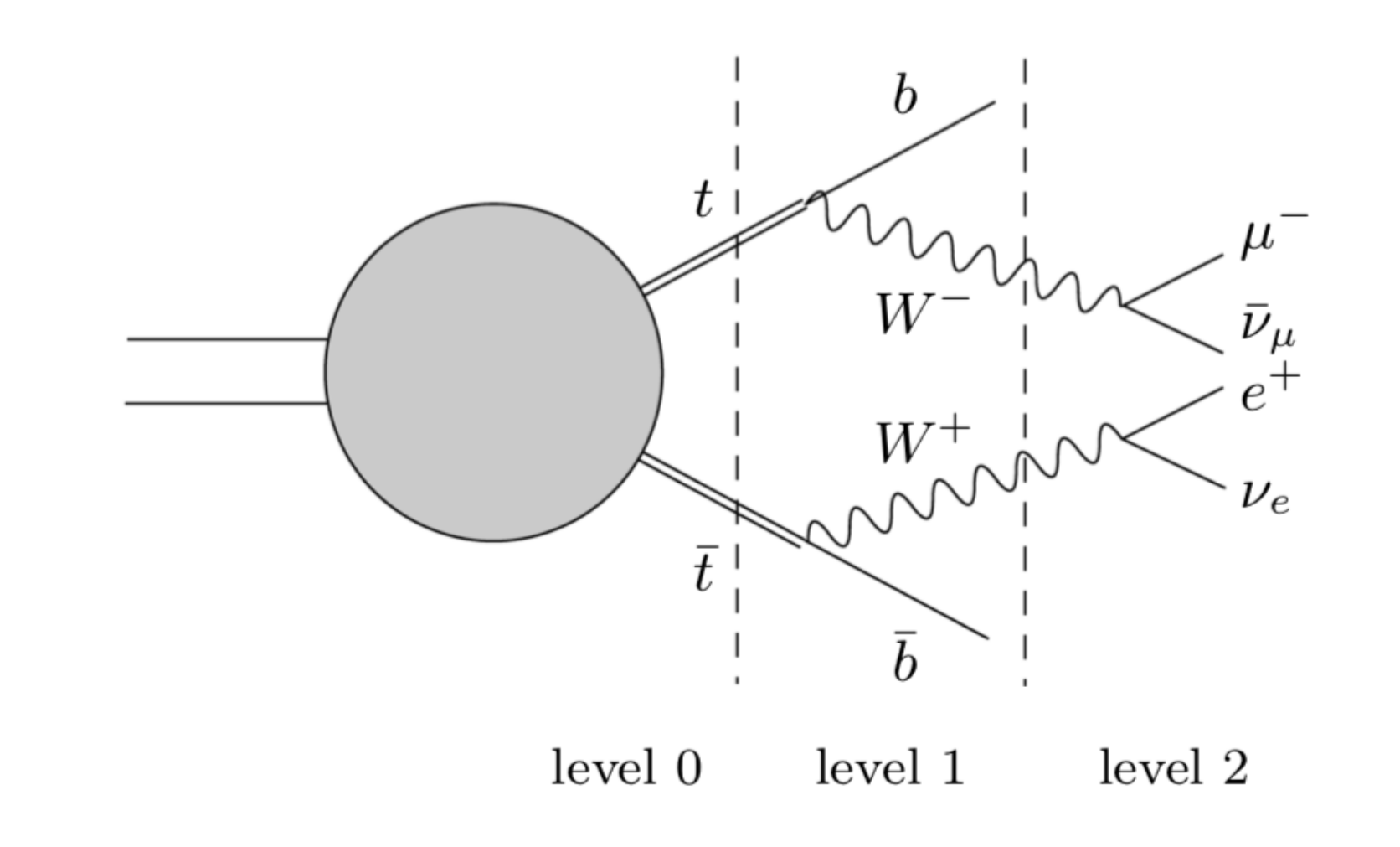}
\end{center}
\caption{\it Schematic illustration of the decay chain for the $pp\to t\bar{t}
\to b\bar{b} W^+W^- \to b\bar{b}e^+ \nu_e \mu^- \bar{\nu}_\mu$
process. The dashed lines  indicate top quark and $W$ gauge boson
propagators which are treated as on-shell particles in the NWA.}
\label{fig:tt}
\end{figure}

In the $\Gamma_t/m_t\to 0$ limit, the differential cross-section for
the top quark pair production in the di-lepton decay channel is given
by
\begin{equation}
d\sigma^{\rm NWA}_{t\bar{t}}
= d\sigma_{t\bar{t}} \, d{\cal B}_{\,t \, \to \,   b e^+\nu_e} \, d{\cal
  B}_{\,\bar{t}  \, \to \,  \bar{b} \mu^-\bar{\nu}_\mu}\,,
\end{equation}  
where ${\cal B}$ stands for the respective branching fraction. The
above equation is valid to all orders in the strong coupling
constant. A graphical representation is given in Figure \ref{fig:tt}
where both top quarks and $W$ bosons are treated in the NWA and the
decay chain is structured into three levels; level $0$ - the
production process of $t\bar{t}$, level $1$ and finally $2$ - the fully
decayed di-lepton final state. The factorisation for $t\bar{t}\gamma$
can be written in the similar manner by inserting a photon in either
of the three terms, giving rise to the following non-overlapping
resonant structures, see e.g. Ref.~\cite{Melnikov:2011ta}
\begin{equation}
  \begin{split}
  d\sigma^{\rm NWA}_{t\bar{t}  \gamma} &= d\sigma_{t\bar{t}\gamma}\, 
  d{\cal B}_{\,t \, \to \,   b e^+\nu_e} \, d{\cal
  B}_{\,\bar{t}  \, \to \,  \bar{b} \mu^-\bar{\nu}_\mu}\\[0.2cm]
&+ d\sigma_{t\bar{t}} \left(
d{\cal B}_{\,t \, \to \,   b e^+\nu_e \gamma} \, d{\cal
  B}_{\,\bar{t}  \, \to \,  \bar{b} \mu^-\bar{\nu}_\mu}
+d{\cal B}_{\,t \, \to \,   b e^+\nu_e} \, d{\cal
  B}_{\,\bar{t}  \, \to \,  \bar{b} \mu^-\bar{\nu}_\mu \gamma}
\right)\,.
\end{split}
\end{equation}  
Accordingly, $t\bar{t}\gamma$ production can be seen as described by
two distinct kinematics. On one hand we have photon emission in the
production part of the process and on the other hand off the top quark
decay products. Although computational complexity is much lower than
for the full off-shell calculation, the number of contributions that
need to be calculated for $t\bar{t}\gamma$ production in NWA increases
rapidly.  The same applies to gluon radiation, which must naturally be
taken into account when NLO QCD corrections are calculated.  Thus,
generally speaking the NWA approach is characterised by a
proliferation of the contributions that need to be put together to
account for all possible resonant structures of $d\sigma^{\rm
NWA}_{t\bar{t} \gamma}$. Several key processes, that are relevant for
top quark physics at the LHC, have already been computed in the NWA
approach. Nevertheless a fully flexible MC generator
capable of performing automated predictions in the NWA at NLO in QCD
for all $t\bar{t}$ plus additional object $(X=\gamma, Z, W^\pm, H,
\dots)$ processes is still missing, even-though  results for a variety of
$t\bar{t}X$ processes exist in the literature \footnote{~NLO
QCD predictions for $t\bar{t}\gamma$ in the NWA are presented in
\cite{Melnikov:2011ta}. For $t\bar{t}Z$ and $t\bar{t}j$ they are
described respectively in \cite{Rontsch:2014cca} and
\cite{Melnikov:2010iu,Melnikov:2011qx}.  Finally, for the
$t\bar{t}W^\pm$ process they are given in \cite{Campbell:2012dh}. In
addition, we note that all these predictions have already been heavily
used in various experimental analyses both at the TeVatron and the
LHC.}. Such a MC program should include NLO QCD corrections to
$t\bar{t}X$ production and top quark decays, retain all $t\bar{t}$
spin correlations and allow for arbitrary cuts on the final states.
Therefore, one of the purposes of the present work is to fill this gap
by extending the \textsc{Helac-Nlo} framework to include the full
NWA. On one hand, having the possibility to provide theoretical
predictions for both approaches within the same tool will facilitate
systematic comparisons. On the other hand, such automation will open a
new path for performing higher order calculations for more complex
processes such as $t\bar{t}b\bar{b}$, $t\bar{t}jj$ or
$t\bar{t}t\bar{t}$, for which only predictions with stable top quarks
are available at NLO in QCD, see e.g.
Refs.~\cite{Bredenstein:2009aj,Bevilacqua:2009zn,
Bredenstein:2010rs,Bevilacqua:2010ve,Bevilacqua:2011aa,
Bevilacqua:2012em,Maltoni:2015ena,Frederix:2017wme}. For
$t\bar{t}b\bar{b}$, there are also predictions with NLO $+$ PS
accuracy, where top quarks are decayed by PS programs
\cite{Garzelli:2014aba,Bevilacqua:2017cru, Jezo:2018yaf}.

%
\subsection{Implementing the NWA in HELAC-NLO}
%

In the conventional implementation of the NWA, the amplitudes of the
various production and decay subprocesses are computed separately and
subsequently combined. In order to preserve the $t\bar{t}$ spin
correlations between the production and decay stages a careful
bookkeeping of all matrix elements corresponding to different
polarisations of the resonant particles is necessary.  We will refer
to this strategy as the bottom-up approach.  The combinatorial burden
increases with the number of resonant particles and with the number of
sequential decays.

In \textsc{Helac-Nlo} we consider, however, a different strategy for
the implementation of the NWA, which we will refer to as the top-down
approach. We construct the fully decayed final state using the
standard Dyson-Schwinger recursive algorithms
\cite{Kanaki:2000ey,Cafarella:2007pc} but restrict the computation to
double resonant topologies only, discarding single- and non-resonant
ones. Furthermore, we introduce additional  modifications to the
fermionic propagator. In case of the resonant propagator
formula \eqref{Eq:NWA} amounts to the following replacement
\begin{equation}
\label{Eq:res_prop}
\frac{\slashed{p}_f + m_f}{\left(p_f^2 - m_f^2\right) + i m_f \Gamma_f} \quad
{\longrightarrow} \quad \left(\slashed{p}_f + m_f \right) \sqrt{\frac{\pi}{m_f \,
    \Gamma_f}}  \,,
\end{equation}
while the on-shell Dirac delta, $\delta(p_t^2-m_t^2)$, is contained in
the phase-space \footnote{~While carefully examining
Eq.~\eqref{Eq:res_prop} it may seem that there is a mismatch of
dimensions on both sides of Eq.~\eqref{Eq:res_prop}.  This is,
however, not the case since the Dirac delta function has been moved to
the phase-space for the right hand side of the equation.}. For
the non-resonant one we have instead
\begin{equation}
\label{Eq:nonres_prop}
\frac{\slashed{p}_f + m_f}{\left(p_f^2 - m_f^2 \right) + i m_f
  \Gamma_f} \quad {\longrightarrow}
\quad  \frac{\slashed{p}_f + m_f}{\left(p_f^2 - m_f^2\right)} \,.
\end{equation}
The numerator in Eq.~(\ref{Eq:res_prop}) can be left
unchanged since in the on-shell limit we can write 
\begin{equation}
(\slashed{p}_f + m_f) = \sum_{s=\pm}
u(p_f,s)\bar{u}(p_f,s) \,.
\end{equation}
A similar modification can be introduced in the case of the $W$ gauge
boson propagator. The main advantage of the top-down approach is that
it reduces the number of contributions to be calculated, and thus
improves bookkeeping  issues. The main challenge of this approach lies
in developing  an efficient algorithm for selecting  double resonant
topologies.

We would like to note here, that both the bottom-up and the top-down
approaches are of course equivalent and should provide the same final
answer.  Different MC programs have, however, distinct internal
structures. Given the level of complexity of such programs the
approach that requires the minimal amount of structural changes in the
code is the one to be incorporated.  Because \textsc{Helac-Nlo} is
designed to efficiently obtain one-loop helicity amplitudes and
calculate total cross sections at NLO in QCD for multi-particle
processes in the SM the top-down approach is more natural and
conceptually easier to implement. Furthermore, it allows us to exploit
highly optimised recursion algorithms that have already been employed
to provide NLO QCD results for $t\bar{t}$, $t\bar{t}j$,
$t\bar{t}\gamma$ and $t\bar{t}Z$ processes with the complete top quark
off-shell effects included \cite{Bevilacqua:2010qb,Bevilacqua:2015qha,
  Bevilacqua:2016jfk,Bevilacqua:2017ipv,Bevilacqua:2018woc,Bevilacqua:2019cvp}.

%
\subsection{Virtual Corrections}
%

\begin{figure}[t]
\begin{center}
  \includegraphics[width=1.0\textwidth]{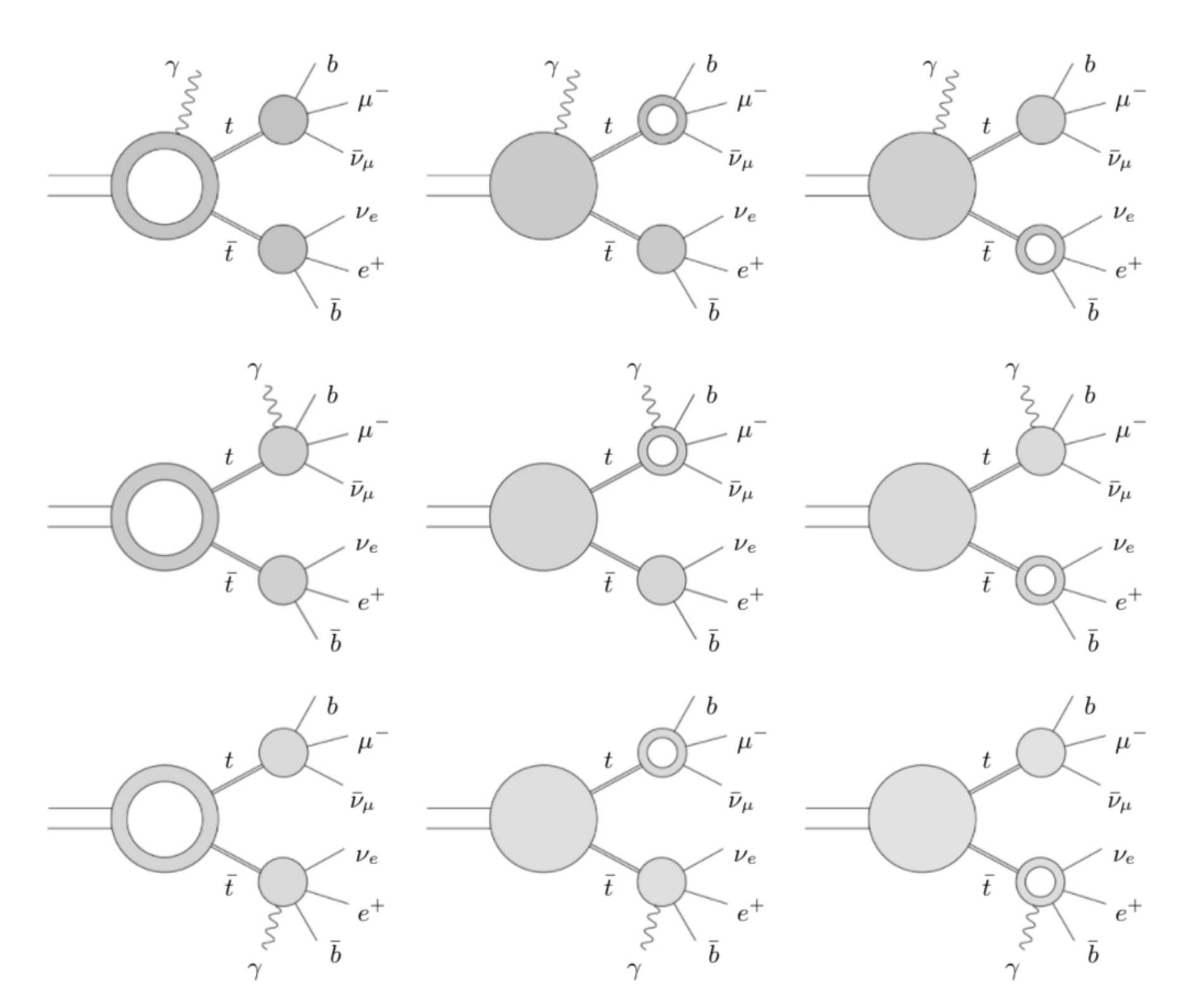}
\end{center}
\caption{\it Schematic representation of one-loop contributions for
the $t\bar{t}\gamma$ production process in the NWA. For simplicity
only top quarks are treated in NWA. The three rows show the three
contributions from  Eq.~\eqref{Eq:NWA_ttA}. The full blobs
represent tree-level sub-amplitudes whereas the blobs with a hole
denote sub-amplitudes with one-loop corrections included.}
\label{Fig:NWA_oneloop}
\end{figure}

From the point of view of the virtual corrections the implementation
of the NWA into the \textsc{Helac-Nlo} framework does not introduce
additional complications. Again the biggest challenge comprises the
efficient selection of the topologies that correspond to factorisable
one-loop contributions.  Schematic representation of one-loop
contributions to the $t\bar{t}\gamma$ production process in the NWA
is shown in Figure~\ref{Fig:NWA_oneloop}. For simplicity we first
assume that only top quarks are treated in the NWA. Consequently,  the
following three contributions  are only considered
\begin{equation}
  \begin{split}
pp & \to  t \bar{t} \gamma \to \left(\, b \, e^+ \nu_e \,\right) \left(\, \bar{b} \,
\mu^- \bar{\nu}_\mu \, \right) \, \gamma \,, \\[0.2cm]
pp & \to
 t \bar{t} \to \left(\, b \, e^+ \nu_e \gamma \, \right) \left(\, \bar{b} \, \mu^-
 \bar{\nu}_\mu \,\right) \,, \\[0.2cm]
 pp & \to  t \bar{t} \to
\left(\, b \, e^+ \nu_e \,\right) \left(\, \bar{b} \, \mu^- \bar{\nu}_\mu \gamma
\,\right) \,,
\end{split}
\label{Eq:NWA_ttA}
\end{equation}
where the parenthesis denotes the resonant structure. The one-loop
amplitudes for each of the three processes listed above contain
corrections to both production and decays. In the case where also the
$W$ gauge boson is treated in the NWA the list of the contributions to
be considered increases and  is given by
\begin{equation}
  \begin{split}
pp & \to  t \bar{t} \gamma \to \left(\,b W^+\right) \,
           \left(\,\bar{b} W^-\,\right) \,
\gamma \to \left[\,b \, (e^+ \nu_e)\, \right] \, \left[\,\bar{b} \, (\mu^-
  \bar{\nu}_\mu)\, \right] \, \gamma \\[0.2cm]
pp & \to  t
\bar{t} \to \left(\,b W^+ \gamma\,\right) \, \left(\,\bar{b} W^-
                                                                       \right) \to
                                                                       \left[\,b \, (e^+
\nu_e) \,\gamma\,\right] \, \left[\,\bar{b} \, (\mu^-
\bar{\nu}_\mu)\,\right] \\[0.2cm]
pp & \to  t \bar{t} \to
\left(\,b W^+\,\right) \, \left(\,\bar{b} W^- \gamma\,\right) \to \left[\,b \,
  (e^+ \nu_e)\,\right] \,
\left[\,\bar{b} \, (\mu^- \bar{\nu}_\mu) \,\gamma\,\right] 
\\[0.2cm]
pp & \to  t \bar{t} \to \left(\,b W^+\, \right) \, \left(\,\bar{b}
  W^-\, \right) \to \left[\,b
\, (e^+ \nu_e \gamma)\, \right] \, \left[\,\bar{b} \, (\mu^-
\bar{\nu}_\mu)\,\right]
\\[0.2cm]
pp & \to  t \bar{t} \to
\left(\,b W^+\,\right) \, \left(\,\bar{b} W^-\,\right) \to \left[\,b
  \, (e^+ \nu_e )\,\right] \, \left[\,\bar{b} \, (\mu^- \bar{\nu}_\mu
  \gamma)\,\right]
\end{split}
\end{equation}
The description outlined above is fully automated and can be used to
tackle the calculation of more complicated processes for which the
full NWA predictions have not yet been computed.

%
\subsection{Real Corrections}
%

\begin{figure}[t]
\begin{center}
  \includegraphics[width=1.0\textwidth]{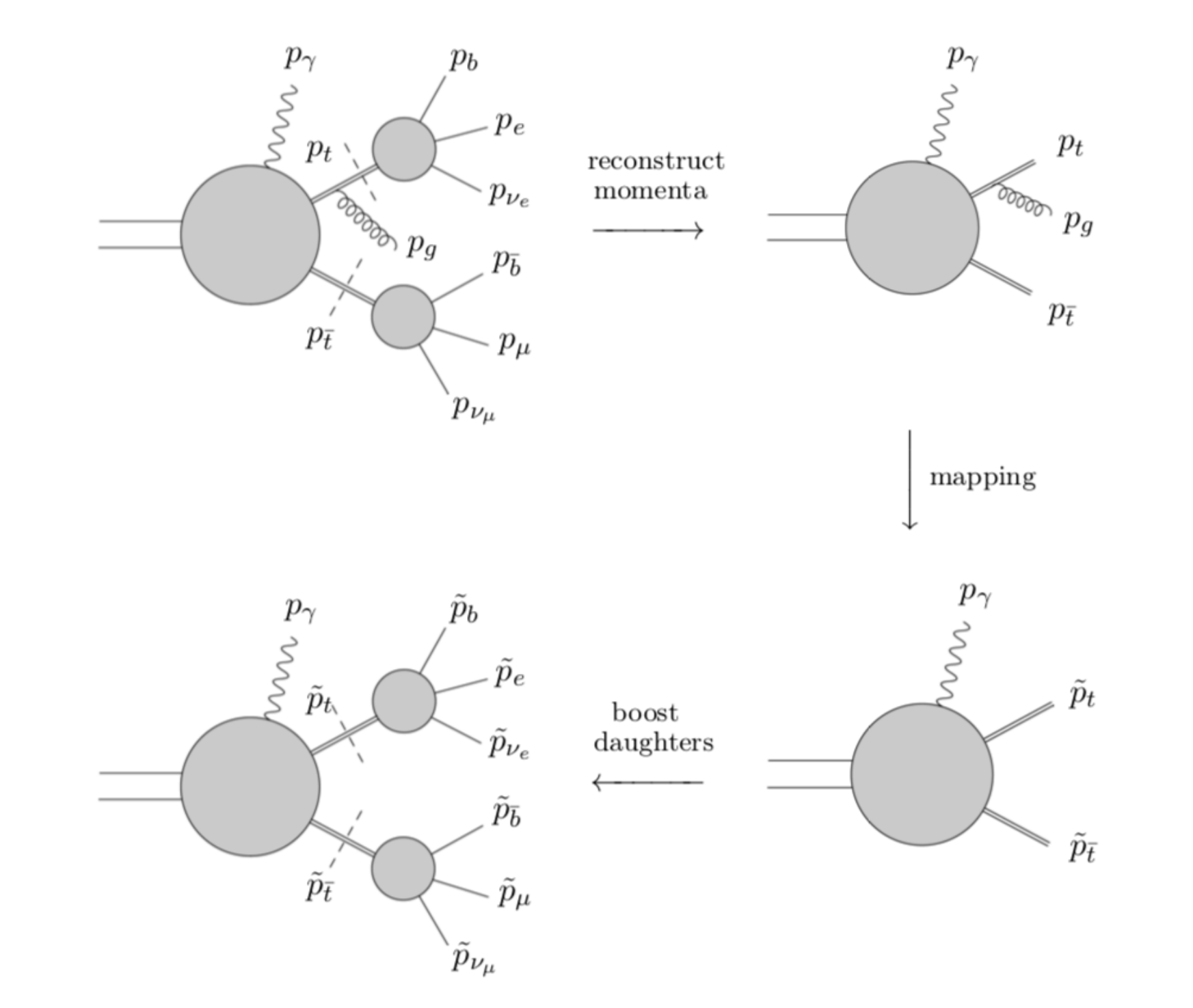}
\end{center}
\caption{\it The phase-space mapping as applied to an intermediate
  emitter. The shown example refers to the case where the emitter is
  the top quark and the spectator is the anti-top quark.}
\label{fig:ttdecay}
\end{figure}

The cancellation of soft and collinear radiation contributions in the
NWA is performed using the Catani-Seymour dipole subtraction scheme
\cite{Catani:1996vz}. Specifically, the formulation presented in
Ref.~\cite{Catani:2002hc} for massive quarks has been used with the
extension to arbitrary helicity eigenstates of the external partons
\cite{Czakon:2009ss}, as implemented in \textsc{Helac-Dipoles}. In
order to deal with gluon radiation in the production and decays of the
on-shell top quarks some modifications have been made that will be
outlined briefly below. For example,  in order to deal with gluon
radiation in the decay stage of the $t\bar{t}\gamma$ production
process we have implemented a modified version of the specialised
subtraction procedure introduced in Ref.~\cite{Campbell:2004ch} for
single top quark production.

At the production level, gluon radiation off the top quarks generates
extra soft divergences that need to be cancelled.  This is in
contrast to the full calculation where intermediate top quark
propagators are not affected by infrared divergences. Thus, we
include the resonant top quarks into the list of emitters and compute
the corresponding dipoles for the cases of final-state and
initial-state spectators. The Catani-Seymour mapping is applied to the
momentum entering the resonant propagator that we labelled as
$p_t$. The latter is reconstructed from its decay products, see
Figure~\ref{fig:ttdecay} for the graphical representation of the
mapping. This mapping must then be carried out onto the decay
products. Since the momenta of the decay products and the intermediate
particle are subject to the mass-shell constraints, the new momenta
can be obtained with a Lorentz transformation. This (ambiguous)
Lorentz transformation is constructed as the product of two boosts,
based on the observation that the momenta of the intermediate particle
before and after the mapping are the same when expressed in their
center-of-mass (CM) frames 
\begin{equation}
\label{Eq:CMframe}
p_t^{\rm CM} = \tilde{p}_t^{\, \rm CM} =
\left( m_t , \vec{0} \right) \,.
\end{equation}
Here $p_t^{\rm CM} \equiv \Lambda \,p_t$ and $\tilde{p}_t^{\,\rm CM}
\equiv \tilde{\Lambda} \,\tilde{p}_t$ indicate the Lorentz
transformations that bring the corresponding momenta into their CM
frames and  a tilde represents the mapped  top quark momentum. From
Eq. \eqref{Eq:CMframe} we obtain
\begin{equation}
  \tilde{p}_t = \left(
    \tilde{\Lambda}^{-1} \Lambda \right) \,p_t\,.
\end{equation}
This Lorentz transformation is applied again into the momenta of the
decay products to construct the mapped final states. Even though the
mapping is not conceptually different from the standard Catani-Seymour
implementation, the form of the new subtraction terms is not so
trivial. Because the top quark propagator is already summed up over
its polarisation states it seems that one cannot just use the
polarised formulae from Ref.~\cite{Czakon:2009ss}. The divergence,
however, has a pure soft nature, so it is independent of the top quark
polarisations. Consequently, standard, non-polarised Catani-Seymour
dipole can be used with an additional symmetry factor of $1/2$ which
compensates for the summation over the two polarisations of the gluon.

In the following we shall shortly discuss the treatment of gluon
radiation in top quark decays. The Catani-Seymour subtraction scheme
can be extended straightforwardly to top quark decays. In
Ref.~\cite{Campbell:2004ch} such extension, that preserves the
momentum of the decaying particle and it is therefore applicable to
the case of top quark decays, has been proposed. It has been later
extended in Ref.~\cite{Melnikov:2011ta} to the case of radiative
decays. We employ the scheme of Ref.~\cite{Campbell:2004ch} to
construct subtraction terms for the final-initial case where initial
state here means the decaying top quark. The latter are available in
the literature only in the unpolarised form. Thus, we have derived the
extension of the subtraction term of Ref.~\cite{Campbell:2004ch} to
the polarised case which, for massless $b$ quarks, reads
\begin{equation}
  \label{Eq:NWA_finalinitial}
\begin{split}
 & D\left(\,(p_t+p_g)^2, (p_b+p_g)^2,
   m_t^2 \,\right)_{\lambda\lambda^\prime\lambda_b\lambda_g}
 =  \\[0.2cm]
 & g^2 \mu^{2\epsilon} C_F \left[
\frac{1}{p_b \cdot p_g} \left( \frac{z^2}{(1-z)} +
\delta_{\lambda_b\lambda_g}(1 + z) \right) -
\frac{1}{2}\frac{m_t^2}{\left(p_t \cdot p_g\right)^2} \right]
\delta_{\lambda\lambda_b}\delta_{\lambda\lambda^\prime} \,.
\end{split}
\end{equation}
Here $\lambda_b, \lambda_g$ are the helicity eigenstates of the
external $b$-quark and gluon respectively and $\lambda,
\lambda^\prime$ are the helicity eigenstates that enter the Born
matrix element. The rest of the notation follows the original
reference. We emphasise that the modifications to the subtraction
terms described above do not affect the analytical structure of the
integrated dipoles. We  used the formulae that were  already
available in the literature without making any additional changes.  We
note here that all other cases have been addressed with the standard
implementation of the Catani-Seymour subtraction scheme that has
already been available in the \textsc{Helac-Dipoles} software.

%
\subsection{Numerical Checks}
%

In order to test our implementation of the NWA in the
\textsc{Helac-Nlo} system we have performed a number of cross-checks.
First of all, the implementation of the real corrections has been
validated for the $pp\to t\bar{t}\gamma$ process in the di-lepton
channel by checking that the subtraction terms match the singular
behaviour of the matrix elements for increasingly collinear and soft
limits. We have explicitly checked the cancellation of the $\epsilon$
poles coming from loop contributions against those of the integrated
subtraction terms for a few phase space points. Additionally, we have
verified that our predictions for the real emission part do not
depend on the particular value of the $\alpha_{max}$ parameter. The
parameter, which controls the size of the subtraction region, has been
first proposed for the Catani-Seymour subtraction scheme in
Ref.~\cite{Nagy:1998bb,Nagy:2003tz}. Furthermore, we have performed
extensive checks of the NWA version of the \textsc{Helac-Nlo}
framework against other publicly available calculations.  The only
publicly available results for the $t\bar{t}\gamma$ production process
in full NWA comprise the  lepton plus jets top quark decay channel
\cite{Melnikov:2011ta}. Thus, we cannot compare to them
directly at the moment. Instead, we focused on the simpler process, i.e the
$t\bar{t}$ production process with full leptonic decays of the top
quarks. As a first step we have checked that our results for the fully
inclusive NLO cross section agree with the NLO part of the
calculation of Ref. \cite{Behring:2019iiv}. On a more exclusive
ground, we have reproduced the NLO differential cross section
distributions that are presented in
Ref. \cite{AlcarazMaestre:2012vp,Denner:2012mx}. Finally, we have
cross-checked our results with NLO QCD corrections included
separately for the production and the decay stages against the results
obtained with the public MC program \textsc{Mcfm}
\cite{Campbell:2012uf}.

%
\section{LHC Setup}
\label{setup}
%

We study $pp \to e^+ \nu_e \mu^- \bar{\nu}_\mu b\bar{b} \gamma +X$ for
the LHC Run II energy of $\sqrt{s}=13$ TeV at ${\cal O}(\alpha_s^3
\alpha^5)$.  Our calculation uses the following parton distribution functions
(PDFs): CT14 (the default PDF set) \cite{Dulat:2015mca}, NNPDF3.0
\cite{Ball:2014uwa} and MMHT14 \cite{Harland-Lang:2014zoa}.  The
number of active flavours is set to $N_F = 5$, however, contributions
induced by the bottom-quark parton density are neglected due to their
numerical insignificance.  We employ the following SM parameters
\begin{equation*}
\begin{array}{lll} m_{W}=80.385 ~{\rm GeV} \,, &\quad \quad \quad
\quad \quad \quad&\Gamma_{W} = 2.0988 ~{\rm GeV}\,, \vspace{0.2cm}\\
m_{Z}=91.1876 ~{\rm GeV} \,, &&\Gamma_{Z} = 2.50782 ~{\rm GeV}\,,
\vspace{0.2cm}\\ G_{ \mu}=1.166378 \times 10^{-5} ~{\rm GeV}^{-2}\,,&
&\sin^2\theta_W =1- m^2_W/m^2_Z\,.
\end{array}
\end{equation*}
Since leptonic $W$ gauge boson decays do not receive NLO QCD
corrections, to account for some higher order effects the NLO QCD
values for the gauge boson widths are used everywhere, i.e. for LO and
NLO matrix elements. The electroweak coupling is derived from the
Fermi constant $G_\mu$ according to
\begin{equation} \alpha= \frac{\sqrt{2} \, G_\mu m^2_W
\sin^2\theta_W}{\pi} \,.
\end{equation}
For the emission of the isolated photon, however, $\alpha_{\rm QED} =
1/137$ is used instead. The top quark mass is set to $m_t = 173.2$
GeV. All other QCD partons including $b$ quarks as well as leptons are
treated as massless.  The final state jets are constructed from the
final state partons $(j)$ with pseudo-rapidity $|\eta(j)| <5$ via the IR-safe
{\it anti}$-k_T$ jet algorithm \cite{Cacciari:2008gp} with $R=0.4$. We
require at least two jets for our process, of which exactly two must
be bottom flavoured jets. We impose the following cuts on the
transverse momenta and the rapidity of two recombined $b$-jets, which we
assume to be always tagged
\begin{equation}
  p_T(b) > 40 ~{\rm GeV} , \quad  \quad \quad \quad
  |y(b)| < 2.5 ,  \quad \quad \quad \quad  \Delta R(bb) > 0.4 \,.
\end{equation}  
The last cut, i.e. the separation between the $b$-jets, is implied by
the jet algorithm.  Furthermore, we request two charged leptons,
missing transverse momentum and an isolated hard photon.  The latter
is defined with $p_T(\gamma) > 25$ GeV and $|y(\gamma)| <
2.5$. To avoid QED collinear singularities from photon
emission that are introduced by the $q\to q\gamma$ splitting, a
separation between quark and photon is needed.  This in turn
introduces a separation between photons and gluons as all partons are
treated in the same way. Therefore, for a given $p_T(\gamma)$ an
angular restriction on the soft gluon emission  is
introduced compromising the cancelation of
infrared divergences.  To ensure infrared safety we use
the Frixione photon isolation prescription described in
Ref.~\cite{Frixione:1998jh}, which is based on a modified cone
approach. The (smooth) photon isolation condition is implemented in
the same way for quarks and gluons. Specifically, we reject the event
unless the following condition is fulfilled
\begin{equation}
\label{Frixione}
  \sum_i E_{T}  (i) \, \Theta \left(
    R - R (\gamma i) 
  \right) \le \epsilon_\gamma \, E_{T} (\gamma) \left(
    \frac{1-\cos(R)}{1-\cos(R (\gamma j)) }
    \right)^n \quad \quad \quad \forall R\le R(\gamma j)\,,
 \end{equation} 
where $E_{T}(i)$ stands for the transverse energy of the parton $i$
and $E_T(\gamma)$ is the transverse energy of the photon.  Because of
the $\Theta \left( R - R (\gamma i) \right) $ condition the sum gets
contributions only when the angular distance of the parton $i$ from the
photon is less than or equal to $R$.  We set the following values for
the parameters of the isolation criterion $\epsilon_\gamma = 1$, $n=
1$ and $ R(\gamma j) = 0.4$. Finally, $R (\gamma i)$ is given by
 \begin{equation}
   R (\gamma i)  = \sqrt{\Big(y(\gamma) -y(i) \Big)^2+
     \Big(\varphi(\gamma)      -\varphi(i) \Big)^2} \,,
 \end{equation}
where $y$ and $\varphi$ are the rapidity and azimuthal angle
respectively. In Eq.~\eqref{Frixione} the $(1-\cos(R))$ term
suppresses the collinear singularities, which are present for $R \to
0$. On the other hand, arbitrarily soft radiation is allowed inside
the $R(\gamma j)$ cone preserving the cancellation of infrared poles
in the calculation. Basic selection cuts are applied to charged
leptons to ensure that they are observed inside the detector and well
separated from each other
\begin{equation}
p_T(\ell) > 30 ~{\rm GeV}\,,  \quad \quad \quad  \quad \Delta R(\ell \ell) >
0.4\,, \quad \quad \quad \quad 
|y(\ell)| < 2.5\,,
\end{equation}  
where $\ell=e^+,\mu^-$. Moreover, we impose that charged leptons are
well separated from  the isolated photon  and from $b$-jets
\begin{equation}
\Delta R (\ell b) > 0.4\,,
\quad \quad \quad  \quad
\Delta R(\ell \gamma) > 0.4\,,
\quad \quad \quad  \quad \Delta R(b \gamma) > 0.4\,.
\end{equation}  
We additionally place a requirement on the missing transverse momentum
$p_T^{miss} > 20$ GeV. Finally, we impose no restrictions on the
kinematics of the extra (light) jet other than it should be separated
from the isolated photon.  In this work we have utilised two different
forms for the factorisation and renormalisation scales for the results
in the NWA: $\mu_0=\mu_R=\mu_F=m_t/2$ and $\mu_0=\mu_{R}=\mu_F
=H_T/4$. We show theoretical predictions for the NWA with the fixed
scale choice mostly for comparison. We note here, however, that the
fixed scale choice is still commonly used in various phenomenological
studies.  In the case of full off-shell results, on the other hand,
only the dynamical scale choice, $\mu_0=H_T/4$, is used since this is
our recommended scale choice for the process under consideration, see
Ref.~\cite{Bevilacqua:2018woc,Bevilacqua:2018dny}. The total
transverse momentum of the system, $H_T $, is defined according to
\begin{equation} H_T = p_T(e^+) + p_T(\mu^-) + p_T^{miss}+p_T(b_1) +
p_T(b_2) + p_T(\gamma)\,,
\end{equation}
where $p_T(b_1)$ and $p_T(b_2) $ are bottom-flavoured jets and
$p_T^{miss}$ is the missing transverse momentum from the two
neutrinos. Even though we assume  that the central scale
$\mu_R=\mu_F=\mu_0$ is the same for both the renormalisation and
factorisation scales the scale systematics  is evaluated by varying
$\mu_R$ and $\mu_F$  independently. Specifically, we vary them  in the
following range
\begin{equation}
  \frac{1}{2} \mu_0 \le  \mu_R,\mu_F\le 2\mu_0\,, \quad\quad
  \quad\quad
  \quad\quad 
  \frac{1}{2} \le \frac{\mu_R}{\mu_F}\le 2\,.
\end{equation}  
The top quark width, as calculated from
\cite{Jezabek:1988iv,Chetyrkin:1999ju,Denner:2012yc}, is taken to be
\begin{equation*}
\begin{array}{lll} \Gamma_{t, {\rm off-shell}}^{\rm LO} = 1.47848
~{\rm GeV}\,, &

  \quad \quad \quad \quad \quad \quad & \Gamma_{t, {\rm
off-shell}}^{\rm NLO} = 1.35159 ~{\rm GeV}\,, \vspace{0.2cm}\\
\Gamma^{\rm LO}_{t,{\rm NWA}} = 1.50176 ~{\rm GeV}\,, && \Gamma^{\rm
NLO}_{t,{\rm NWA}} = 1.37289 ~{\rm GeV}\,.
\end{array}
\end{equation*}
The value of $\alpha_s$ used for the top quark width $\Gamma^{\rm
NLO}_{t}$ calculation is taken at $\alpha_s(m_t)$.  This $\alpha_s$ is
independent of $\alpha_s(\mu_0)$ that goes into the matrix element and
PDF calculations. The latter is used to describe the dynamics of the
whole process, while the former only the top quark decays. For more details
we refer the reader to our previous publications
\cite{Bevilacqua:2018woc,Bevilacqua:2018dny}.

%
\section{Stability of the Full Off-shell Result}
\label{stability}
%

\begin{table}[t!]
\begin{center}
\begin{tabular}{cccccccc}
  \hline \hline
  &&&&&&&\\
  PDF &$p_{T}(b)$ & $\sigma^{\rm LO}$  [fb]  & $\delta_{scale}$ & $\sigma^{\rm 
NLO}$ [fb] & $\delta_{scale}$ & $\delta_{\rm PDF}$ & ${\cal K} $
 \\
  &&&&&&&\\
  \hline
  \hline
  &&&&&&&\\
  CT14 & $25$ & $10.68$ & $^{+3.54~(33\%)}_{-2.49~(23\%)}$ & $11.19$ & $^{+0.16~(1
\%)}_{-0.54~(5\%)}$ & $^{+0.32~(3\%)}_{-0.35~(3\%)}$ & $1.05$\\[0.2cm]
     & $30$ &  $9.58$ & $^{+3.18~(33\%)}_{-2.24~(23\%)}$ & $9.93$  & $^{+0.14~(1
\%)}_{-0.54~(5\%)}$ & $^{+0.28~(3\%)}_{-0.31~(3\%)}$ & $1.04$\\[0.2cm]
     & $35$ &  $8.44$ & $^{+2.80~(33\%)}_{-1.97~(23\%)}$ & $8.69$  & $^{+0.12~(1
\%)}_{-0.50~(6\%)}$ & $^{+0.25~(3\%)}_{-0.27~(3\%)}$ & $1.03$\\[0.2cm]
     & $40$ &  $7.32$ & $^{+2.45~(33\%)}_{-1.71~(23\%)}$ & $7.50$  & $^{+0.11~(1
\%)}_{-0.45~(6\%)}$ & $^{+0.22~(3\%)}_{-0.23~(3\%)}$ & $1.02$\\
  &&&&&&&\\
  \hline 
  \hline
  &&&&&&&\\
  MMHT14 & $25$ & $11.59$ & $^{+4.22~(36\%)}_{-2.88~(25\%)}$ & $11.29$ & $^{+0.16~
(1\%)}_{-0.57~(5\%)}$ & $^{+0.24~(2\%)}_{-0.22~(2\%)}$ & $0.97$\\[0.2cm]
  & $30$ & $10.38$ & $^{+3.78~(36\%)}_{-2.58~(25\%)}$ & $10.02$
      & $^{+0.13~(1\%)}_{-0.58~(6\%)}$ & $^{+0.22~(2\%)}_{-0.19~(2\%)}$ & $0.97$\\[0.2cm]
  & $35$ &  $9.12$ & $^{+3.33~(36\%)}_{-2.26~(25\%)}$ & $8.77$
      & $^{+0.11~(1\%)}_{-0.54~(6\%)}$ & $^{+0.19~(2\%)}_{-0.17~(2\%)}$ & $0.96$\\[0.2cm]
  & $40$ &  $7.90$ & $^{+2.89~(37\%)}_{-1.96~(25\%)}$ & $7.57$
      & $^{+0.09~(1\%)}_{-0.48~(6\%)}$ & $^{+0.16~(2\%)}_{-0.15~(2\%)}$ & $0.96$\\
  &&&&&&&\\
  \hline
  \hline
  &&&&&&&\\
  NNPDF3.0 & $25$ & $10.78$ & $^{+3.82~(35\%)}_{-2.62~(24\%)}$ & $11.62$
      & $^{+0.17~(1\%)}_{-0.58~(5\%)}$ & $^{+0.16~(1\%)}_{-0.16~(1\%)}$ & $1.08$\\[0.2cm]
  & $30$ &  $9.65$ & $^{+3.42~(35\%)}_{-2.34~(24\%)}$ & $10.31$
      & $^{+0.14~(1\%)}_{-0.58~(6\%)}$ & $^{+0.14~(1\%)}_{-0.14~(1\%)}$ & $1.07$\\[0.2cm]
  & $35$ &  $8.48$ & $^{+3.01~(35\%)}_{-2.05~(24\%)}$ & $9.02$
      & $^{+0.12~(1\%)}_{-0.53~(6\%)}$ & $^{+0.12~(1\%)}_{-0.12~(1\%)}$ & $1.06$\\[0.2cm]
  & $40$ &  $7.34$ & $^{+2.61~(36\%)}_{-1.78~(24\%)}$ & $7.79$
      & $^{+0.10~(1\%)}_{-0.48~(6\%)}$ & $^{+0.11~(1\%)}_{-0.11~(1\%)}$ & $1.06$\\
  &&&&&&&\\
  \hline
  \hline
\end{tabular}
\end{center}
\caption{ \label{tab:LO_NLO_HT}\it
Integrated cross sections for the $pp\to e^+\nu_e \mu^-
  \bar{\nu}_\mu b \bar{b} \gamma +X$ production process at the LHC
  with $\sqrt{s}=13$ TeV. Results are evaluated using $\mu_R = \mu_F =
  \mu_0 = H_T/4$ for three different PDF sets and four different
  $p_T(b)$ cuts for the  $b$-jets. Also given are theoretical
  uncertainties coming from scale variation, $\delta_{scale}$, and
  from PDFs, $\delta_{PDF}$. In the last column a ${\cal K}$-factor,
  that is defined as ${\cal K}=\sigma^{\rm NLO}/\sigma^{\rm LO}$, is
  shown.}
\end{table}
%
Before comparing various approaches for the top quark decay modelling
for the $pp\to e^+\nu_e \mu^-\bar{\nu}_\mu b\bar{b}\gamma$ process,
we first investigate the stability of the full off-shell results.
Because they constitute the most realistic NLO
computation for top quark pair production with an isolated photon in
hadronic collision in the di-lepton top quark decay channel we would
like to analyse  them more precisely. In Table~\ref{tab:LO_NLO_HT} we
present LO and NLO QCD predictions for the integrated cross sections
for three PDF sets and for different values of the cut on the
transverse momentum of the $b$-jet. We vary the cut in the range of
$25-40$ GeV in steps of $5$ GeV. The values of $\sigma^{\rm LO}$
and $\sigma^{\rm NLO}$ are evaluated using $\mu_0=H_T/4$. Theoretical
uncertainties coming from scale variation are denoted by
$\delta_{scale}$ and from PDFs as $\delta_{\rm PDF}$.  Finally, in the
last column the ${\cal K}$-factor, defined as ${\cal
K}=\sigma^{\rm NLO}/\sigma^{\rm LO}$, is shown. The scale dependence
is derived with the standard seven-point variation around the central
value of the scale $\mu_0$ and indicated by the upper and lower
indices. They correspond to the minimum and maximum of the resulting
cross sections. For the PDF uncertainties we use the corresponding
prescription from each group to provide the $68\%$ confidence level
(C.L.) PDF uncertainties. Both CT14 PDFs and MMHT14 PDFs include a
central set and error sets in the Hessian representation. In that case
we use the asymmetric expression for PDF uncertainties. Additionally,
the CT14 errors are rescaled by a factor $1/1.645$ since they are
provided at $90\%$ C.L. On the other hand, for the NNPDF3.0 PDF sets
PDF uncertainties are obtained using the replicas method.

We can observe that the scale uncertainty is reduced considerably
through the inclusion of the NLO QCD corrections. PDF uncertainties
are of the order of a few \%.  These findings do not take into account
additional systematics coming from the underlying assumptions that
enter the parametrisation of different PDF sets. They simply cannot be
quantified within a given scheme. We therefore additionally present
results for other PDF sets. All three are recommended to be used for
applications at the LHC Run II \cite{Butterworth:2015oua}. We see that
CT14, MMHT14 and NNPDF3.0 NLO results differ at most by $4\%$, which
is comparable to the individual estimates of PDF systematics. Overall,
the PDF uncertainties for the process under scrutiny are below the
theoretical uncertainties due to the scale dependence, which remain
the dominant source of the theoretical systematics.

The most important message here, however, is that the above findings
for $\delta_{scale}$ are rather insensitive to the chosen $p_T(b)$ cut
value.  We could uncover only variations at the percent level.  In
particular, there is no big difference for the lowest cut value of the
$p_T(b)$ cut, $25$ GeV,  and the value we use as the default one in our
analysis, i.e. $40$ GeV.  This suggests that the perturbative
expansion is not spoiled by the appearance of large logarithms, thus,
under excellent theoretical control. Finally, even though NLO QCD
corrections for different PDF sets and for all values of the $p_T(b)$
cut are varying between positive or negative ones, they are all
consistently below $10\%$.

Having established stability of the full off-shell results with
respect to the $p_T(b)$ cut we move on to the main part of the paper
and investigate differences between full off-shell results and the
calculations in the NWA.

%
\section{Phenomenological Results for  Integrated Cross Sections}
\label{integrated}
%

\begin{table}[t!]
  \begin{center}
\begin{tabular}{lcc}
  \hline \hline
  &&\\
  \textsc{Modelling Approach} & $\sigma^{\rm LO}$ [{\rm fb}]
                              & $\sigma^{\rm NLO}$ [{\rm  fb}]
  \\[0.2cm]
  \hline \hline
  &&\\
 full off-shell $(\mu_0=H_T/4)$ & ${7.32}^{+2.45\, (33\%)}_{-1.71\, (23\%)}$
 & ${7.50}^{+0.11\,(1\%)}_{-0.45\, (6\%)}$  \\[0.2cm]\hline\hline
 &&\\ 
        NWA  $(\mu_0=m_t/2)$ &
                               ${8.08}^{+2.84\,(35\%)}_{-1.96\,(24\%)}$
                              & ${7.28}_{-0.03\,(0.4\%)}^{-0.99\,(13\%)}$   \\[0.2cm]
        NWA  $(\mu_0=H_T/4)$ &
                               ${7.18}^{+2.39\,(33\%)}_{-1.68\,(23\%)}$
                              & ${7.33}_{-0.24\,(3.3\%)}^{-0.43\,(5.9\%)}$
  \\[0.2cm]
   \hline \hline
  &&\\
  NWA${}_{\gamma-{\rm prod}}$ $(\mu_0=m_t/2)$
  & ${4.52}^{+1.63\,(36\%)}_{-1.11\,(24\%)}$  &
                                                ${4.13}_{-0.05\,(1.2\%)}^{-0.53\,(13\%)}$
  \\[0.2cm]
  NWA${}_{\gamma-{\rm prod}}$ $(\mu_0=H_T/4)$
  & ${3.85}^{+1.29\,(33\%)}_{-0.90\,(23\%)}$  &
    ${4.15}_{-0.21\,(5.1\%)}^{-0.12\,(2.3\%)}$   \\[0.2cm]
   \hline \hline
  &&\\
  NWA${}_{\gamma-{\rm decay}}$ $(\mu_0=m_t/2)$
  & ${3.56}^{+1.20\,(34\%)}_{-0.85\,(24\%)}$  &
                                                ${3.15}_{+0.03\,(0.9\%)}^{-0.46\,(15\%)}$
  \\[0.2cm]
  NWA${}_{\gamma-{\rm decay}}$ $(\mu_0=H_T/4)$
  & ${3.33}^{+1.10\,(33\%)}_{-0.77\,(23\%)}$  &
                                                ${3.18}^{-0.31\,(9.7\%)}_{-0.03\,(0.9\%)}$
  \\[0.2cm]
   \hline \hline
  &&\\
  NWA${}_{\rm LOdecay}$ $(\mu_0=m_t/2)$  &
                             & ${4.85}^{+0.26\,(5.4\%)}_{-0.48\,(9.9\%)}$   \\[0.2cm]
        NWA${}_{\rm LOdecay}$ $(\mu_0=H_T/4)$  &
                                                 & ${4.63}^{+0.44\,(9.5\%)}_{-0.52\,(11\%)}$\\[0.2cm]
 \hline     \hline                                             
\end{tabular}
\end{center}
\caption{\label{tab:integarted}\it Integrated cross sections for the
$pp\to e^+\nu_e \mu^- \bar{\nu}_\mu b \bar{b} \gamma +X$ production
process at the LHC with $\sqrt{s}=13$ TeV. Results for various
approaches for the modelling of top quark decays are listed. Also
given is the full off-shell result. We additionally
provide theoretical uncertainties as obtained from the scale
dependence. The CT14 PDF set is employed.}
\label{tab:moddeling}
\end{table}

In the following we compare the full off-shell results with the
calculations in the NWA.  In the latter case two versions will be
examined: the full NWA and the NWA${}_{\rm LOdecay}$ (NWA with LO decays of
top quarks and photon radiation in the production stage
only). Theoretical predictions for these three cases, that have been
evaluated for the choice of the kinematic cuts and SM parameters as
described in the previous Section, are listed in
Table~\ref{tab:integarted}. We additionally provide theoretical
uncertainties from the scale dependence. Moreover, all
results are evaluated with the CT14 PDF sets. We note here, that
results for top quark pair production and also for $t\bar{t}$ process
with an additional jet or the gauge boson in the full NWA are usually
provided as a consistent expansion in $\alpha_s$. More explicitly, the
NLO top quark decay width, that appears as $\Gamma_{t, {\rm
NLO}}^{-2}$ and is a part of the branching fractions for the
corresponding $t\,(\bar{t})$ decay, has usually been expanded in
powers of $\alpha_s$. For the comparison at hand, however, such
expansion has not been performed and $\Gamma_{t, {\rm NLO}}^{-2}$ in
the theoretical prediction in the NWA is valid to all orders in the
strong coupling constant. The reason for not using this expansion for
the results in the NWA should be clear, namely such a procedure can
not be directly applied to the full off-shell calculations. Because
the main purpose of the paper is a consistent comparison between the
NWA and the full off-shell results such approach seems to be more
appropriate. Nevertheless, we have checked that the difference
between the expanded and unexpanded full NWA results is at the level of
$1\%$ independently of the scale choice.
  
We first assess the size of the non-factorisable corrections for our setup.
Finite top quark width effects change the NLO cross section
by less than $3\%$ independent of the scale choice. This finding is
consistent with the expected uncertainty of the NWA that is of the
order of ${\cal O}({\Gamma_t/m_t})\approx 0.8\%$. At LO we have
received $2\%$ corrections for $\mu_0=H_T/4$ and $10\%$ for
$\mu_0=m_t/2$. We note, however, that should we compare the LO NWA
result with $\mu_0=m_t/2$ to the full off-shell one with the same
scale choice we would also get only $2\%$ corrections. We would like
to add at this point, that in Ref.~\cite{Bevilacqua:2018woc} the size of the top
quark non-factorisable corrections has been estimated for the $pp\to
e^+ \nu_e \mu^- \bar{\nu}_\mu b\bar{b} \gamma$ process from the full
off-shell result by rescaling the coupling of the top quark to the $W$
boson and the $b$ quark as well as the coupling of $W$ and the leptons
by several large factors, as described in
Ref.~\cite{Bevilacqua:2010qb}. This approach should mimic the
$\Gamma_t \to 0$ limit when the scattering cross section factorises into
on-shell production and decay. Indeed, using this method we reported
$1.5\%$ for LO and $2.5\%$ for NLO with $\mu_0$ set to
$m_t/2$. Our current findings confirm that rescaling works very
accurately for the process at hand where rather inclusive  cuts on the
final states have been applied. 

In Table~\ref{tab:integarted} we additionally quote results for the LO
and NLO QCD cross sections where photon radiation occurs either in the
production $(pp\to t\bar{t}\gamma)$ or in the decay stage
($t\to b W^+\gamma\to b e^+\nu_e \gamma$,  
$t\to b W^+\to b e^+\nu_e \gamma$,  
$\bar{t}\to \bar{b} W^- \gamma \to \bar{b} \mu^- \bar{\nu}_\mu \gamma$,
$\bar{t}\to \bar{b} W^- \to \bar{b} \mu^- \bar{\nu}_\mu \gamma$) 
processes. In this way we can
estimate the importance of the photon emission in top quark decays.
Using results from Table~\ref{tab:integarted}, we have calculated that
at NLO more than $57\%$ of photons are emitted either from the initial
state light quarks or off-shell top quarks that afterwards go
on-shell \footnote{~At the central value of the scale $gg$ channel
dominates the total LO $pp \to e^+\nu_e \mu^- \bar{\nu}_\mu b \bar{b}
\gamma$ cross section by $79\%$ and it is followed by the
$q\bar{q}+\bar{q}q$ channel with $21\%$.}. This means that $43\%$ of
all isolated photons are emitted in the decay stage, i.e. either from
on-shell top quarks or its decay products ($b$-jets, $W$ gauge bosons
and/or charged leptons).  This conclusion is independent of the scale
choice. Similar estimates can be obtained at LO.  Consequently, the
radiative decay of the top quark must be incorporated into theoretical
predictions for $t\bar{t}\gamma$ production at the LHC since it yields
a significant contribution to the cross section. Once the $p_T(b)$ cut
is lowered to $25$ GeV, i.e. to the value that is currently used in
measurements of inclusive and differential cross-sections of
$t\bar{t}\gamma$ production in the $e\mu$ channel at $13$ TeV by the
ATLAS collaboration \cite{ATLAS:2019gkg}, the photon contribution in
the decay stage increases up to almost $50\%$. This means that photon
radiation is distributed evenly between the $t\bar{t}\gamma$
production process and the leptonic top quark decay stages.

Table~\ref{tab:integarted} also shows results for the special
case of the NWA, i.e. for the NWA${}_{\rm LOdecay}$. They comprise NLO
QCD corrections to the production of $t\bar{t}\gamma$ and LO top quark
decays. Furthermore, photon radiation is restricted only to the production
stage . Such a prediction should mimic the computation of Ref.~\cite{Kardos:2014zba}
where the NLO QCD corrections are computed for
the $t\bar{t}\gamma$ production  stage but include neither exact LO spin
correlations nor radiative corrections to decays. On top of it, Ref.~\cite{Kardos:2014zba}
omits photon emission in the parton shower evolution.  Because the
contribution from photon emission in top quark decays is large and NLO
QCD corrections to decays are also relevant it is not surprising that
NWA${}_{\rm LOdecay}$ result can not reproduce the correct
normalisation. The discrepancy to the
NWA approach amounts to $50\% \, (58\%)$ for $\mu_0=m_t/2\,
(\mu_0=H_T/4)$. NLO QCD corrections to the top quark decays are
negative and at the level of $17\%$ $(12\%)$ when $\mu_0=m_t/2\,
(\mu_0=H_T/4)$ is employed in the NWA.

Finally, in Table~\ref{tab:integarted} theoretical uncertainties as
obtained from the scale dependence are provided for all cases that we
have considered up until now. When comparing the full off-shell case
with the full NWA one at NLO in QCD we observe that theoretical
uncertainties are not underestimated when the NWA is employed.
Instead, they are consistent at the level of $6\%$ for $\mu_0=H_T/4$
and $13\%$ for $\mu_0=m_t/2$, see Ref.~\cite{Bevilacqua:2018woc} for
the full off-shell results at NLO in QCD with the fixed scale choice.

%
\section{Phenomenological Results for Differential Cross Sections}
\label{differential}
%
%
%
\subsection{Off-shell vs On-shell Top Quark Modelling}
%
%
\begin{figure}[t!]
\begin{center}
 \includegraphics[width=0.49\textwidth]{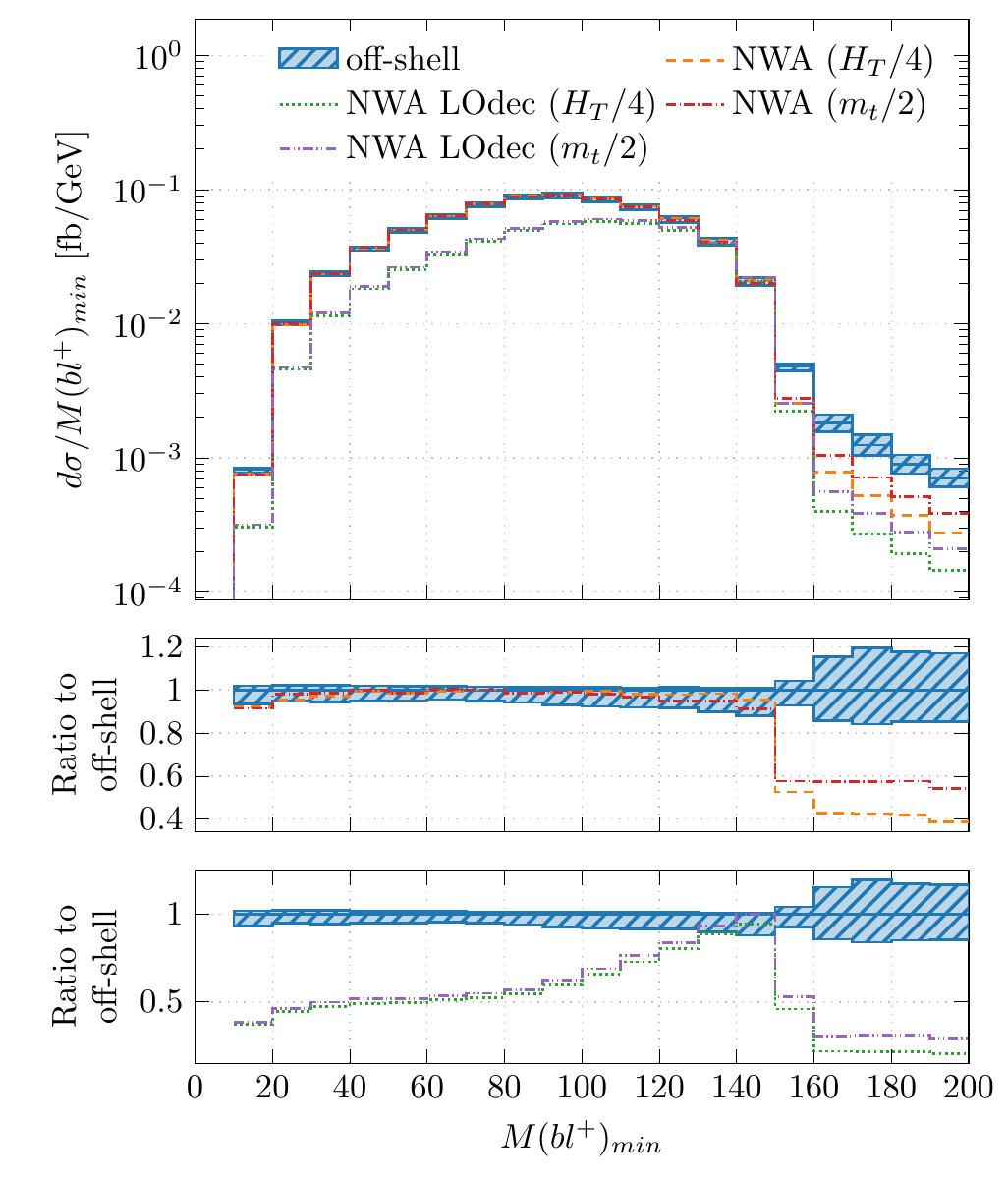}
  \includegraphics[width=0.49\textwidth]{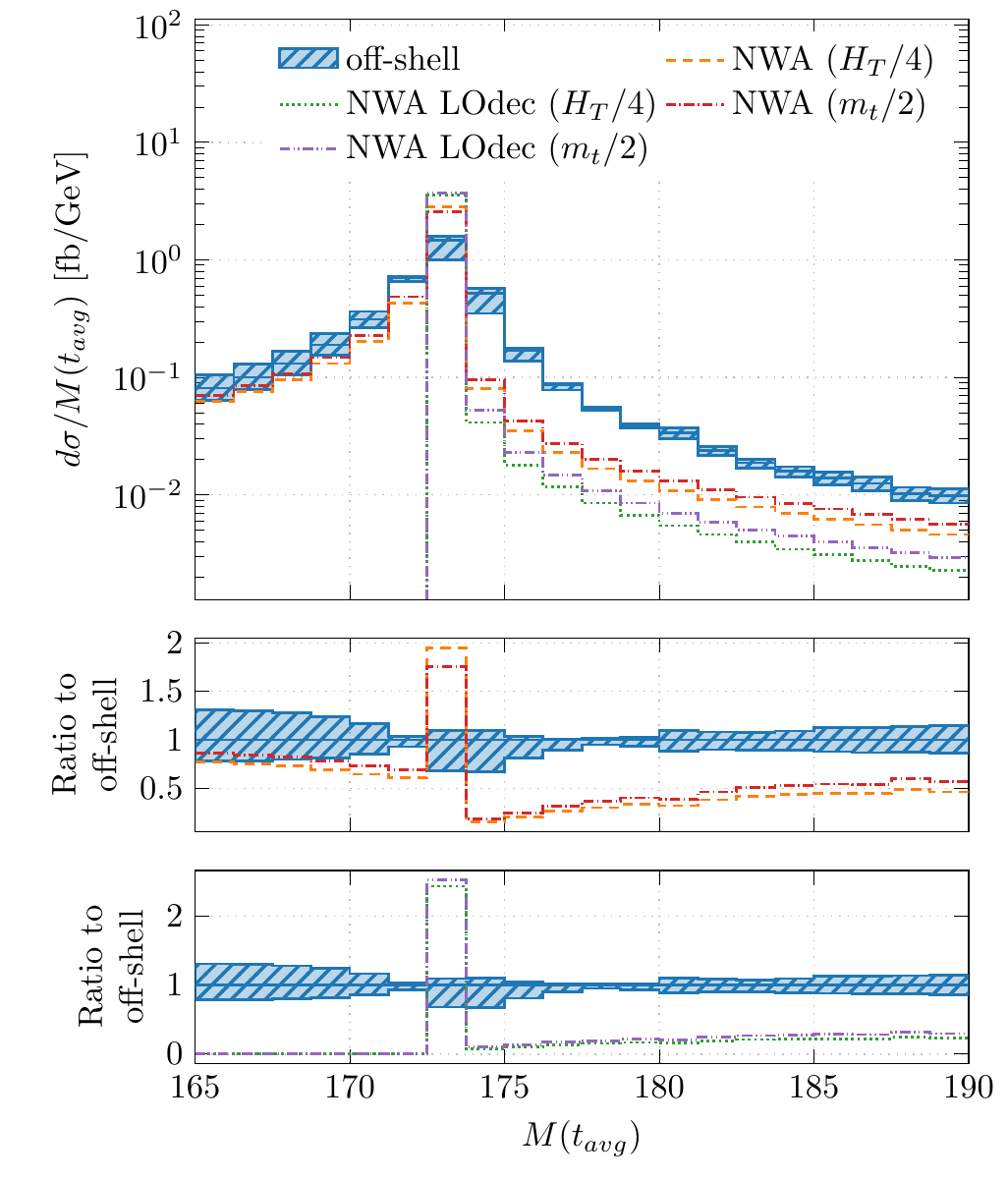}
 \end{center}
 \caption{\it
Differential cross section distribution as a function of the minimum
invariant mass of the positron and bottom-jet, $M(be^+)_{\rm min}$,
and the (averaged) invariant mass of the top quark, $M(t_{\rm avg})$,
for the $pp\to e^+\nu_e \mu^- \bar{\nu}_\mu b\bar{b} \gamma$ process
at the LHC run II with $\sqrt{s}=13$ TeV. The CT14 PDF set is
employed.}
\label{fig:offshell1}
\end{figure}
\begin{figure}[t!]
\begin{center}
  \includegraphics[width=0.49\textwidth]{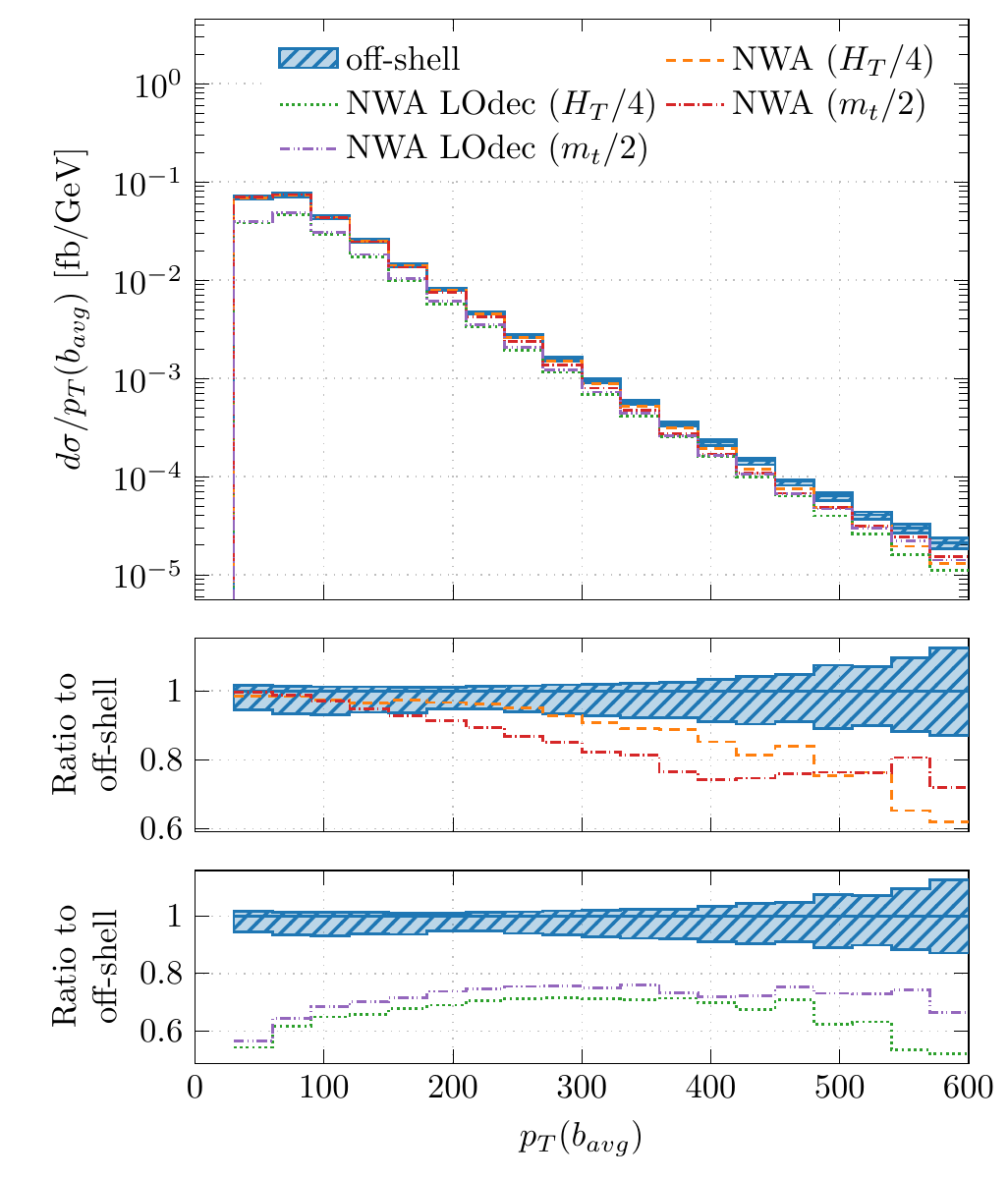}
  \includegraphics[width=0.49\textwidth]{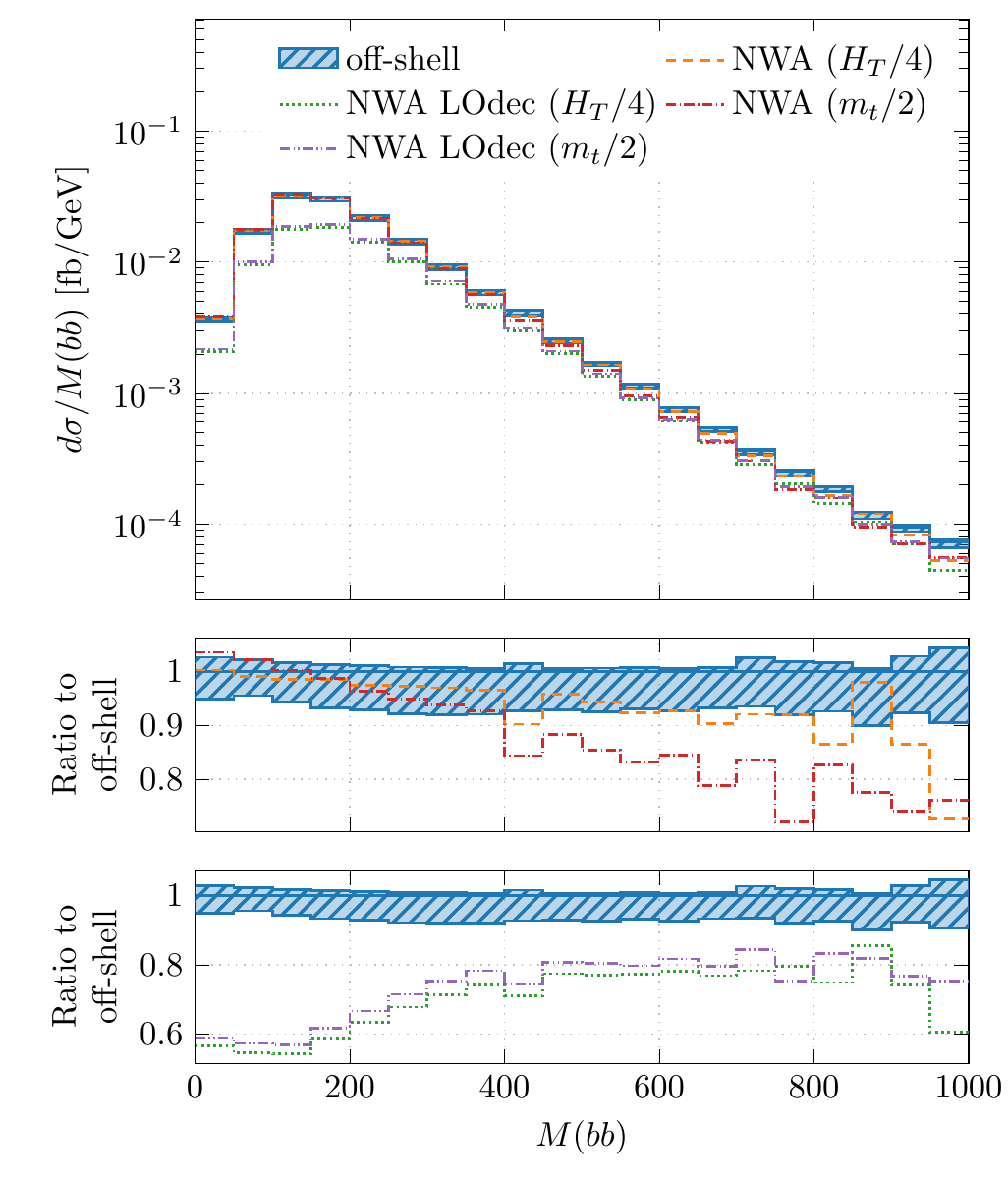} 
  \includegraphics[width=0.49\textwidth]{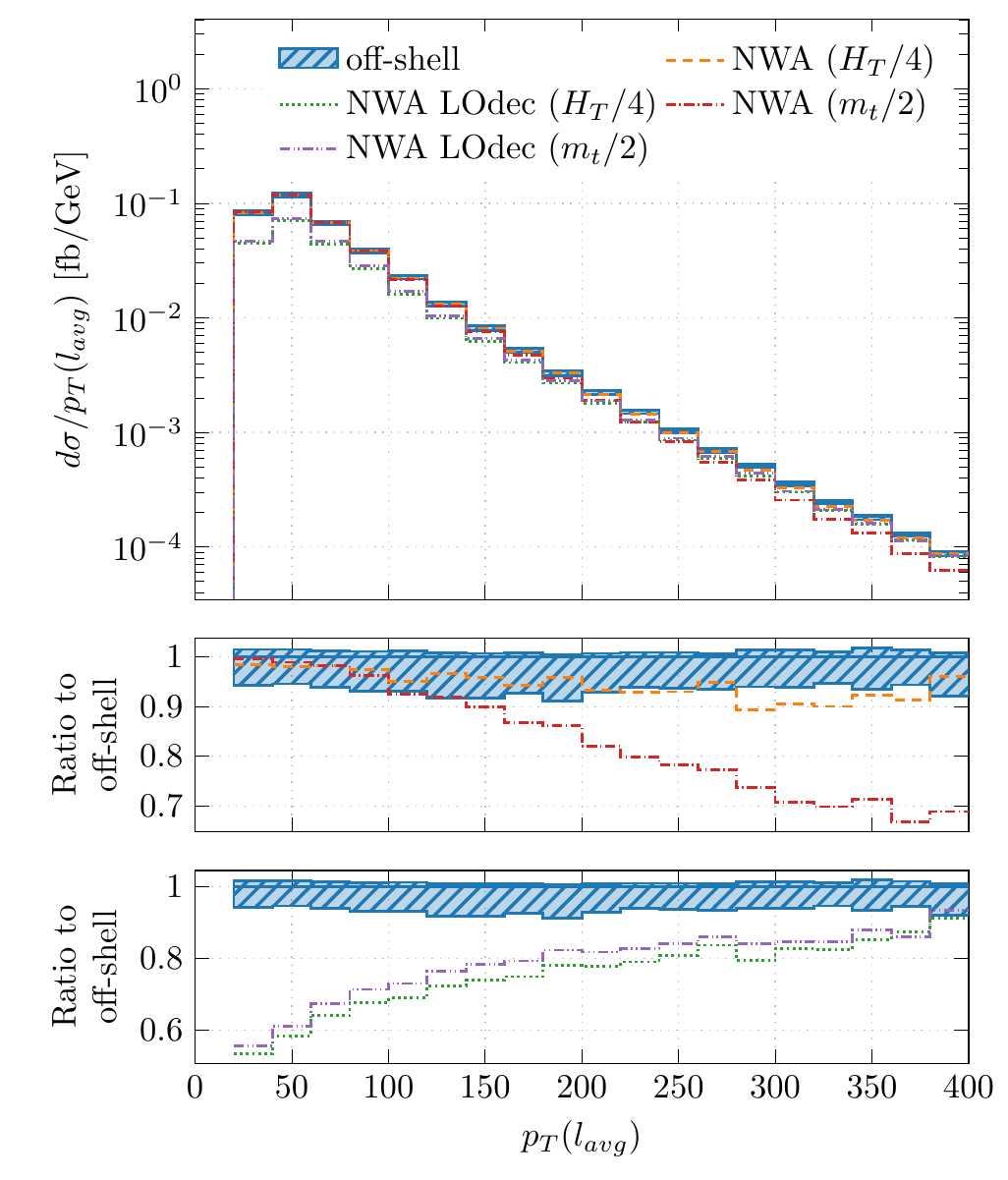}
  \includegraphics[width=0.49\textwidth]{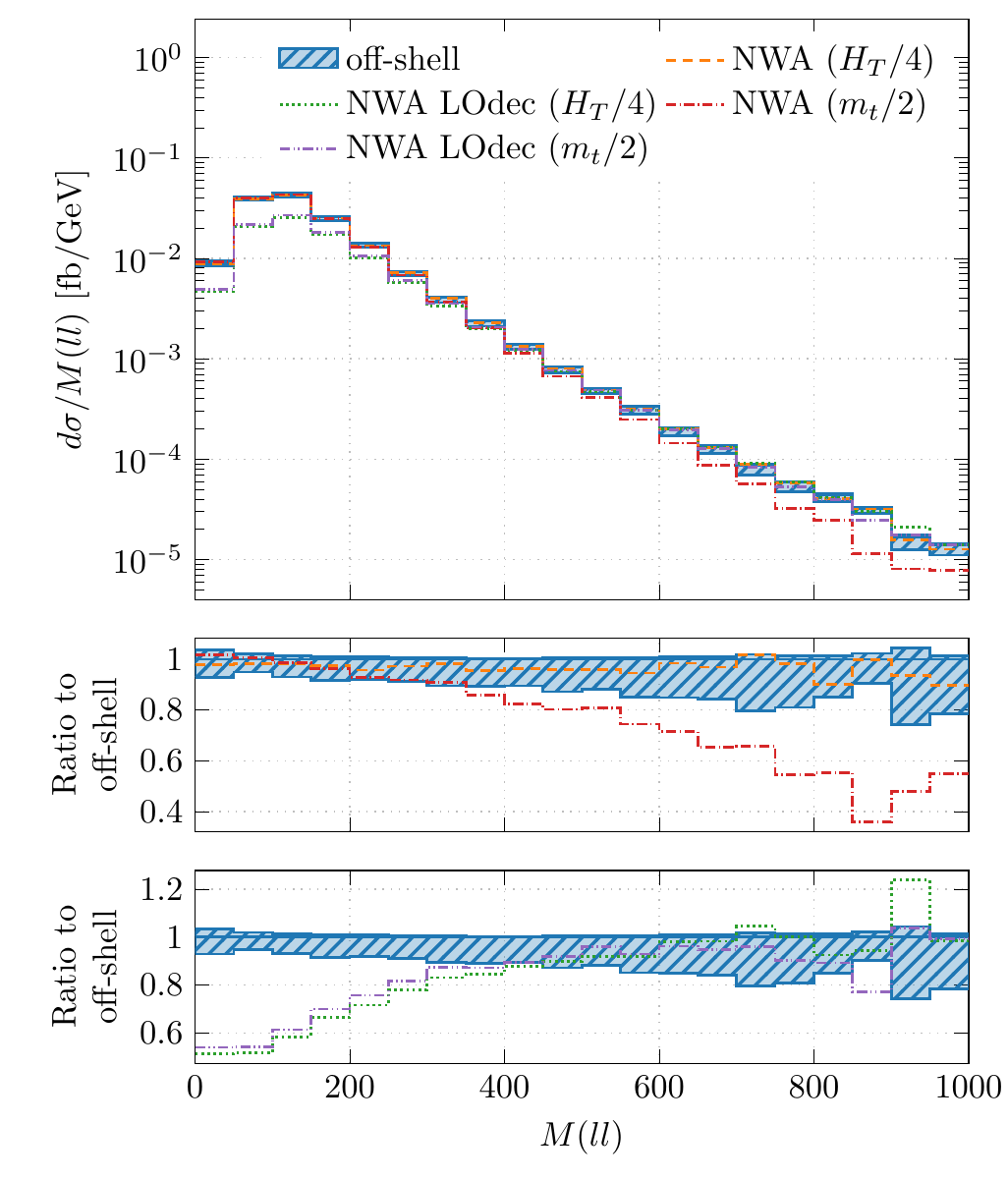}
 \end{center}
\caption{\it  Differential cross section distribution as a function of
the (averaged) transverse momentum of the $b$-jet, $p_T(b_{\rm avg})$,
and charged lepton, $p_T(\ell_{\rm avg})$, as well as the invariant
mass of the two $b$-jets, $M(bb)$, and two charged leptons system,
$M(\ell\ell)$, for the $pp\to e^+\nu_e \mu^- \bar{\nu}_\mu b\bar{b}
\gamma$ process at the LHC run II with $\sqrt{s}=13$ TeV. The CT14 PDF
set is employed.}
\label{fig:offshell2}
\end{figure}
\begin{figure}[t!]
\begin{center}
  \includegraphics[width=0.49\textwidth]{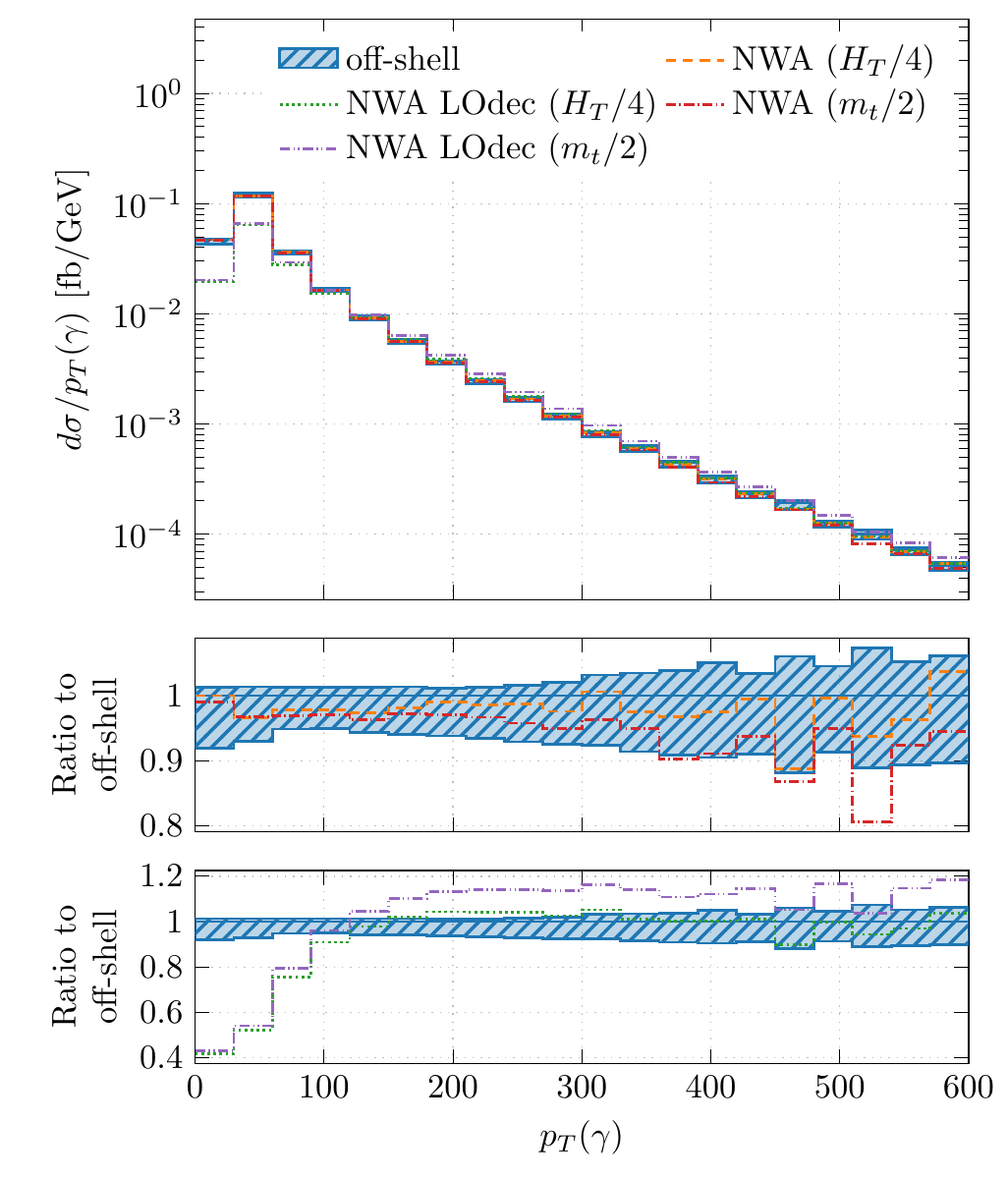}
  \includegraphics[width=0.49\textwidth]{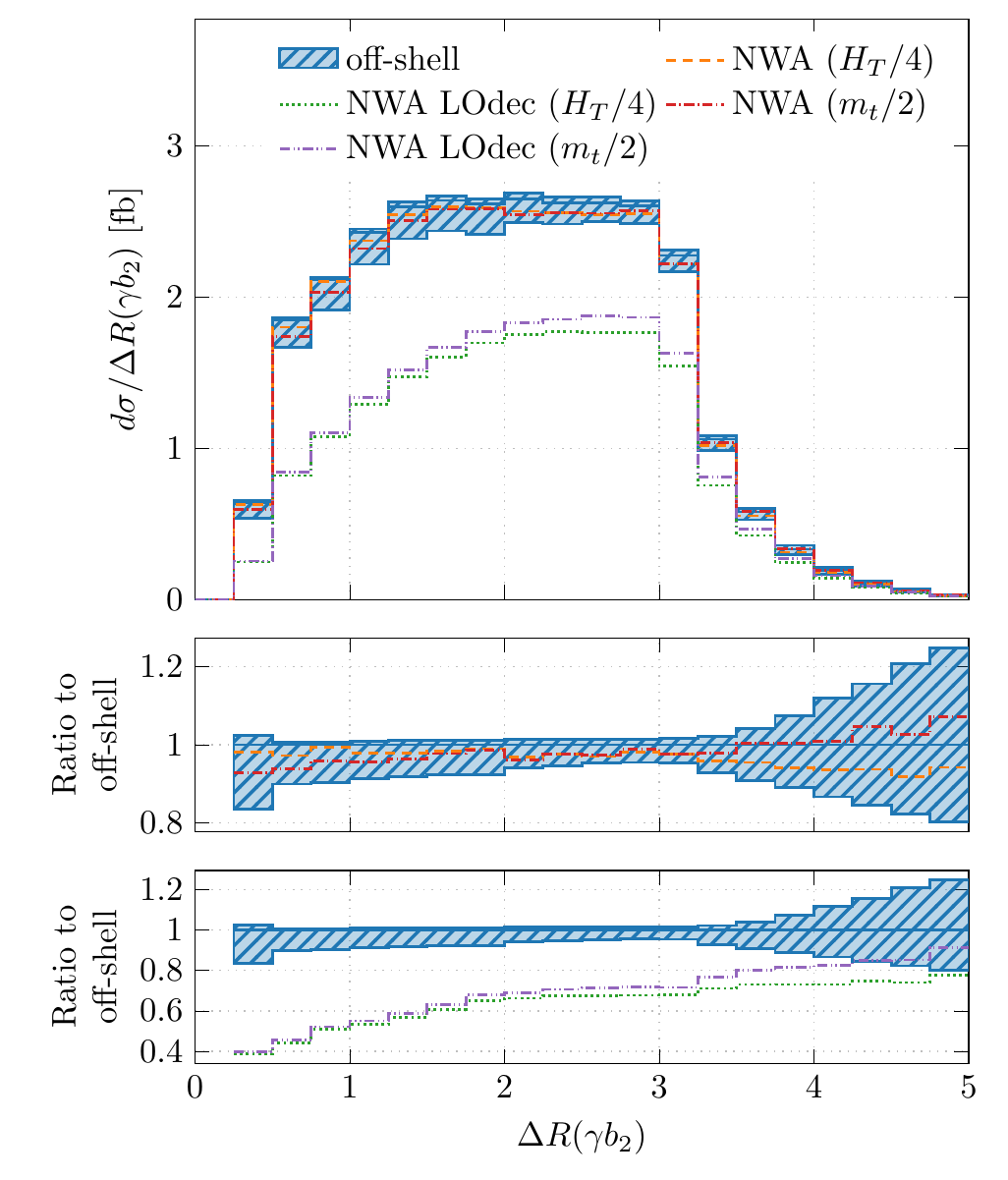}
   \includegraphics[width=0.49\textwidth]{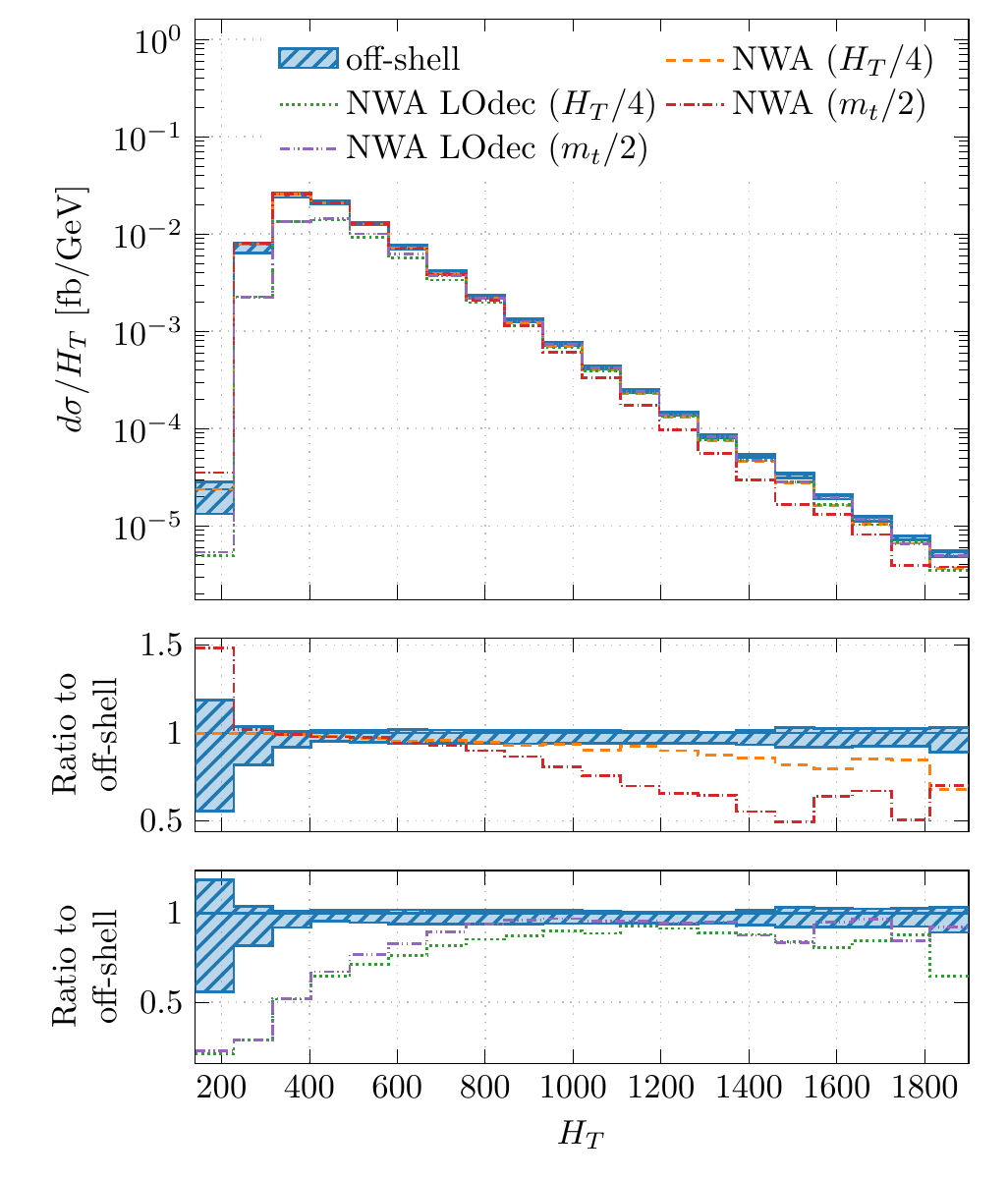}
 \end{center}
\caption{\it  Differential cross section distribution as a function of
the transverse momentum of the photon, $p_T(\gamma)$, the separation
of the photon and the softest $b$-jet in the rapidity-azimuthal angle
plane, $\Delta R(\gamma b_2)$, and the total transverse momentum of
the $e^+\nu_e \mu^- \bar{\nu}_\mu b\bar{b} \gamma$ system, $H_T$ for
the $pp\to e^+\nu_e \mu^- \bar{\nu}_\mu b\bar{b} \gamma$ process at
the LHC run II with $\sqrt{s}=13$ TeV.  The CT14 PDF set is employed.}
\label{fig:offshell3}
\end{figure}

In the following we examine the size of top quark off-shell effects at
the differential level. To this end we compare differential cross
sections for a few observables at NLO in QCD using three different
theoretical descriptions: the NWA, the NWA${}_{\rm
LOdecay}$ and results with the full off-shell effects. In
the case of NWA, two scale choices, $\mu_0=m_t/2$ and
$\mu_0=H_T/4$, are used, whereas for the full off-shell case only the
latter is utilised. We show theoretical uncertainties as obtained from the
scale dependence for the full off-shell case since
we are only interested in effects that exceed the
theoretical uncertainties. For all observables we employed the CT14
PDF set. The upper plots will show absolute NLO QCD predictions
for three different theoretical descriptions. The ratios to the full
off-shell result including its scale uncertainty band will  be
plotted in the middle and bottom plots.

In Figure~\ref{fig:offshell1} we start with the two observables that
are very well known from the top quark mass measurement in the
$t\bar{t}(j)$ production process, see e.g.
Refs.~\cite{Heinrich:2013qaa,Heinrich:2017bqp,Bevilacqua:2017ipv,
Ravasio:2018lzi}. Specifically, we plot the minimum invariant mass of
the positron and $b$-jet, $M(be^+)_{\rm min}$, and the (averaged)
invariant mass of the top quark, $M(t_{\rm avg})$. Because generally
one can not distinguish which $b$-jet should be paired with the
positron we define the $M(be^+)_{\rm min}$ observable as $M(be^+)_{\rm
min}=\min\left\{M(b_1e^+),M(b_2e^+)\right\}$, where $b_1$ and $b_2$
are bottom jets. Such criterion selects the correct pairing in
approximately $85\%$ of the cases \cite{Beneke:2000hk}. At lowest
order in perturbative expansion when both top quarks and $W$ gauge
bosons are treated as on-shell particles there is a strict kinematical
limit for $M(be^+)$ given by $M(be^+) = \sqrt{ m_t^2-m_W^2} \approx
153$ GeV. Due to the matching ambiguity stemming from the presence of
two $b$-jets the upper bound of $153$ GeV does not necessarily need to
be obeyed. Choosing, however, the smallest $M(be^+)$ for each event
will guarantee that $M(be^+)_{\rm min} \le M(be^+)$ and that the
kinematical endpoint of the distribution is preserved. For off-shell
top quarks this kinematic limit is smeared, and furthermore, additional
NLO QCD radiation and photon emission affect this
region. Nevertheless there is a sharp fall of the cross section around
$153$ GeV.  At NLO in QCD depending on the scale choice we can
observe large and negative finite top quark width effects of the
order of $40\%-60\%$, which are way above the theoretical
uncertainties in that region. At the same time we can see that the
NWA$_{\rm LOdecay}$ predictions are unable to correctly describe the
observable in the whole plotted region. Not only the overall
normalisation, but also the shape of the $M(be^+)_{\rm min}$
distribution can not be predicted by this approach. In the
same manner the peak of the $M(t_{\rm avg})$ distribution is smeared
by the off-shell top quark effects and additional gluon and photon
radiations. We note here, that the top and anti-top quarks are
reconstructed from their decay products assuming  the exact
$W$ gauge boson reconstruction and perfect $b$ and $\bar{b}$ tagging
efficiency. The full NWA results are consistently outside of the
theoretical uncertainty band and finite top quark width effects are
ranging from almost $+100\%$ to $-100\%$. For the NWA$_{\rm LOdecay}$
case non-factorisable corrections are even larger reaching $150\%$.  
$M(be^+)_{\rm min}$ and $M(t_{\rm avg})$ belong
to the first class of observables potentially susceptible to the
modelling of the top quark decays, i.e.  the observables with
kinematical threshold and edges.

In Figure~\ref{fig:offshell2} we exhibit differential cross section
distribution as a function of the (averaged) transverse momentum of
the $b$-jet, $p_T(b_{\rm avg})$, and charged lepton, $p_T(\ell_{\rm
avg})$. Also shown are the invariant mass of the two $b$-jets,
$M(b\bar{b})$, and the invariant mass of the two charged leptons,
$M(\ell\ell)$. For this class of observables the finite top quark
width effects will appear in the high $p_T$ tails and for large
values of the corresponding invariant mass differential
distributions. Specifically, for the $b$-jet kinematics we have
non-factorisable corrections of the order of $30\%-40\%$ independently
of the scale choice. Similar effects are observed for the charged leptons for
$\mu_0=m_t/2$. On the other hand, for the dynamical scale choice they
are negligible and within the theoretical uncertainties.  Once again
the NWA${}_{\rm LOdecay}$ is not adequate  to describe these
observables. They fail already in the low $p_T$ regions.

Finally, in Figure~\ref{fig:offshell3} we display differential cross
section distribution as a function of the transverse momentum of the
photon, $p_T(\gamma)$, the separation of the photon and the softest
$b$-jet in the rapidity-azimuthal angle plane, $\Delta R(\gamma b_2)$,
and the total transverse momentum of the $e^+\nu_e \mu^- \bar{\nu}_\mu
b\bar{b} \gamma$ system, $H_T$. These three observables are
well known for being sensitive to physics beyond the SM, see
e.g. Ref.~\cite{Baur:2001si,Baur:2004uw}.  For the $p_T(\gamma)$
distribution the finite top quark width effects are small and within
theoretical uncertainties. Thus, for this particular differential
cross section the full NWA description would be sufficient. Likewise
for the angular distribution $\Delta R(\gamma b_2)$. Generally
speaking, for all dimensionless observables, that we have studied,
negligible non-factorisable corrections have been observed. Large top
quark off-shell effects are estimated for the last observable $H_T$.
They are in the range of $-50\%$ to $+50\%$ for $\mu_0=m_t/2$ and up to
$25\%$ for $\mu_0=H_T/4$. This of course is not surprising since
$H_T$ comprises both $p_T(b_1), \,p_T(b_2)$ and
$p_T(\ell_1)$, $p_T(\ell_2)$ among others. On the other hand, the
NWA${}_{\rm LOdecay}$ can be disregard because it is consistently
unable to describe correctly the shape of various observables.

We can summarise this part by concluding that among all observables that
we have examined only dimensionful observables are sensitive to
non-factorizable top quark corrections that imply a cross-talk between
production and decays of top quarks. We could identify two
categories. The first class comprises observables with kinematical
thresholds or edges. Such observables should be carefully
examined in the vicinity of these thresholds/edges. The second class
consists of dimensionful observables in the high $p_T$ regions.  On
the contrary, dimensionless observables like angular distributions 
do not seem to be very sensitive to the top quark
off-shell effects.

%
\subsection{Double-, Single- and Non-resonant Phase-space Regions}
%
%
%
\begin{figure}[t!]
\begin{center}
  \includegraphics[width=0.49\textwidth]{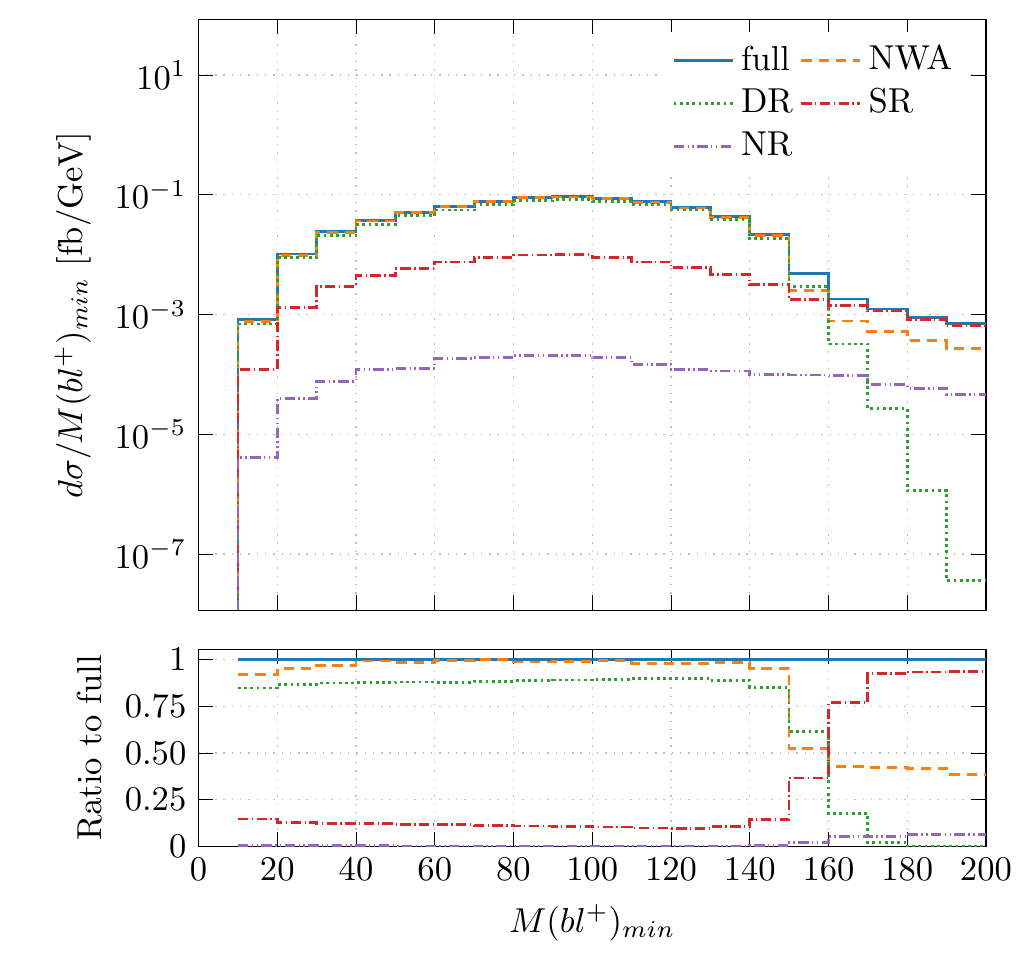}
  \includegraphics[width=0.49\textwidth]{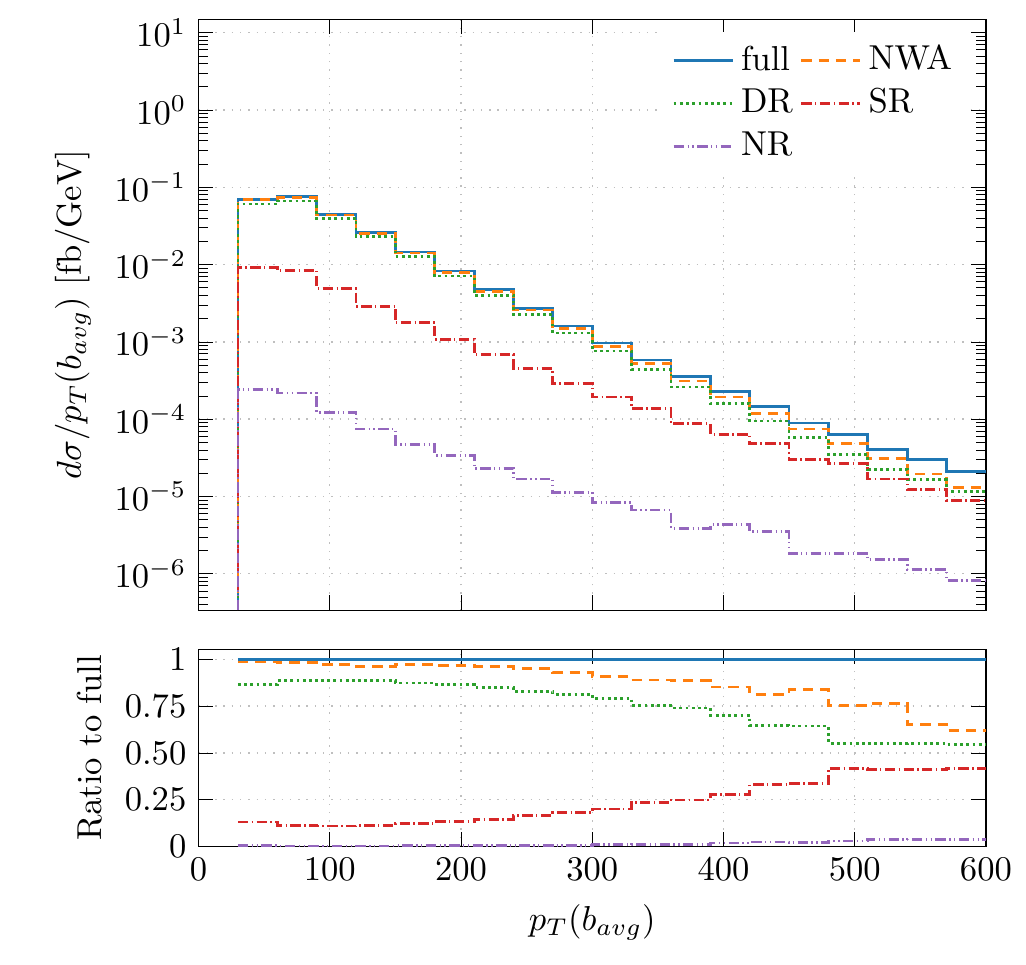}
      \includegraphics[width=0.49\textwidth]{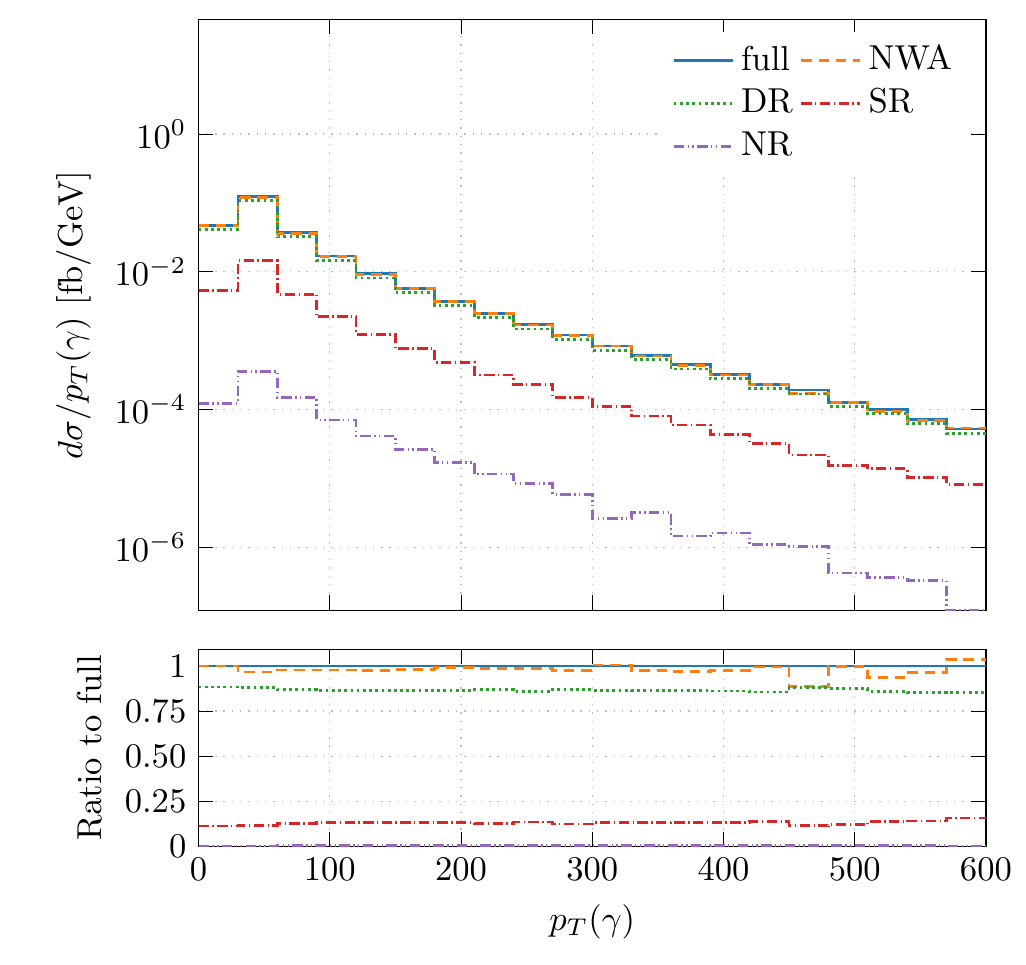}
  \includegraphics[width=0.49\textwidth]{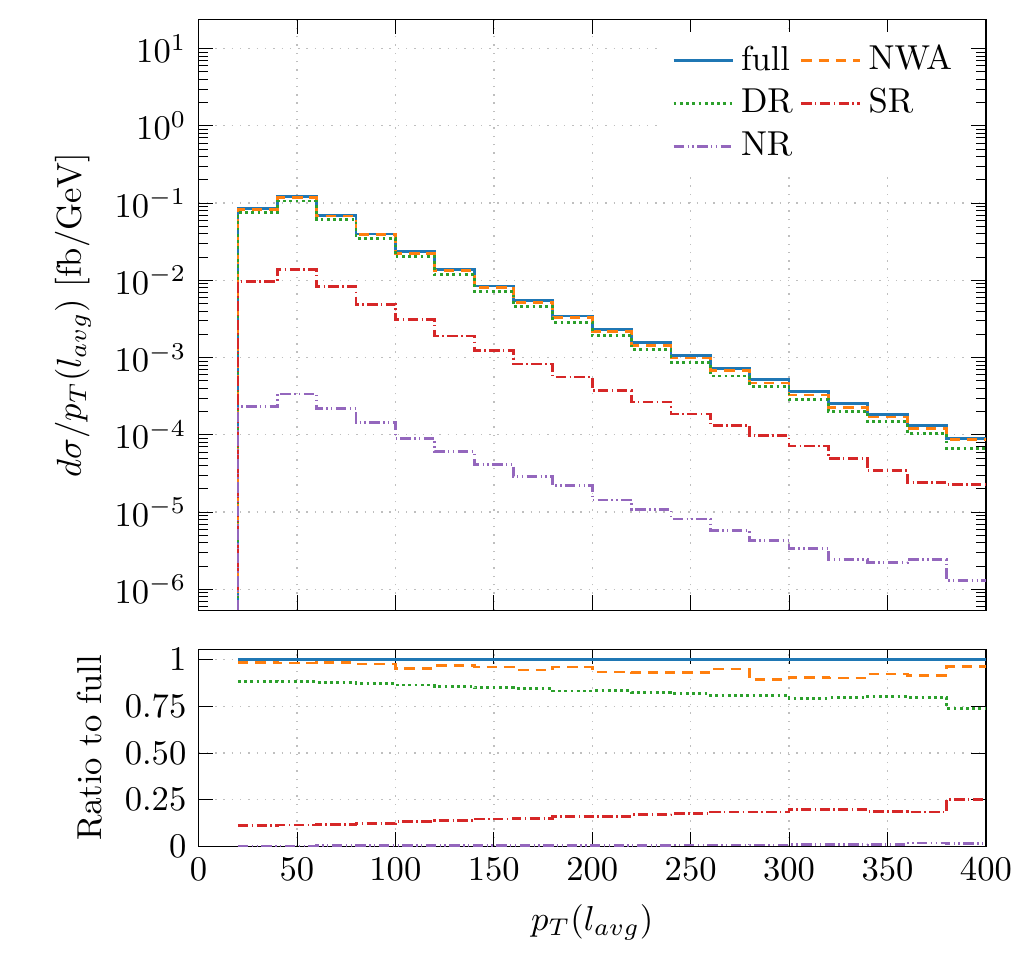}
 \end{center}
\caption{\it The $pp\to e^+\nu_e \mu^- \bar{\nu}_\mu b\bar{b} \gamma$
differential cross section distribution as a function of the
minimum invariant mass of the $b$-jet and the positron, the (averaged)
transverse momentum of the $b$-jet, the hard photon and the (averaged)
charged lepton at the LHC run II with $\sqrt{s}=13$ TeV.  The upper
plots show absolute NLO QCD predictions for DR, SR and NR
regions. Also shown are NLO results for full off-shell and NWA
case. The ratios of these contributions to the full off-shell result
are also shown. Results are given for $\mu_0=H_T/4$ and the CT14 PDF
set.}
\label{fig:top1}
\end{figure}
\begin{figure}[t!]
  \begin{center}
  \includegraphics[width=0.49\textwidth]{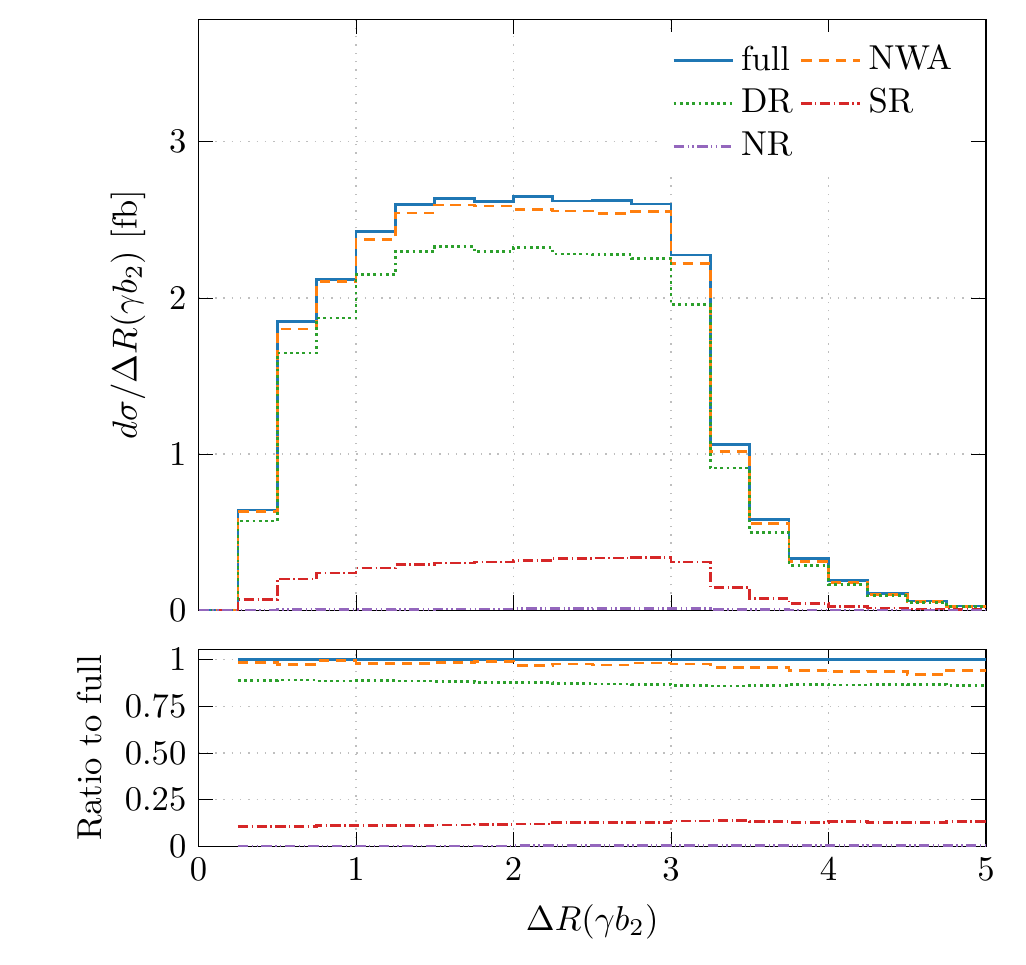}
\includegraphics[width=0.49\textwidth]{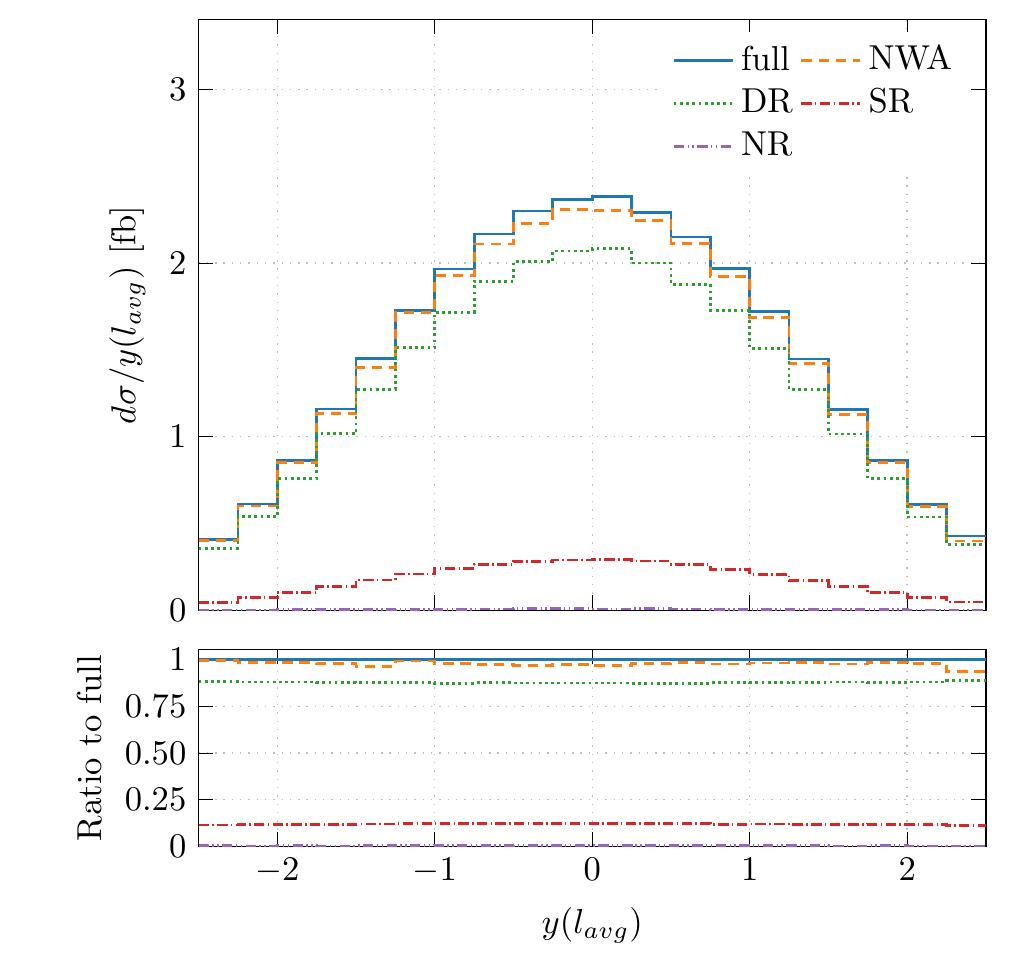}
 \end{center}
\caption{\it The $pp\to e^+\nu_e \mu^- \bar{\nu}_\mu b\bar{b} \gamma$
differential cross section distribution as a function of the
rapidity-azimuthal angle separation between the hard photon and the
softest $b$-jet and the (averaged) rapidity of the charged lepton at
the LHC run II with $\sqrt{s}=13$ TeV.  The upper plots show absolute
NLO QCD predictions for DR, SR and NR regions. Also shown are NLO
results for full off-shell and NWA case. The ratios of these
contributions to the full off-shell result are also shown. Results are
given for $\mu_0=H_T/4$ and the CT14 PDF set.}
\label{fig:top2}
\end{figure}

To understand better why some observables are more sensitive to the
top quark off-shell effects than the other we investigate the
contribution of double-, single- and non-resonant regions to the
integrated and differential cross sections for the full off-shell
case.  This should qualitatively show us the importance of particular
contributions and their overall distribution for particular phase space
regions.  To identify these contributions we have generalised the
method introduced in Ref.~\cite{Kauer:2001sp} and further discussed in
e.g. \cite{Liebler:2015ipp,Baskakov:2018huw}. We first identify the
following three different resonance histories
\begin{equation}
\begin{array}{llll}
  (\mathrm{i}) \quad & t = W^+(\to e^+ \nu_e) \, b  &
                                                  \quad    \quad \quad
                                                      \mathrm{and}
                                                      \quad \quad
                                                      \quad
  & \bar{t}  =W^-(\to \mu^- \bar{\nu}_\mu) \,  \bar{b} \,,\\[0.2cm]
  (\mathrm{ii}) \quad &   t = W^+( \to e^+ \nu_e ) \,b \gamma &
                                                             \quad
                                                                \quad
                                                                \quad
                                                                \mathrm{and}  \quad \quad \quad
  & \bar{t} = W^-(\to \mu^- \bar{\nu}_\mu)\,  \bar{b} \,,\\[0.2cm]
  (\mathrm{iii}) \quad &  t = W^+( \to e^+ \nu_e) \,b &
                                                 \quad       \quad
                                                        \quad
                                                        \mathrm{and}
                                                        \quad \quad
                                                        \quad
  & \bar{t} = W^-(\to \mu^- \bar{\nu}_\mu)\,  \bar{b} \gamma\,.\\
\end{array}
\end{equation}
These three categories are not sufficient if NLO QCD calculations are
considered. Therefore, in this case an additional resolved light jet
(if present) has to be incorporated into the list.  In practice, to
closely mimic what is done on the experimental side only the light jet
that passes all the cuts, that we have also applied for the $b$-jets,
is added to the resonance history. In this case a total of $9$ different
possibilities have to be considered. We compute for each history the
following quantity
\begin{equation}
Q = |M(t) - m_t| + |M(\,\bar{t}\,) - m_t|\,,
\end{equation}  
where $M(t)$ and $M(\,\bar{t}\,)$ are invariant masses of the top and
anti-top quark respectively, where their momenta are reconstructed from
the decay products assuming exact $W$ gauge boson reconstruction and
perfect $b$ and $\bar{b}$-tagging efficiencies. Finally, we pick the history
that minimises the $Q$ value.  Once the history is determined, the
check whether $t$ and $\bar{t}$ are off-shell or on-shell is
performed. Specifically, we define the double-resonant (DR) region,
when both $t$ and $\bar{t}$ are on-shell, via the following condition
\begin{equation}
  |M(t) -m_t|< n \, \Gamma_t \,, \quad \quad \quad {\rm and}
  \quad \quad \quad |M(\,\bar{t}\,) -m_t|< n \, \Gamma_t \,.
\end{equation}  
There are two  single-resonant (SR)  regions that are  given by
\begin{equation}
|M(t) -m_t|< n \, \Gamma_t \,, \quad \quad \quad {\rm and}
  \quad \quad \quad |M(\,\bar{t}\,) -m_t|> n \, \Gamma_t \,,
\end{equation}
or
\begin{equation}
 |M(t) -m_t|> n \, \Gamma_t \,, \quad \quad \quad {\rm and}
  \quad \quad \quad |M(\,\bar{t}\,) -m_t|< n \, \Gamma_t \,.
\end{equation}
Finally,  the non-resonant (NR) region is chosen according to 
\begin{equation}
  |M(t) -m_t|> n \, \Gamma_t \,, \quad \quad \quad {\rm and}
  \quad \quad \quad |M(\,\bar{t}\,) -m_t|> n \, \Gamma_t \,.
\end{equation}  
The boundary parameter, which determines the size of the resonant
region for each reconstructed top quark, has been set to $n=15$. This
corresponds to the following condition for the DR region: $M(t) \in
(152.9,193.5)$ GeV and $M(\, \bar{t}\,) \in (152.9,193.5)$ GeV. The
exact value of the boundary parameter is of course arbitrary.  In the 
literature more stringent conditions, like for example $n=10$, $n=5$,
have also been applied, see
e.g. \cite{Kauer:2001sp,Liebler:2015ipp,Baskakov:2018huw}. Having used
the above outlined procedure the contributions at the integrated cross
section level for these three regions are given by
\begin{equation}
  \sigma^{\rm NLO}_{\rm DR}= 6.57 ~{\rm fb}\,,
  \quad  \quad  \quad  \quad \sigma^{\rm NLO}_{\rm SR}=
  0.91 ~{\rm fb}\,, \quad  \quad  \quad  \quad
  \sigma^{\rm NLO}_{\rm NR}= 0.02~{\rm fb}\,.
\end{equation}  
The integrated fiducial cross section is dominated by the DR
contributions. More than $88\%$ of $\sigma^{\rm NLO}_{\rm full
~off-shell}$ comes from the DR contribution. Thus, it is not
surprising that the integrated cross section is not really sensitive
to SR and NR contributions and therefore also to the top quark
off-shell effects. The SR comprises only $12\%$ of the full off-shell
cross section at NLO in QCD whereas the contribution from the NR
regions of the phase space is negligible, below
$0.5\%$ \footnote{~Should we instead use $n=5$ we would get
$\sigma^{\rm NLO}_{\rm DR}= 4.82 ~{\rm fb}$, $\sigma^{\rm NLO}_{\rm
SR}= 2.50 ~{\rm fb}$ and $\sigma^{\rm NLO}_{\rm NR}= 0.18~{\rm fb}$.
}. The situation, however, looks quite different at the differential
level.

In Figure~\ref{fig:top1} we show the differential cross section
distribution as a function of $M(be^+)_{\rm min}$, $M(t_{\rm avg})$,
$p_T(b_{\rm avg})$ and $p_T(\ell_{\rm avg})$. Also given are NLO
results for the full off-shell and the full NWA cases. In the case of
the $M(be^+)_{\rm min}$ distribution we observe that in the region
that is not sensitive to the finite top quark width effects,
i.e. $M(be^+)_{\rm min} < 153$ GeV, the DR contribution is almost
indistinguishable from the full off-shell and the full NWA
result. However, once the vicinity of the kinematical cut-off is
reached, these particular contributions begin to spread out. For
$M(be^+)_{\rm min} \ge 153$ GeV the SR part rapidly starts to dominate
the full off-shell result and even the input from the NR regions of
the phase space is larger than from the DR one.  A similar effect can
be observed for $p_T(b_{\rm avg})$. Even though in the high $p_T$
tail of the differential cross section distribution the SR part does
not dominate the full off-shell result, its contribution increases to
almost $50\%$. On the other hand, the DR part contribution is greatly
reduced, from around $85\%$ down to about $50\%$. For $p_T(\gamma)$
and $p_T(\ell_{\rm avg})$, which are evaluated with $\mu_0=H_T/4$, in
the whole plotted range the contribution from the SR part of the cross
section is rather constant, of the order of $10\%$ for $p_T(\gamma)$
and below $25\%$ for $p_T(\ell_{\rm avg})$. Thus, it is consistently
smaller than the DR one. This is obviously reflected in the small
sensitivity of $d\sigma^{\rm NLO}_{\rm full ~off-shell}/dX,$
$X=p_T(\gamma)$, $p_T(\ell_{\rm avg})$ to non-factorisable top quark
corrections. We note at this point that all differential cross section
distributions that are dimensionless in nature, which we have
examined, received large $(90\%)$ and   constant
contributions from the DR regions of the phase space in the whole
kinematical range. At the same time the SR contribution was rather
small  $(10\%)$, whereas the NR regions of the phase space were
negligible. To illustrate these findings we present in
Figure~\ref{fig:top2} two examples, namely $\Delta R(\gamma b_2)$ and
$y(\ell_{\rm avg})$. Consequently, dimensionless observables are
rather insensitive to the finite top quark width  effects.

We can conclude this part by stressing that observables that are
sensitive to top quark off-shell effects have substantial
contributions from the single top quark process. In these cases the
best description is provided by the full off-shell calculation since
it is free of ambiguities related to disentangling   single and double
resonant contributions.

%
\subsection{Photon Radiation in the Production and Decays}
%
%
%
\begin{figure}[t!]
\begin{center}
  \includegraphics[width=0.49\textwidth]{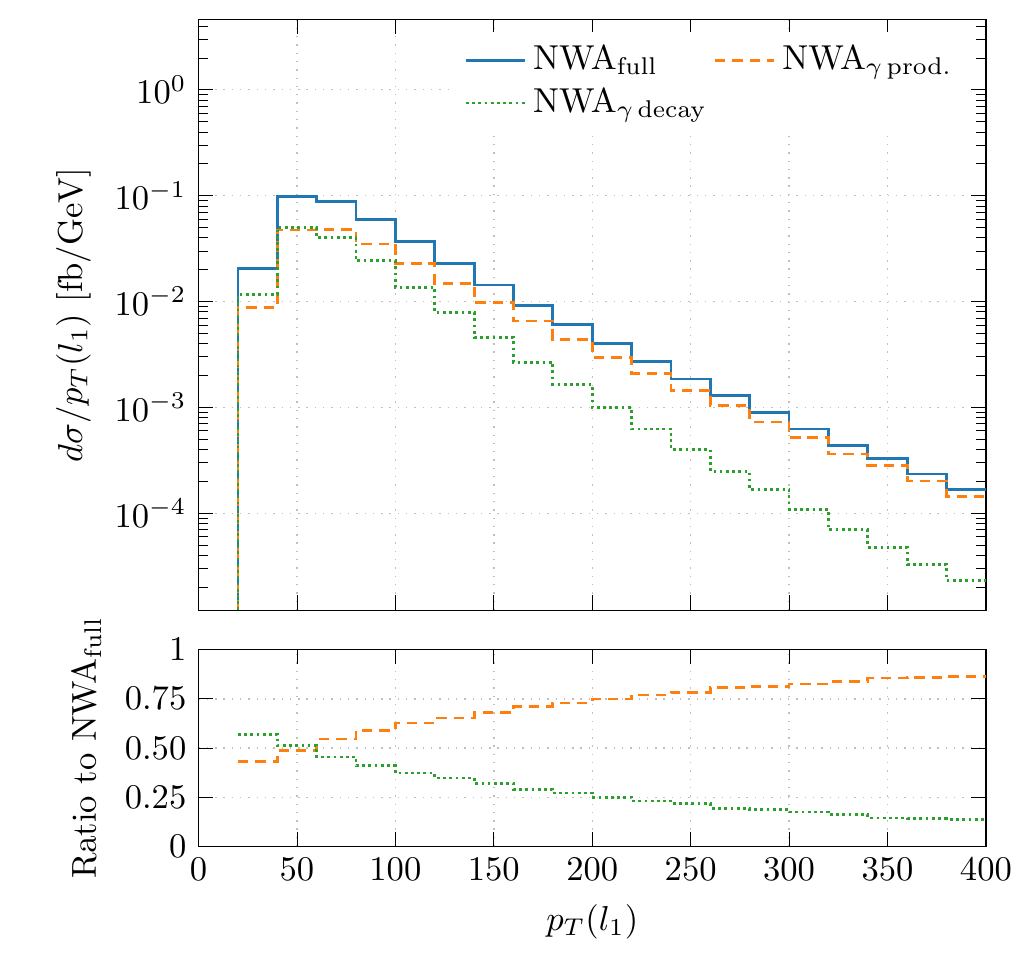}
  \includegraphics[width=0.49\textwidth]{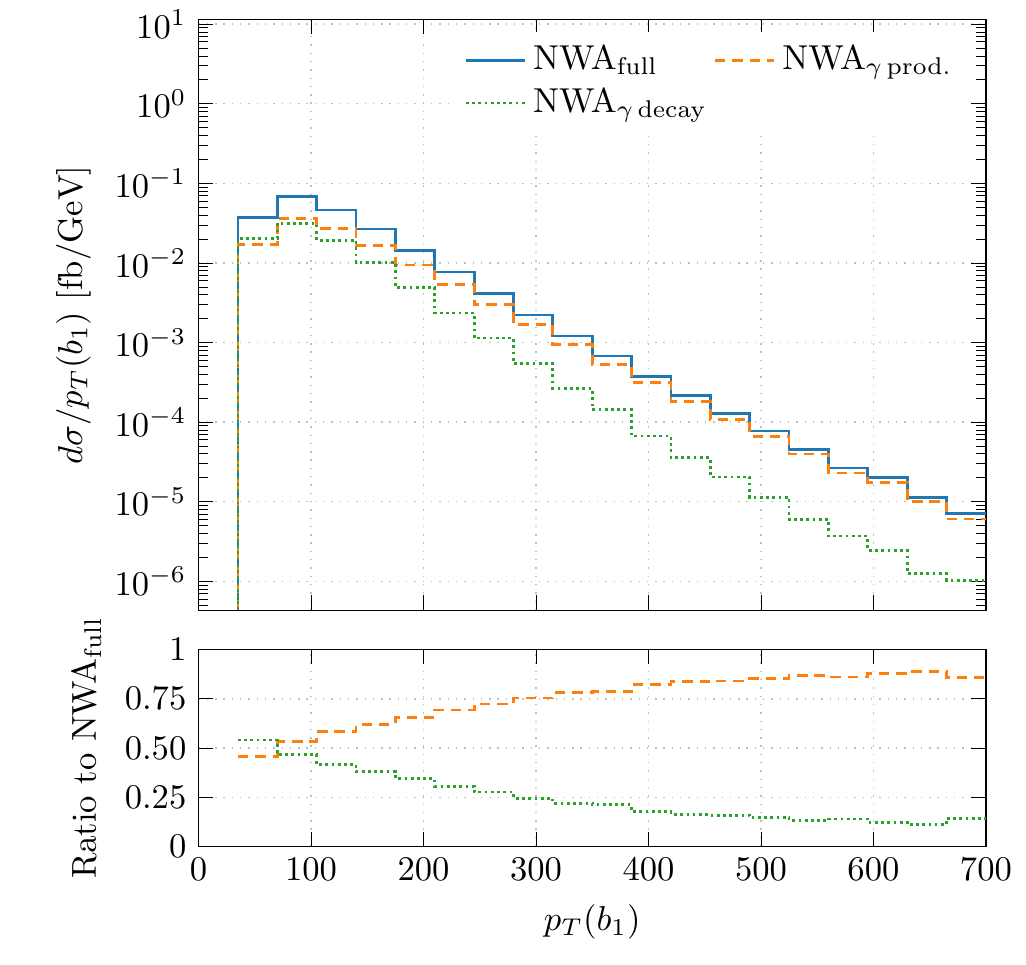}
  \includegraphics[width=0.49\textwidth]{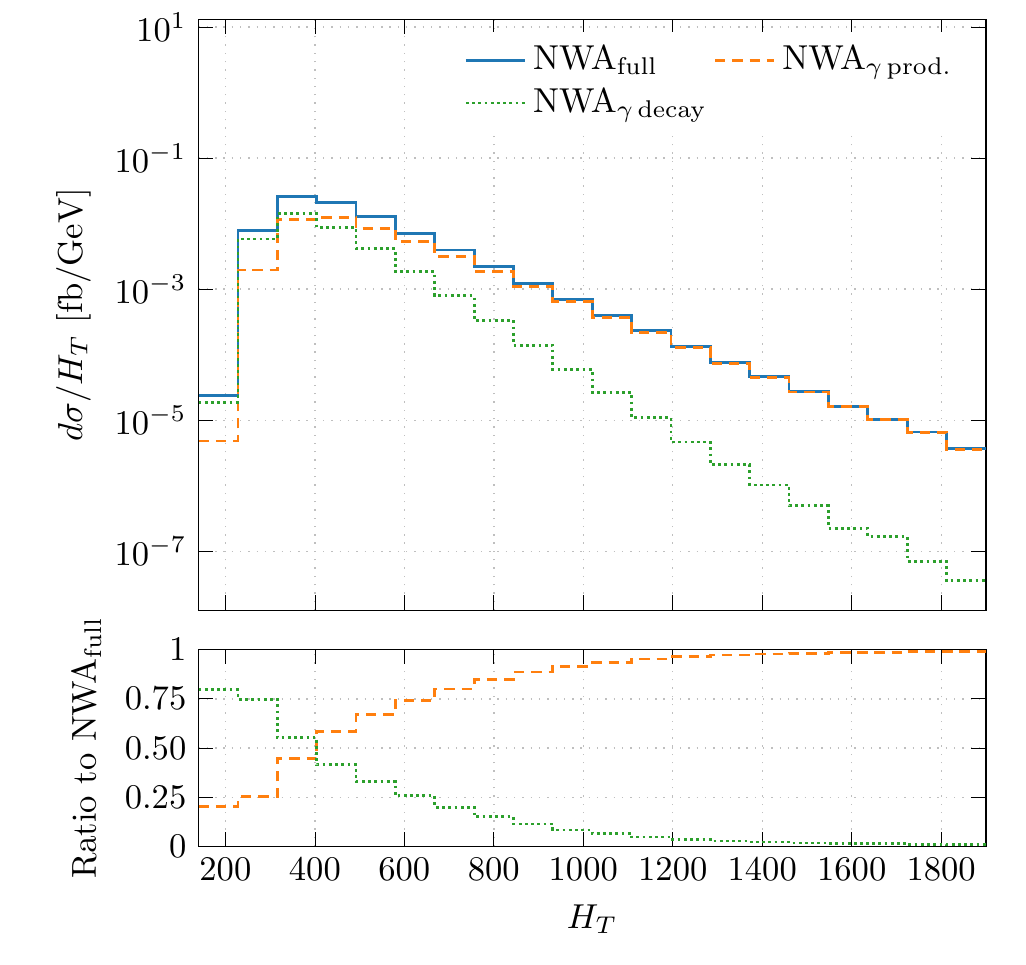}
    \includegraphics[width=0.49\textwidth]{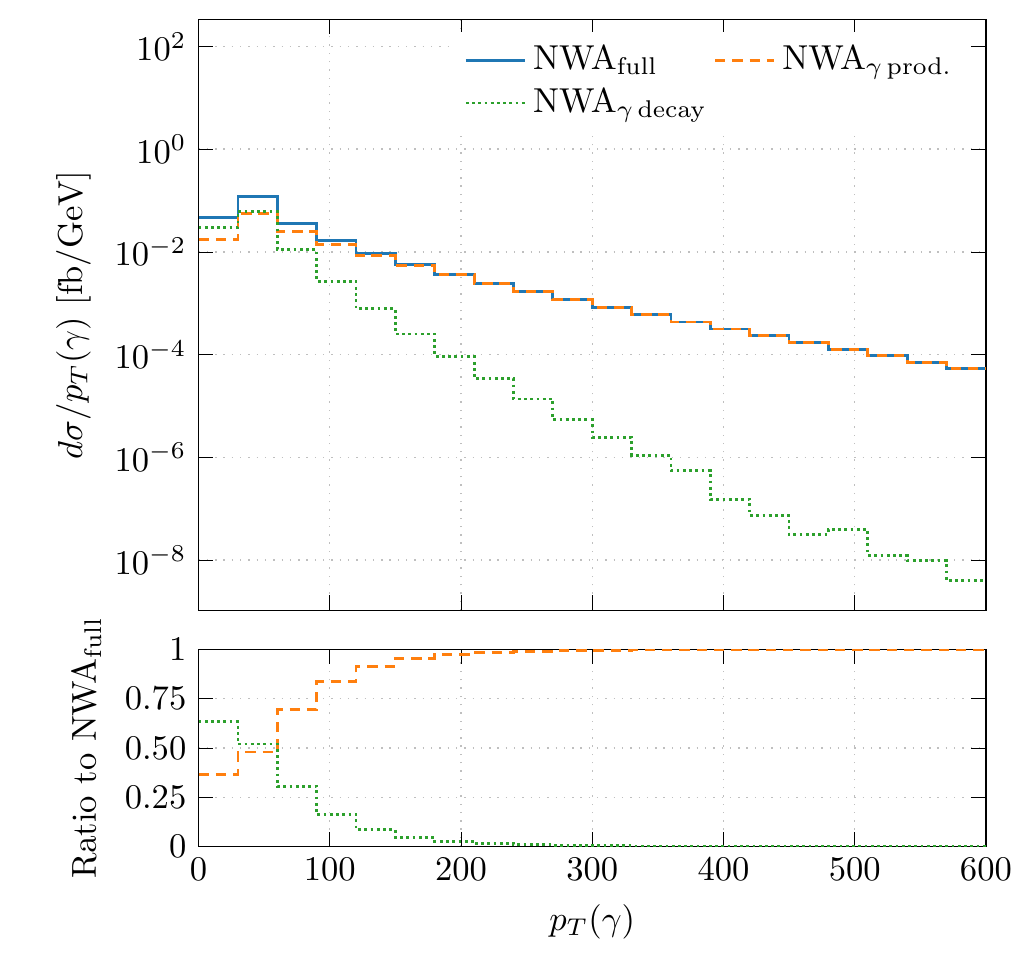}
 \end{center}
 \caption{\it
The $pp\to e^+\nu_e \mu^- \bar{\nu}_\mu b\bar{b} \gamma$ differential
cross section distribution as a function of the transverse momentum of
the hardest charged lepton, $p_T(\ell_1)$, and $b$-jet, $p_T(b_1)$,
the total transverse momentum of the $e^+\nu_e \mu^- \bar{\nu}_\mu
b\bar{b} \gamma$ system, $H_T$, and the transverse momentum of the
isolated photon, $p_T(\gamma)$, at the LHC run II with $\sqrt{s}=13$
TeV. The upper plots show absolute NLO QCD predictions in the full NWA
together with fraction of events originating from photon radiation in
the production, NWA${}_{\gamma \, {\rm prod}}$, and in decays,
NWA${}_{\gamma \, {\rm decay}}$. The ratios of these contributions to
the full NWA result are also shown. Results are given for
$\mu_0=H_T/4$ and the CT14 PDF sets are employed.}
\label{fig:photon1}
\end{figure}
%
\begin{figure}[t!]
  \begin{center}
    \includegraphics[width=0.49\textwidth]{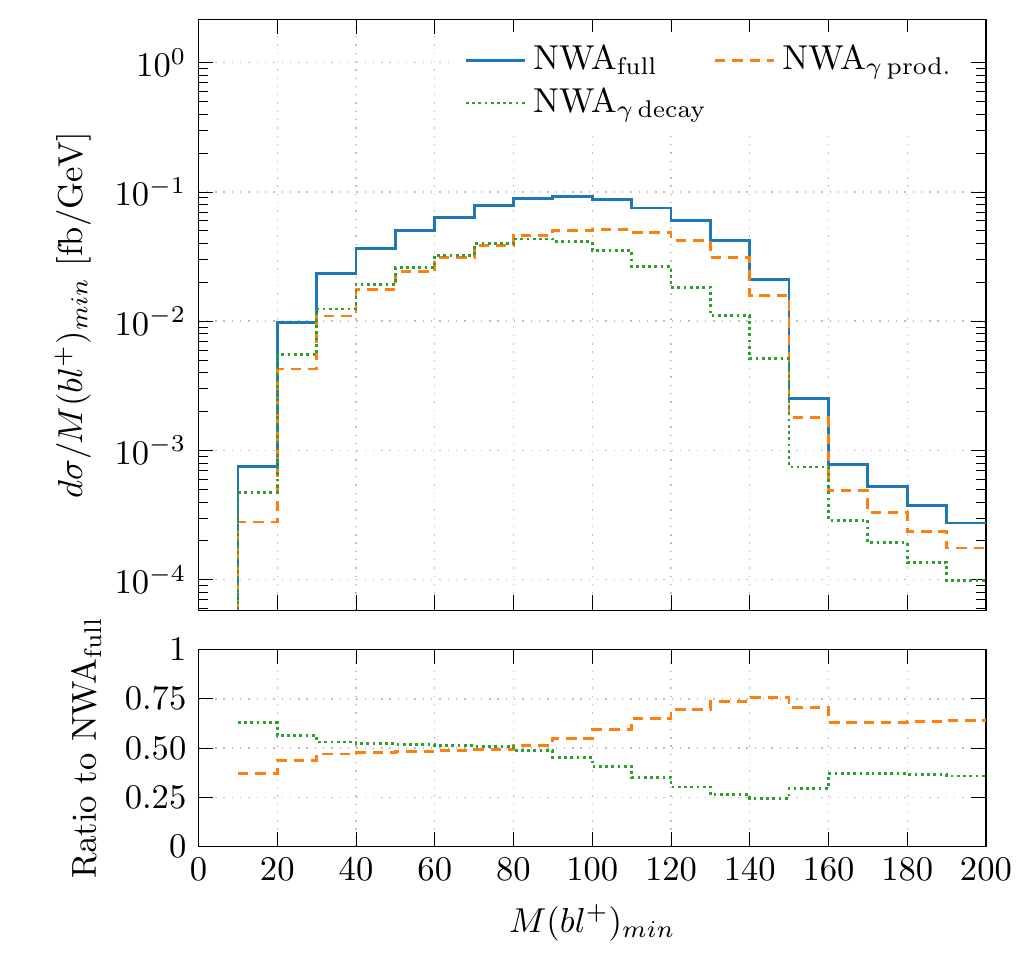}
    \includegraphics[width=0.49\textwidth]{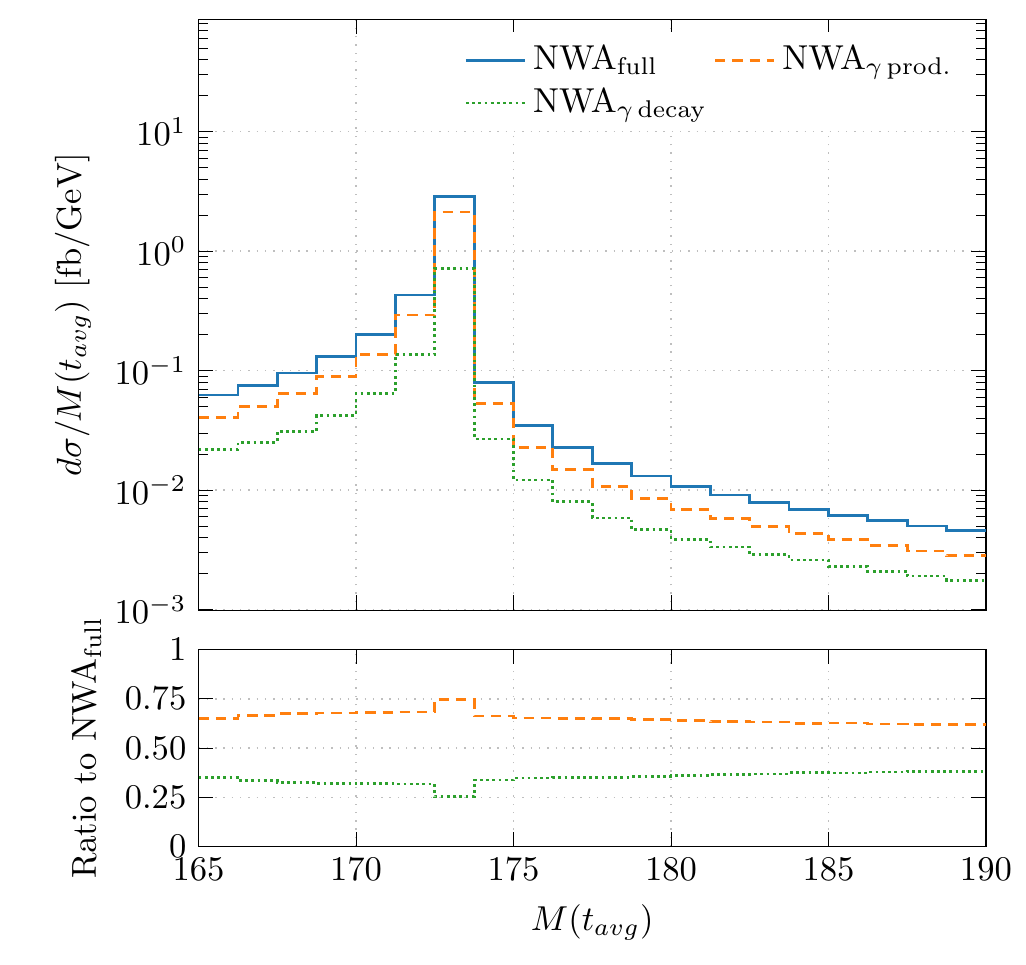}
 \end{center}
\caption{\it The $pp\to e^+\nu_e \mu^- \bar{\nu}_\mu b\bar{b} \gamma$
differential cross section distribution as a function of the minimum
invariant mass of the positron and $b$-jet, $M(b\ell^+)_{\rm min}$,
and the (averaged) invariant mass of the reconstructed top quark,
$M(t_{\rm avg})$, at the LHC run II with $\sqrt{s}=13$ TeV. The upper
plots show absolute NLO QCD predictions in the full NWA together with
fraction of events originating from photon radiation in the
production, NWA${}_{\gamma \, {\rm prod}}$, and in decays,
NWA${}_{\gamma \, {\rm decay}}$. The ratios of these contributions to
the full NWA result are also shown. Results are given for
$\mu_0=H_T/4$ and the CT14 PDF sets are employed.}
\label{fig:photon2}
\end{figure}
%
\begin{figure}[t!]
\begin{center}
  \includegraphics[width=0.49\textwidth]{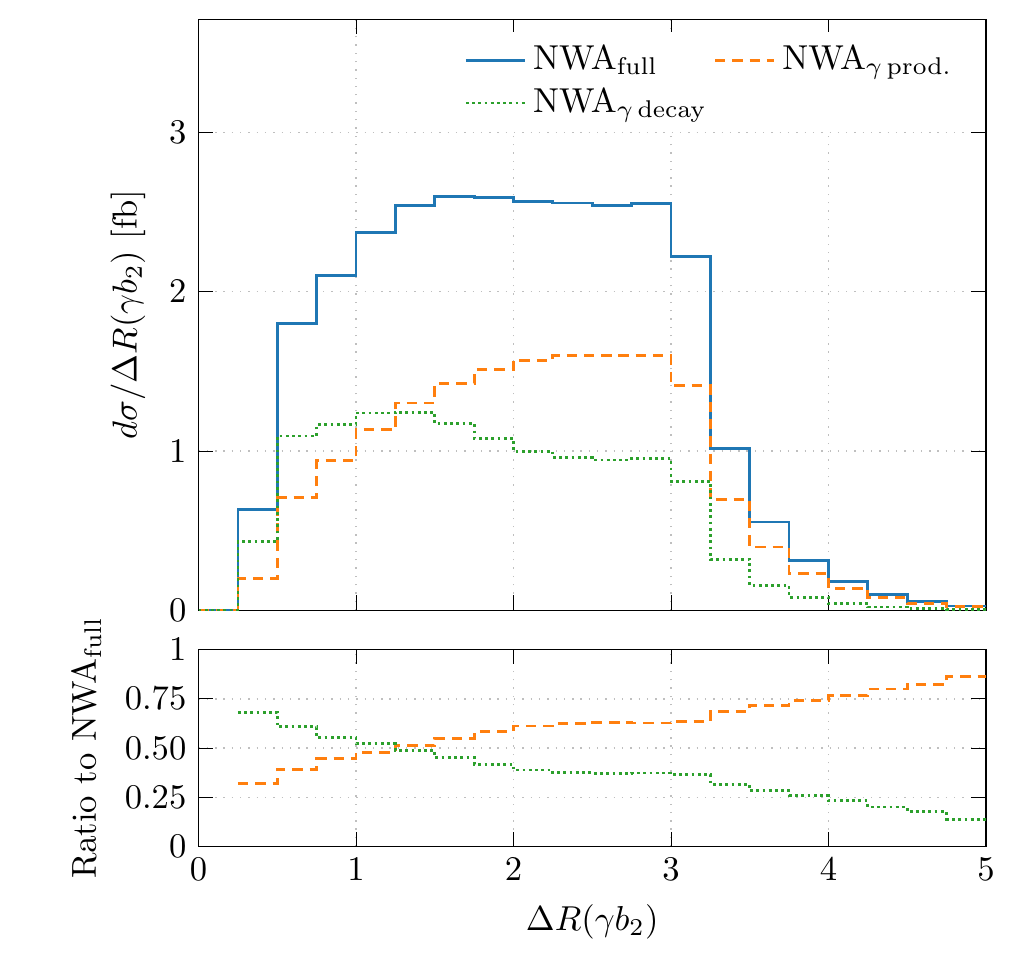}
  \includegraphics[width=0.49\textwidth]{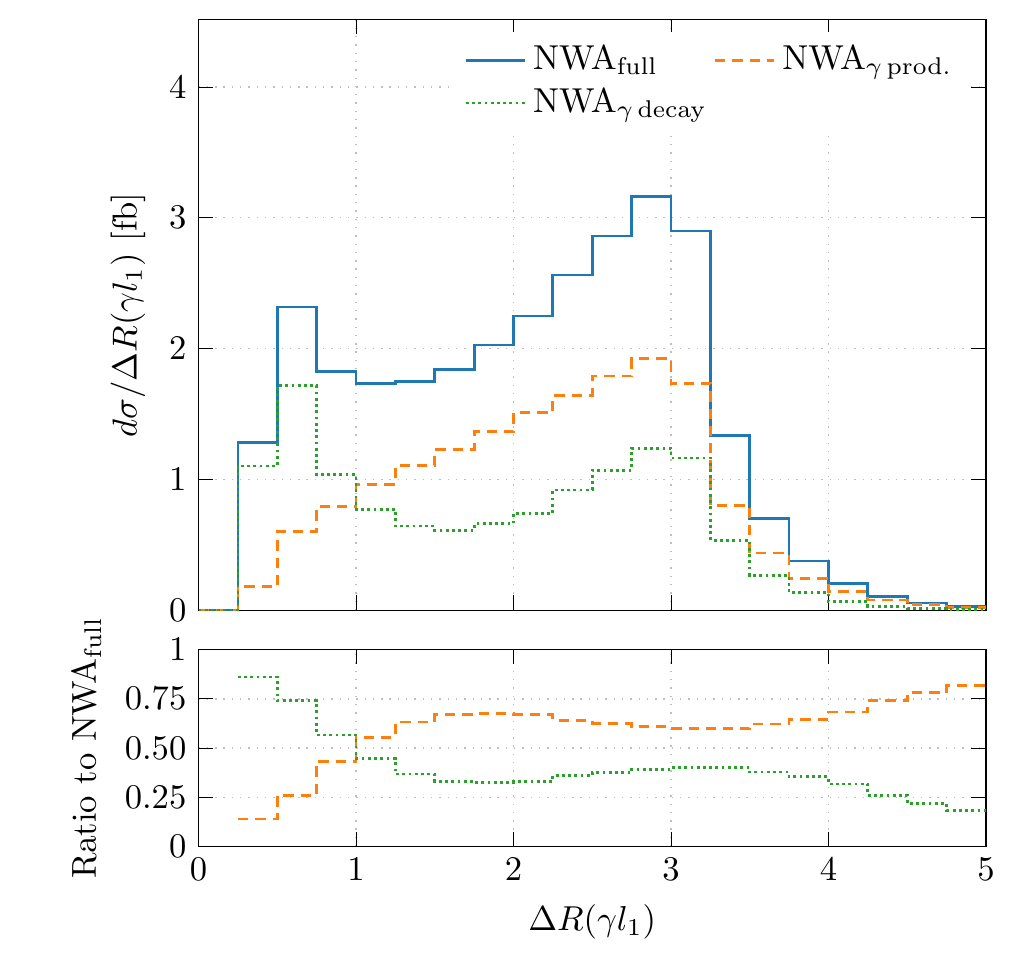}
    \includegraphics[width=0.49\textwidth]{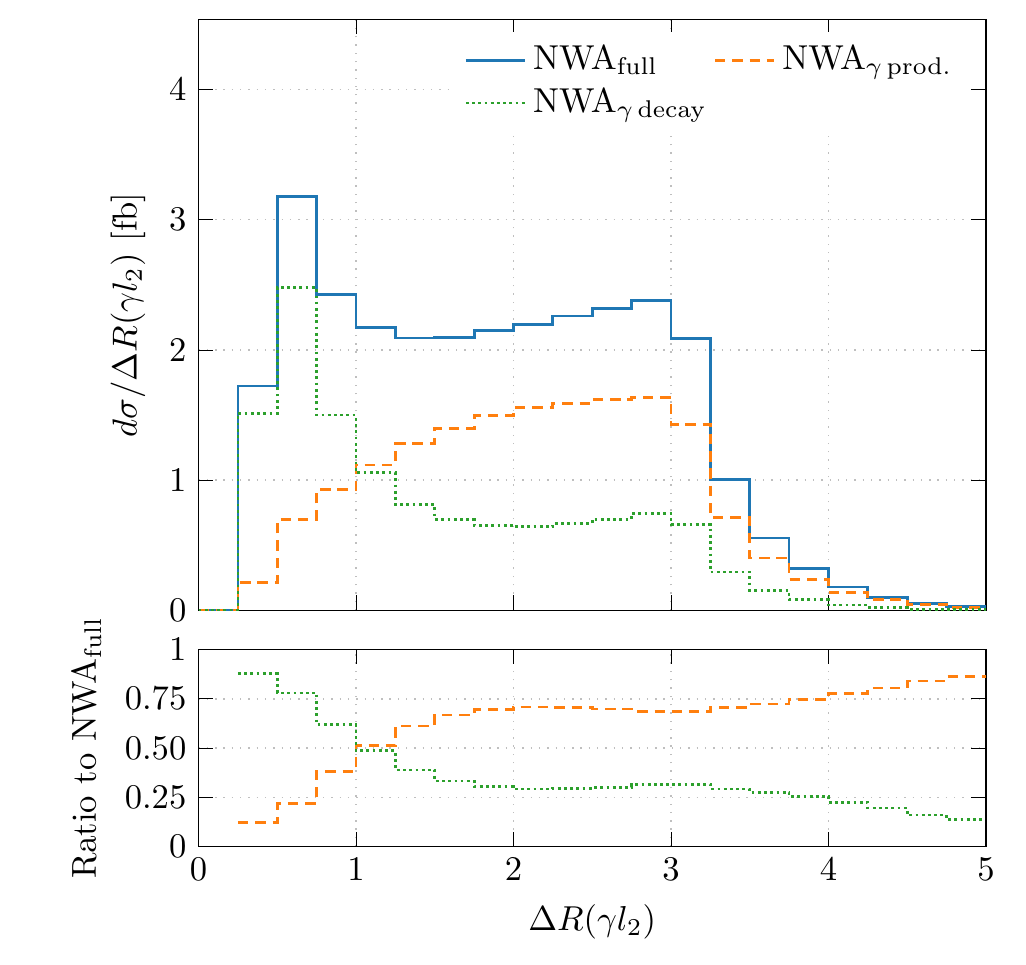}
    \includegraphics[width=0.49\textwidth]{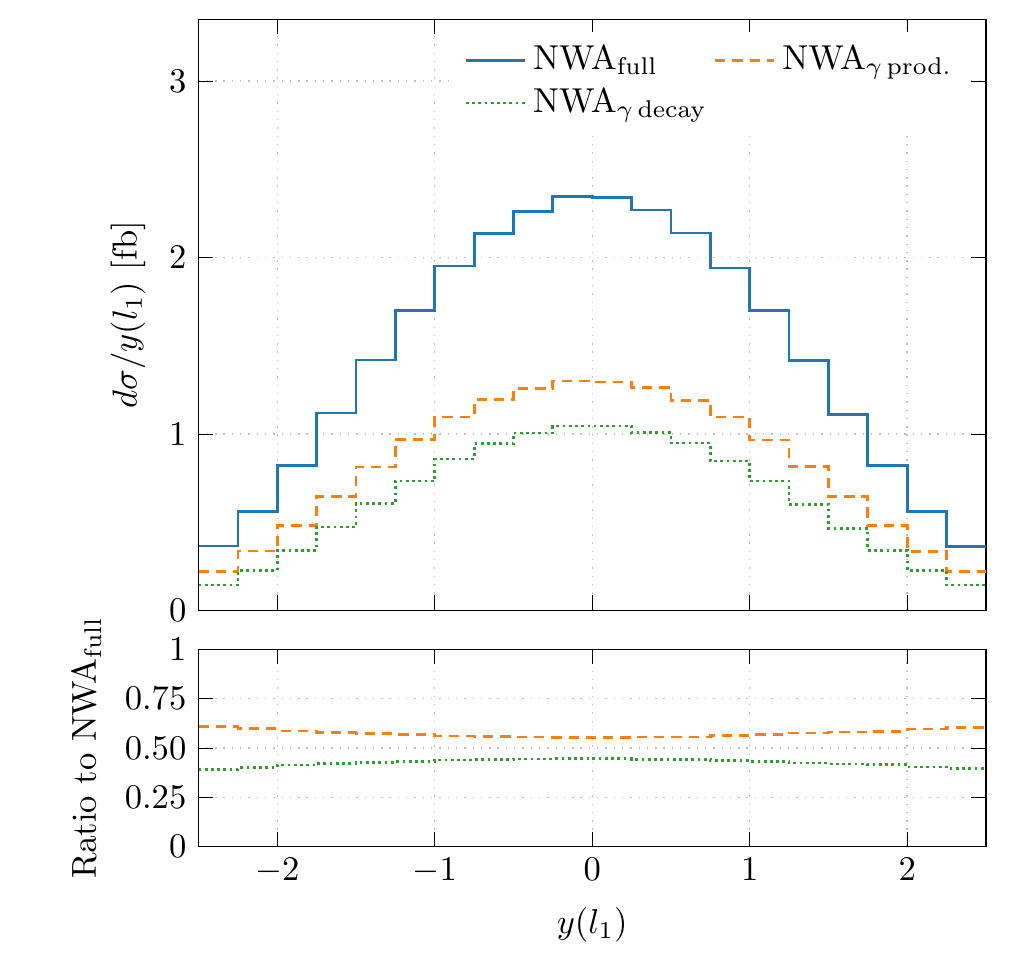}
 \end{center}
\caption{\it The $pp\to e^+\nu_e \mu^- \bar{\nu}_\mu b\bar{b} \gamma$
differential cross section distribution as a function of the
rapidity-azimuthal angle separation between: the photon and the softer
b-jet, $\Delta R(\gamma b_2)$, the photon and the hardest charged
lepton, $\Delta R(\gamma \ell_1)$, as well as the photon and the
softest charged lepton, $\Delta R(\gamma \ell_2)$ at the LHC run II
with $\sqrt{s}=13$ TeV.  Also shown is the differential cross section
distribution as a function of the rapidity of the hardest
charged lepton, $y(\ell_1)$, The upper plots show absolute NLO QCD
predictions in the full NWA together with fraction of events
originating from photon radiation in the production, NWA${}_{\gamma \,
{\rm prod}}$, and in decays, NWA${}_{\gamma \, {\rm decay}}$. The
ratios of these contributions to the full NWA result are also
shown. Results are given for $\mu_0=H_T/4$ and the CT14 PDF sets are
employed.}
\label{fig:photon3}
\end{figure}

Finally, we would like to investigate the composition of photon emissions
in the $t\bar{t}\gamma$ process, whether they come from production or decay stage.
Having the full NWA implemented at
the NLO in QCD level in the \textsc{Helac-Nlo} software such study can
be straightforwardly performed. First, we discuss dimensionful
observables like the transverse momentum of the hardest (in $p_T$)
charged lepton, $p_T(\ell_1)$, and $b$-jet, $p_T(b_1)$, the total
transverse momentum of the $e^+\nu_e \mu^- \bar{\nu}_\mu b\bar{b}
\gamma$ system, $H_T$, and the transverse momentum of the hard photon,
$p_T(\gamma)$. They are all exhibited in Figure~\ref{fig:photon1}. We
notice that in all four cases for the low values of the transverse
momentum the differential distributions are dominated by photon
emission in the decay stage. Specifically, in these regions more than
$50\%$ of photons come from top quark decays. For the $H_T$ observable
this contribution reaches even $75\%$. However, once the high $p_T$
region of these observables are probed, photon emission from the
production part of the $t\bar{t}\gamma$ process dominates completely
the full results. Thus, for all four observables setting high $p_T$ cut
would eliminate or at least substantially diminish the contribution
from the hard photon in top quark decays.

In Figure~\ref{fig:photon2} we show differential cross section
distribution as a function of $M(b\ell^+)_{\rm min}$ and $M(t_{\rm
avg})$. Even though these two also belong to dimensionful observables
photon radiation is distributed almost evenly between the production
and decay part of the $t\bar{t}\gamma$ process. In the case of
$M(b\ell^+)_{\rm min}$ the contribution from the NWA${}_{\gamma \,
{\rm decay}}$ part is at least $25\%$ and can go up to about
$60\%$. Instead, for $M(t_{\rm avg})$, rather constant $65\%$ and
$35\%$ contributions can be seen respectively for the NWA${}_{\gamma \,
{\rm prod}}$ and NWA${}_{\gamma \, {\rm decay}}$ case. Therefore, for both
observable there is no straightforward way to decrease the
contribution from the latter.

At last, in Figure~\ref{fig:photon3} dimensionless observables are
displayed. In particular, we provide differential cross section
distribution as a function of the rapidity-azimuthal angle separation
between: the photon and the softer $b$-jet, $\Delta R(\gamma b_2)$,
the photon and the hardest charged lepton, $\Delta R(\gamma \ell_1)$,
and the photon and the softest charged lepton, $\Delta R(\gamma
\ell_2)$. Also shown is the rapidity of the hardest charged lepton,
$y(\ell_1)$. For all three rapidity-azimuthal angle separations up to
about $\Delta R \lesssim 1$ photon radiation in top quark decays
dominates the corresponding differential cross section in that region
of the phase-space. For $\Delta R(\gamma \ell_1)$ and $\Delta R(\gamma
\ell_2)$ for example more than $80\%$ of all emitted photons originate in
$t\to b e^+ \nu_e \gamma$ and $\bar{t}\to \bar{b} \mu^- \bar{\nu}_\mu
\gamma$ processes. For $\Delta R \gtrsim 1$ the NWA${}_{\gamma \,
{\rm prod}}$ contribution surpasses the NWA${}_{\gamma \, {\rm
decay}}$ one, although the latter stays at the substantial
level. Specifically, we can observe contributions of the order of
$20\%-40\%$. Finally, as far as $y(\ell_1)$ is concerned rather
constant contributions from both parts are visible for   $y(\ell_1) \in
\left<-2.5,2.5\right>$, i.e.  about $40\%$ for  the  NWA${}_{\gamma
  \, {\rm decay}}$ case and  around $60\%$  for  NWA${}_{\gamma \,
  {\rm prod}}$.

%
\section{Conclusions}
\label{conclusions}
%

In this paper we have presented a comparative study of various
approaches for modelling of the $e^+ \nu_e \mu^- \bar{\nu}_\mu
b\bar{b} \gamma$ final state in $t\bar{t}\gamma$ production at the
LHC. We compared the fully realistic description as given by a
complete calculation with the one provided by the NWA.  In the latter case two
versions have been examined: the full NWA and the NWA${}_{\rm LOdecay}$
(i.e. NWA with LO decays of top quarks and photon radiation in the production
stage only). When comparing full off-shell and full NWA results we
confirmed that for the integrated cross sections the finite top quark
width effects are small, of the order of ${\cal O}(\Gamma_t/m_t)$. We
have shown, however, that they are strongly enhanced for more
exclusive (dimensionful) observables even up to $60\%$.  On the
contrary, dimensionless observables like  angular
differential cross section distributions appear to be relatively insensitive
to the top quark off-shell effects. Furthermore, we have revealed that
the NWA${}_{\rm LOdecay}$ approach is simply not adequate in
describing  the $e^+ \nu_e \mu^- \bar{\nu}_\mu b\bar{b} \gamma$
process neither at the integrated level nor at the differential
one. Not only the NLO QCD corrections to top quark decays have to be
incorporated but also hard photon emission from the top quark decays
must be included. 

To better understand the sensitivity of kinematic observables to the
non-factorisable top quark corrections, we have devised the procedure
to divide the full fiducial phase space of the $e^+ \nu_e \mu^-
\bar{\nu}_\mu b\bar{b} \gamma$ process into double-, single- and
non-resonant parts. We concluded that observables that are sensitive
to top quark off-shell effects have substantial contribution from the
single top quark process. In these cases the best description is
provided by the full off-shell calculation since it is free of
ambiguities related to disentangling single and double resonant
contributions.

In addition, we investigate fractions of events where the photon is
radiated either in the production or in the decay stage. We find that
large fraction of isolated photons comes from the decays of top
quarks. Based on our findings, selection criteria might be developed
to reduce such contributions that constitute a background for the
measurement of the anomalous couplings in the $t\bar{t}\gamma$
vertex. For example, for the transverse momentum of the hardest lepton,
$b$-jet and the hard photon as well as the total transverse momentum
of the $e^+ \nu_e \mu^- \bar{\nu}_\mu b\bar{b} \gamma$ system such 
kinematical cuts could be introduced. However, many other observables
have rather constant and substantial contribution from the hard photon
in the  top quark decays. Thus, a simple  procedure to decrease such
contributions would not be possible for them. Here, the most important
examples are the minimum invariant mass of the $b$-jet and the
positron as well as the (averaged) invariant mass of the top quark.  

Last but not least, on the technical side we have implemented the full
NWA approach into the \textsc{Helac-Nlo} Monte Carlo program.  This
has helped us to provide theoretical predictions for the full NWA and
the NWA${}_{\rm LOdecay}$ cases for the $t\bar{t}\gamma$ production at
the LHC.  Such automation opens a new path for performing higher order
calculations for more complex processes at the LHC such as
$t\bar{t}b\bar{b}$, $t\bar{t}jj$ and $t\bar{t}t\bar{t}$ where the
top-quark decays are realistically simulated through the NWA approach.

\acknowledgments{
We  thank Rene Poncelet  for a cross-check with the results of
Ref. \cite{Behring:2019iiv}.
  
The research of G.B. was supported by grant K 125105 of the National
Research, Development and Innovation Office in Hungary. G.B. also
thanks the Institute for Theoretical Particle Physics and Cosmology of
RWTH Aachen University for financial support and hospitality during
the completion of this work.

The work of H.B.H. has received funding from the European Research
Council (ERC) under the European Union's Horizon 2020 research and
innovation programme (grant agree- ment No 772099).

The work of M.W. and T.W. was supported in part by the German Research
Foundation (DFG) Individual Research Grant: {\it Top-Quarks under the
LHCs Magnifying Glass: From Process Modelling to Parameter Extraction}
and in part by the DFG Collaborative Research Centre/Transregio
project CRC/TRR 257: {\it P3H - Particle Physics Phenomenology after
the Higgs Discovery}.

Support by a grant of the Bundesministerium f\"ur Bildung und
Forschung (BMBF) is additionally acknowledged.

Simulations were performed with computing resources granted by RWTH
Aachen University under project {\tt rwth0414.}}

%
%


\begin{thebibliography}{99}

\bibitem{Aaltonen:2011sp}
  T.~Aaltonen {\it et al.} [CDF Collaboration],
  {\it Evidence for $t\bar{t}\gamma$ Production and Measurement of
    $\sigma_{t\bar{t}\gamma} / \sigma_{t\bar{t}}$},
  Phys.\ Rev.\ D {\bf 84} (2011) 031104
\href{http://arxiv.org/abs/arXiv:1106.3970}{\tt [arXiv:1106.3970 [hep-ex]]}.

\bibitem{Aad:2015uwa}
  G.~Aad {\it et al.} [ATLAS Collaboration],
  {\it Observation of top-quark pair production in association with a
    photon and measurement of the $t\bar{t}\gamma$ production cross
    section in pp collisions at $\sqrt{s}=7$ TeV using the ATLAS
    detector},
  Phys.\ Rev.\ D {\bf 91} (2015) no.7,  072007
  \href{http://arxiv.org/abs/arXiv:1502.00586}{\tt [arXiv:1502.00586 [hep-ex]]}.

\bibitem{Aaboud:2017era}
  M.~Aaboud {\it et al.} [ATLAS Collaboration],
  {\it Measurement of the $ t\overline{t}\gamma $ production cross
    section in proton-proton collisions at $ \sqrt{s}=8 $ TeV with the
    ATLAS detector},
  JHEP {\bf 1711} (2017) 086
  \href{http://arxiv.org/abs/arXiv:1706.03046}{\tt [arXiv:1706.03046 [hep-ex]]}.

\bibitem{Sirunyan:2017iyh}
  A.~M.~Sirunyan {\it et al.} [CMS Collaboration],
  {\it Measurement of the semileptonic $
    t \bar{t} + \gamma$ production cross section in
    pp collisions at $ \sqrt{s}=8 $ TeV},
  JHEP {\bf 1710} (2017) 006
  \href{http://arxiv.org/abs/arXiv:1706.08128}{\tt [arXiv:1706.08128 [hep-ex]]}.

\bibitem{Aaboud:2018hip}
  M.~Aaboud {\it et al.} [ATLAS Collaboration],
  {\it Measurements of inclusive and differential fiducial
    cross-sections of $t\bar{t}\gamma $ production in leptonic final
    states at $\sqrt{s}=13~\text {TeV}$ in ATLAS},
  Eur.\ Phys.\ J.\ C {\bf 79} (2019) no.5,  382
 \href{http://arxiv.org/abs/arXiv:1812.01697}{\tt [arXiv:1812.01697 [hep-ex]]}.
 
\bibitem{ATLAS:2019gkg}
  The ATLAS collaboration [ATLAS Collaboration],
  {\it Measurements of inclusive and differential cross-sections of
    $t\bar{t}\gamma$ production in the $e\mu$ channel at 13
    TeV with the ATLAS detector},
  \href{http://cds.cern.ch/record/2690350}{\tt ATLAS-CONF-2019-042}.

\bibitem{Baur:2001si}
  U.~Baur, M.~Buice and L.~H.~Orr,
  {\it Direct measurement of the top quark charge at hadron
    colliders}, 
  Phys.\ Rev.\ D {\bf 64} (2001) 094019
  \href{http://arxiv.org/abs/hep-ph/0106341}{\tt [hep-ph/0106341]}.
  
\bibitem{Aaltonen:2013sgl}
  T.~Aaltonen {\it et al.} [CDF Collaboration],
  {\it Exclusion of exotic top-like quarks with $-4/3$ electric charge
    using jet-charge tagging in single-lepton ttbar events at CDF}
  Phys.\ Rev.\ D {\bf 88} (2013) no.3,  032003
  \href{http://arxiv.org/abs/arXiv:1304.4141}{\tt  [arXiv:1304.4141 [hep-ex]]}.
  
\bibitem{Aad:2013uza}
  G.~Aad {\it et al.} [ATLAS Collaboration],
  {\it Measurement of the top quark charge in $pp$ collisions at
    $\sqrt{s} =$ 7 TeV with the ATLAS detector},
  JHEP {\bf 1311} (2013) 031
  \href{http://arxiv.org/abs/arXiv:1307.4568}{\tt [arXiv:1307.4568 [hep-ex]]}.
  
\bibitem{AguilarSaavedra:2008zc}
  J.~A.~Aguilar-Saavedra,
  {\it A Minimal set of top anomalous couplings}, 
  Nucl.\ Phys.\ B {\bf 812} (2009) 181
  \href{http://arxiv.org/abs/arXiv:0811.3842}{\tt [arXiv:0811.3842 [hep-ph]]}.

  \bibitem{Baur:2004uw}
  U.~Baur, A.~Juste, L.~H.~Orr and D.~Rainwater,
  {\it Probing electroweak top quark couplings at hadron colliders},
  Phys.\ Rev.\ D {\bf 71} (2005) 054013
  \href{http://arxiv.org/abs/hep-ph/0412021}{\tt [hep-ph/0412021]}.

\bibitem{Bouzas:2012av}
  A.~O.~Bouzas and F.~Larios,
  {\it Electromagnetic dipole moments of the Top quark}, 
  Phys.\ Rev.\ D {\bf 87} (2013) no.7,  074015
 \href{http://arxiv.org/abs/arXiv:1212.6575}{\tt [arXiv:1212.6575 [hep-ph]]}.
    
 \bibitem{Schulze:2016qas}
  M.~Schulze and Y.~Soreq,
  {\it Pinning down electroweak dipole operators of the top quark}, 
  Eur.\ Phys.\ J.\ C {\bf 76} (2016) no.8,  466
  \href{http://arxiv.org/abs/arXiv:1603.08911}{\tt [arXiv:1603.08911 [hep-ph]]}.

\bibitem{Bylund:2016phk}
  O.~Bessidskaia Bylund, F.~Maltoni, I.~Tsinikos, E.~Vryonidou and C.~Zhang,
  {\it Probing top quark neutral couplings in the Standard Model
    Effective Field Theory at NLO in QCD},
  JHEP {\bf 1605} (2016) 052
\href{http://arxiv.org/abs/arXiv:1601.08193}{\tt [arXiv:1601.08193 [hep-ph]]}.
    
  \bibitem{Bevilacqua:2018dny}
  G.~Bevilacqua, H.~B.~Hartanto, M.~Kraus, T.~Weber and M.~Worek,
  {\it Precise predictions for $t\bar{t}\gamma/t\bar{t}$ cross section
    ratios at the LHC},    
  JHEP {\bf 1901} (2019) 188
 \href{http://arxiv.org/abs/arXiv:1809.08562}{\tt [arXiv:1809.08562 [hep-ph]]}.
 
\bibitem{Aguilar-Saavedra:2014vta}
  J.~A.~Aguilar-Saavedra, E.~Alvarez, A.~Juste and F.~Rubbo,
  {\it Shedding light on the $t \bar t$ asymmetry: the photon handle},
  JHEP {\bf 1404} (2014) 188
  \href{http://arxiv.org/abs/arXiv:1402.3598}{\tt [arXiv:1402.3598 [hep-ph]]}.

\bibitem{Aguilar-Saavedra:2018gfv}
  J.~A.~Aguilar-Saavedra,
  {\it Single lepton charge asymmetries in $t \bar t$ and $t \bar t
    \gamma$ production at the LHC},
  Eur.\ Phys.\ J.\ C {\bf 78} (2018) no.6,  434
  \href{http://arxiv.org/abs/arXiv:1802.05721}{\tt [arXiv:1802.05721 [hep-ph]]}.

\bibitem{Bergner:2018lgm}
  J.~Bergner and M.~Schulze,
  {\it The top quark charge asymmetry in $t\bar{t}\gamma$ production
    at the LHC},
  Eur.\ Phys.\ J.\ C {\bf 79} (2019) no.3,  189
  \href{http://arxiv.org/abs/arXiv:1812.10535}{\tt [arXiv:1812.10535 [hep-ph]]}.

\bibitem{PengFei:2009ph}
  P.~F.~Duan, W.~G.~Ma, R.~Y.~Zhang, L.~Han, L.~Guo and S.~M.~Wang,
  {\it QCD corrections to associated production of $t\bar t\gamma$ at
    hadron colliders},
  Phys.\ Rev.\ D {\bf 80} (2009) 014022
  \href{http://arxiv.org/abs/arXiv:0907.1324}{\tt [arXiv:0907.1324 [hep-ph]]}.

\bibitem{PengFei:2011qg}
  P.~F.~Duan, R.~Y.~Zhang, W.~G.~Ma, L.~Han, L.~Guo and S.~M.~Wang,
  {\it Next-to-leading order QCD corrections to $t\bar t\gamma$
    production at the 7 TeV LHC},
  Chin.\ Phys.\ Lett.\  {\bf 28} (2011) 111401
  \href{http://arxiv.org/abs/arXiv:1110.2315}{\tt [arXiv:1110.2315 [hep-ph]]}.

\bibitem{Maltoni:2015ena}
  F.~Maltoni, D.~Pagani and I.~Tsinikos,
  {\it Associated production of a top-quark pair with vector bosons at
    NLO in QCD: impact on $ t\bar{t}H$ searches at the LHC},
  JHEP {\bf 1602} (2016) 113
  \href{http://arxiv.org/abs/arXiv:1507.05640}{\tt [arXiv:1507.05640 [hep-ph]]}.
  
\bibitem{Duan:2016qlc}
  P.~F.~Duan, Y.~Zhang, Y.~Wang, M.~Song and G.~Li,
  {\it Electroweak corrections to top quark pair production in
    association with a hard photon at hadron colliders},
  Phys.\ Lett.\ B {\bf 766} (2017) 102
  \href{http://arxiv.org/abs/arXiv:1612.00248}{\tt [arXiv:1612.00248 [hep-ph]]}.
  
  \bibitem{Kardos:2014zba}
  A.~Kardos and Z.~Trocsanyi,
  {\it Hadroproduction of t anti-t pair in association with an
    isolated photon at NLO accuracy matched with parton shower},
  JHEP {\bf 1505} (2015) 090
  \href{http://arxiv.org/abs/arXiv:1406.2324}{\tt [arXiv:1406.2324 [hep-ph]]}.

 \bibitem{Frixione:2007vw}
  S.~Frixione, P.~Nason and C.~Oleari,
  {\it Matching NLO QCD computations with Parton Shower simulations:
    the POWHEG method},
  JHEP {\bf 0711} (2007) 070
  \href{http://arxiv.org/abs/arXiv:0709.2092}{\tt [arXiv:0709.2092 [hep-ph]]}.

\bibitem{Alioli:2010xd}
  S.~Alioli, P.~Nason, C.~Oleari and E.~Re,
  {\it A general framework for implementing NLO calculations in shower
    Monte Carlo programs: the POWHEG BOX},
  JHEP {\bf 1006} (2010) 043
  \href{https://arxiv.org/abs/1002.2581}{\tt [arXiv:1002.2581 [hep-ph]]}.

\bibitem{Bevilacqua:2011xh}
  G.~Bevilacqua, M.~Czakon, M.~V.~Garzelli, A.~van Hameren, A.~Kardos,
  C.~G.~Papadopoulos, R.~Pittau and M.~Worek,
  {\it Helac-NLO},
  Comput.\ Phys.\ Commun.\  {\bf 184} (2013) 986
  \href{https://arxiv.org/abs/1110.1499}{\tt [arXiv:1110.1499 [hep-ph]]}.

 \bibitem{Sjostrand:2014zea}
  T.~Sj\"ostrand {\it et al.},
  {\it An Introduction to PYTHIA 8.2},
  Comput.\ Phys.\ Commun.\  {\bf 191} (2015) 159
  \href{http://arxiv.org/abs/arXiv:1410.3012}{\tt [arXiv:1410.3012 [hep-ph]]}.
  
\bibitem{Bahr:2008pv}
  M.~Bahr {\it et al.},
  {\it Herwig++ Physics and Manual},
  Eur.\ Phys.\ J.\ C {\bf 58} (2008) 639
 \href{https://arxiv.org/abs/0803.0883}{\tt [arXiv:0803.0883 [hep-ph]]}.
  
  \bibitem{Melnikov:2011ta}
  K.~Melnikov, M.~Schulze and A.~Scharf,
  {\it QCD corrections to top quark pair production in association
    with a photon at hadron colliders},
  Phys.\ Rev.\ D {\bf 83} (2011) 074013
  \href{http://arxiv.org/abs/arXiv:1102.1967}{\tt [arXiv:1102.1967 [hep-ph]]}.

\bibitem{Bevilacqua:2018woc}
  G.~Bevilacqua, H.~B.~Hartanto, M.~Kraus, T.~Weber and M.~Worek,
  {\it Hard Photons in Hadroproduction of Top Quarks with Realistic
    Final States},
  JHEP {\bf 1810} (2018) 158
  \href{http://arxiv.org/abs/arXiv:1803.09916}{\tt [arXiv:1803.09916 [hep-ph]]}.
 
\bibitem{vanHameren:2009dr}
  A.~van Hameren, C.~G.~Papadopoulos and R.~Pittau,
  {\it Automated one-loop calculations: A Proof of concept},
  JHEP {\bf 0909} (2009) 106
   \href{https://arxiv.org/abs/0903.4665}{\tt [arXiv:0903.4665 [hep-ph]]}.

\bibitem{Czakon:2009ss}
  M.~Czakon, C.~G.~Papadopoulos and M.~Worek,
  {\it Polarizing the Dipoles},
  JHEP {\bf 0908} (2009) 085
 \href{https://arxiv.org/abs/0905.0883}{\tt [arXiv:0905.0883 [hep-ph]]}.

\bibitem{Kauer:2001sp}
  N.~Kauer and D.~Zeppenfeld,
  {\it Finite width effects in top quark production at hadron colliders},
  Phys.\ Rev.\ D {\bf 65} (2002) 014021
  \href{http://arxiv.org/abs/hep-ph/0107181}{\tt [hep-ph/0107181]}.
    
\bibitem{Uhlemann:2008pm}
  C.~F.~Uhlemann and N.~Kauer,
  {\it Narrow-width approximation accuracy}, 
  Nucl.\ Phys.\ B {\bf 814} (2009) 195
  \href{http://arxiv.org/abs/arXiv:0807.4112}{\tt [arXiv:0807.4112 [hep-ph]]}.

\bibitem{Bernreuther:2004jv}
  W.~Bernreuther, A.~Brandenburg, Z.~G.~Si and P.~Uwer,
  {\it Top quark pair production and decay at hadron colliders}
  Nucl.\ Phys.\ B {\bf 690} (2004) 81
 \href{http://arxiv.org/abs/hep-ph/0403035}{\tt [hep-ph/0403035]}.
  
\bibitem{Melnikov:2009dn}
  K.~Melnikov and M.~Schulze,
  {\it NLO QCD corrections to top quark pair production and decay at
    hadron colliders},
  JHEP {\bf 0908} (2009) 049
  \href{http://arxiv.org/abs/arXiv:0907.3090}{\tt [arXiv:0907.3090 [hep-ph]]}.

\bibitem{Melnikov:2011qx}
  K.~Melnikov, A.~Scharf and M.~Schulze,
  {\it Top quark pair production in association with a jet: QCD
  corrections and jet radiation in top quark decays}, 
 Phys.\ Rev.\ D {\bf 85} (2012) 054002
 \href{http://arxiv.org/abs/arXiv:1111.4991}{\tt [arXiv:1111.4991 [hep-ph]]}.

\bibitem{Campbell:2012uf}
  J.~M.~Campbell and R.~K.~Ellis,
  {\it Top-Quark Processes at NLO in Production and Decay},
  J.\ Phys.\ G {\bf 42} (2015) no.1,  015005
  \href{http://arxiv.org/abs/arXiv:1204.1513}{\tt [arXiv:1204.1513 [hep-ph]]}.

 \bibitem{Behring:2019iiv}
  A.~Behring, M.~Czakon, A.~Mitov, A.~S.~Papanastasiou and R.~Poncelet,
  {\it Higher order corrections to spin correlations in top quark pair
    production at the LHC},
  Phys.\ Rev.\ Lett.\  {\bf 123} (2019) no.8,  082001
  \href{http://arxiv.org/abs/arXiv:1901.05407}{\tt [arXiv:1901.05407 [hep-ph]]}.

\bibitem{Fadin:1993kt}
  V.~S.~Fadin, V.~A.~Khoze and A.~D.~Martin,
  {\it How suppressed are the radiative interference effects in heavy
    instable particle production $?$},
  Phys.\ Lett.\ B {\bf 320} (1994) 141
  \href{http://arxiv.org/abs/hep-ph/9309234}{\tt [hep-ph/9309234]}.

\bibitem{AlcarazMaestre:2012vp}
  J.~Alcaraz Maestre {\it et al.} [SM and NLO MULTILEG Working Group
  and SM MC Working Group],
  {\it The SM and NLO Multi-leg and SM MC Working Groups: Summary
    Report},    \href{http://arxiv.org/abs/arXiv:1203.6803}{\tt
    arXiv:1203.6803 [hep-ph]}.
  
\bibitem{Denner:2012mx}
  A.~Denner, S.~Dittmaier, S.~Kallweit and S.~Pozzorini,
  {\it NLO QCD corrections to off-shell ttbar production at hadron
    colliders},   PoS LL {\bf 2012} (2012) 015
  \href{http://arxiv.org/abs/arXiv:1208.4053}{\tt [arXiv:1208.4053 [hep-ph]]}.

\bibitem{Rontsch:2014cca}
  R.~R\"ontsch and M.~Schulze,
  {\it Constraining couplings of top quarks to the Z boson in $
    t\overline{t} $ + Z production at the LHC},
  JHEP {\bf 1407} (2014) 091,
   Erratum: [JHEP {\bf 1509} (2015) 132]
 \href{https://arxiv.org/abs/1404.1005}{\tt [arXiv:1404.1005 [hep-ph]]}.
  
\bibitem{Melnikov:2010iu}
  K.~Melnikov and M.~Schulze,
  {\it NLO QCD corrections to top quark pair production in association
    with one hard jet at hadron colliders},
  Nucl.\ Phys.\ B {\bf 840} (2010) 129
 \href{https://arxiv.org/abs/1004.3284}{\tt [arXiv:1004.3284 [hep-ph]]}.

\bibitem{Campbell:2012dh}
  J.~M.~Campbell and R.~K.~Ellis,
  {\it $t \bar{t} W^\pm$ production and decay at NLO},
  JHEP {\bf 1207} (2012) 052
  \href{https://arxiv.org/abs/1204.5678}{\tt [arXiv:1204.5678 [hep-ph]]}.
 
\bibitem{Bredenstein:2009aj}
  A.~Bredenstein, A.~Denner, S.~Dittmaier and S.~Pozzorini,
  {\it NLO QCD corrections to $pp \to t \bar{t} b \bar{b} + X$ at the
    LHC}, 
  Phys.\ Rev.\ Lett.\  {\bf 103} (2009) 012002
 \href{http://arxiv.org/abs/arXiv:0905.0110}{\tt [arXiv:0905.0110 [hep-ph]]}.

\bibitem{Bevilacqua:2009zn}
  G.~Bevilacqua, M.~Czakon, C.~G.~Papadopoulos, R.~Pittau and M.~Worek,
  {\it Assault on the NLO Wishlist: $pp\to t\bar{t}b\bar{b}$},
  JHEP {\bf 0909} (2009) 109
  \href{http://arxiv.org/abs/arXiv:0907.4723}{\tt [arXiv:0907.4723 [hep-ph]]}.

  \bibitem{Bredenstein:2010rs}
  A.~Bredenstein, A.~Denner, S.~Dittmaier and S.~Pozzorini,
  {\it NLO QCD Corrections to Top Anti-Top Bottom Anti-Bottom
    Production at the LHC: 2. full hadronic results},
  JHEP {\bf 1003} (2010) 021
  \href{http://arxiv.org/abs/arXiv:1001.4006}{\tt [arXiv:1001.4006 [hep-ph]]}.
  
\bibitem{Bevilacqua:2010ve}
  G.~Bevilacqua, M.~Czakon, C.~G.~Papadopoulos and M.~Worek,
  {\it Dominant QCD Backgrounds in Higgs Boson Analyses at the LHC: A
    Study of $pp \to t\bar{t} + 2$ jets at Next-To-Leading Order},
  Phys.\ Rev.\ Lett.\  {\bf 104} (2010) 162002
 \href{http://arxiv.org/abs/arXiv:1002.4009}{\tt [arXiv:1002.4009 [hep-ph]]}.
  
\bibitem{Bevilacqua:2011aa}
  G.~Bevilacqua, M.~Czakon, C.~G.~Papadopoulos and M.~Worek,
  {\it Hadronic top-quark pair production in association with two jets
    at Next-to-Leading Order QCD},    
  Phys.\ Rev.\ D {\bf 84} (2011) 114017
  \href{http://arxiv.org/abs/arXiv:1108.2851}{\tt [arXiv:1108.2851 [hep-ph]]}.

\bibitem{Bevilacqua:2012em}
  G.~Bevilacqua and M.~Worek,
  {\it Constraining BSM Physics at the LHC: Four top final states with
    NLO accuracy in perturbative QCD},
  JHEP {\bf 1207} (2012) 111
 \href{http://arxiv.org/abs/arXiv:1206.3064}{\tt [arXiv:1206.3064 [hep-ph]]}.

\bibitem{Frederix:2017wme}
  R.~Frederix, D.~Pagani and M.~Zaro,
  {\it Large NLO corrections in $t\bar{t}W^{\pm}$ and
    $t\bar{t}t\bar{t}$ hadroproduction from supposedly subleading EW
    contributions},
  JHEP {\bf 1802} (2018) 031
  \href{http://arxiv.org/abs/arXiv:1711.02116}{\tt [arXiv:1711.02116 [hep-ph]]}.

\bibitem{Garzelli:2014aba}
  M.~V.~Garzelli, A.~Kardos and Z.~Trocsanyi,
  {\it Hadroproduction of $t\bar{t}b\bar{b}$ final states at LHC:
    predictions at NLO accuracy matched with Parton Shower},
  JHEP {\bf 1503} (2015) 083
  \href{https://arxiv.org/abs/1408.0266}{\tt [arXiv:1408.0266 [hep-ph]]}.
  
\bibitem{Bevilacqua:2017cru}
  G.~Bevilacqua, M.~V.~Garzelli and A.~Kardos,
  {\it $t\bar{t}b\bar{b}$ hadroproduction with massive bottom quarks
    with \textsc{PowHel}},
  \href{http://arxiv.org/abs/arXiv:1709.06915}{\tt arXiv:1709.06915 [hep-ph]}.
 
\bibitem{Jezo:2018yaf}
  T.~Jezo, J.~M.~Lindert, N.~Moretti and S.~Pozzorini,
  {\it New NLOPS predictions for $t \bar{t} +b$-jet
    production at the LHC},  
  Eur.\ Phys.\ J.\ C {\bf 78} (2018) no.6,  502
  \href{http://arxiv.org/abs/arXiv:1802.00426}{\tt [arXiv:1802.00426 [hep-ph]]}.
   
\bibitem{Kanaki:2000ey}
  A.~Kanaki and C.~G.~Papadopoulos,
  {\it HELAC: A Package to compute electroweak helicity amplitudes},
  Comput.\ Phys.\ Commun.\  {\bf 132} (2000) 306
  \href{https://arxiv.org/abs/hep-ph/0002082}{\tt [hep-ph/0002082]}.

  \bibitem{Cafarella:2007pc}
  A.~Cafarella, C.~G.~Papadopoulos and M.~Worek,
  {\it Helac-Phegas: A Generator for all parton level processes},
  Comput.\ Phys.\ Commun.\  {\bf 180} (2009) 1941
  \href{https://arxiv.org/abs/0710.2427}{\tt [arXiv:0710.2427 [hep-ph]]}.
  
\bibitem{Bevilacqua:2010qb}
  G.~Bevilacqua, M.~Czakon, A.~van Hameren, C.~G.~Papadopoulos and M.~Worek,
  {\it Complete off-shell effects in top quark pair hadroproduction
    with leptonic decay at next-to-leading order},
  JHEP {\bf 1102} (2011) 083
  \href{https://arxiv.org/abs/1012.4230}{\tt [arXiv:1012.4230 [hep-ph]]}.
  
\bibitem{Bevilacqua:2015qha}
  G.~Bevilacqua, H.~B.~Hartanto, M.~Kraus and M.~Worek,
  {\it Top Quark Pair Production in Association with a Jet with
    Next-to-Leading-Order QCD Off-Shell Effects at the Large Hadron
    Collider},
  Phys.\ Rev.\ Lett.\  {\bf 116} (2016) no.5,  052003
  \href{https://arxiv.org/abs/1509.09242}{\tt [arXiv:1509.09242 [hep-ph]]}.
  
\bibitem{Bevilacqua:2016jfk}
  G.~Bevilacqua, H.~B.~Hartanto, M.~Kraus and M.~Worek,
  {\it Off-shell Top Quarks with One Jet at the LHC: A comprehensive
    analysis at NLO QCD},
  JHEP {\bf 1611} (2016) 098
  \href{http://arxiv.org/abs/arXiv:1609.01659}{\tt [arXiv:1609.01659 [hep-ph]]}.
  
\bibitem{Bevilacqua:2017ipv}
  G.~Bevilacqua, H.~B.~Hartanto, M.~Kraus, M.~Schulze and M.~Worek,
  {\it Top quark mass studies with $ t\overline{t}j $ at the LHC},
  JHEP {\bf 1803} (2018) 169
\href{http://arxiv.org/abs/arXiv:1710.07515}{\tt [arXiv:1710.07515 [hep-ph]]}.
    
\bibitem{Bevilacqua:2019cvp}
  G.~Bevilacqua, H.~B.~Hartanto, M.~Kraus, T.~Weber and M.~Worek,
  {\it Towards constraining Dark Matter at the LHC: Higher order QCD
    predictions for $t\bar{t}+Z(Z\to \nu_\ell \bar{\nu}_\ell)$},
  JHEP {\bf 1911} (2019) 001
 \href{https://arxiv.org/abs/1907.09359}{\tt [arXiv:1907.09359 [hep-ph]]}.

\bibitem{Catani:1996vz}
  S.~Catani and M.~H.~Seymour,
  {\it A General algorithm for calculating jet cross-sections in NLO
    QCD}, Nucl.\ Phys.\ B {\bf 485} (1997) 291
   Erratum: [Nucl.\ Phys.\ B {\bf 510} (1998) 503]
  \href{https://arxiv.org/abs/hep-ph/9605323}{\tt [hep-ph/9605323]}.
  
\bibitem{Catani:2002hc}
  S.~Catani, S.~Dittmaier, M.~H.~Seymour and Z.~Trocsanyi,
  {\it The Dipole formalism for next-to-leading order QCD calculations
    with massive partons},
  Nucl.\ Phys.\ B {\bf 627} (2002) 189
  \href{https://arxiv.org/abs/hep-ph/0201036}{\tt [hep-ph/0201036]}.

\bibitem{Campbell:2004ch}
  J.~M.~Campbell, R.~K.~Ellis and F.~Tramontano,
  {\it Single top production and decay at next-to-leading order},
  Phys.\ Rev.\ D {\bf 70} (2004) 094012
  \href{https://arxiv.org/abs/hep-ph/0408158}{\tt [hep-ph/0408158]}.

\bibitem{Nagy:1998bb}
  Z.~Nagy and Z.~Trocsanyi,
  {\it Next-to-leading order calculation of four jet observables in
    electron positron annihilation},
  Phys.\ Rev.\ D {\bf 59} (1999) 014020
   Erratum: [Phys.\ Rev.\ D {\bf 62} (2000) 099902]
  \href{https://arxiv.org/abs/hep-ph/9806317}{\tt [hep-ph/9806317]}.

\bibitem{Nagy:2003tz}
  Z.~Nagy,
  {\it Next-to-leading order calculation of three jet observables in
    hadron hadron collision}, 
  Phys.\ Rev.\ D {\bf 68} (2003) 094002
  \href{https://arxiv.org/abs/hep-ph/0307268}{\tt [hep-ph/0307268]}.
  
  \bibitem{Dulat:2015mca}
  S.~Dulat {\it et al.},
  {\it New parton distribution functions from a global analysis of
    quantum chromodynamics},
  Phys.\ Rev.\ D {\bf 93} (2016) no.3,  033006
  \href{https://arxiv.org/abs/1506.07443}{\tt [arXiv:1506.07443 [hep-ph]]}.

\bibitem{Ball:2014uwa}
  R.~D.~Ball {\it et al.} [NNPDF Collaboration],
  {\it Parton distributions for the LHC Run II}, 
  JHEP {\bf 1504} (2015) 040
  \href{https://arxiv.org/abs/1410.8849}{\tt [arXiv:1410.8849 [hep-ph]]}.

\bibitem{Harland-Lang:2014zoa}
  L.~A.~Harland-Lang, A.~D.~Martin, P.~Motylinski and R.~S.~Thorne,
  {\it Parton distributions in the LHC era: MMHT 2014 PDFs}, 
  Eur.\ Phys.\ J.\ C {\bf 75} (2015) no.5,  204
   \href{https://arxiv.org/abs/1412.3989}{\tt [arXiv:1412.3989 [hep-ph]]}.

   \bibitem{Cacciari:2008gp}
  M.~Cacciari, G.~P.~Salam and G.~Soyez,
  {\it The anti-$k_T$ jet clustering algorithm}
  JHEP {\bf 0804} (2008) 063,
  \href{http://arxiv.org/abs/arXiv:0802.1189}{\tt [arXiv:0802.1189 [hep-ph]]}.
 
\bibitem{Frixione:1998jh}
  S.~Frixione,
  {\it Isolated photons in perturbative QCD}, 
  Phys.\ Lett.\ B {\bf 429} (1998) 369
  \href{http://arxiv.org/abs/hep-ph/9801442}{\tt [hep-ph/9801442]}.
  
\bibitem{Jezabek:1988iv}
  M.~Jezabek and J.~H.~Kuhn,
  {\it QCD Corrections to Semileptonic Decays of Heavy Quarks}, 
  Nucl.\ Phys.\ B {\bf 314} (1989) 1.

\bibitem{Chetyrkin:1999ju}
  K.~G.~Chetyrkin, R.~Harlander, T.~Seidensticker and M.~Steinhauser,
  {\it Second order QCD corrections to $\Gamma(t  \to W b)$}, 
  Phys.\ Rev.\ D {\bf 60} (1999) 114015
 \href{http://arxiv.org/abs/hep-ph/9906273}{\tt [hep-ph/9906273]}.

\bibitem{Denner:2012yc}
  A.~Denner, S.~Dittmaier, S.~Kallweit and S.~Pozzorini,
 {\it NLO QCD corrections to off-shell top-antitop production with
 leptonic decays at hadron colliders},
JHEP {\bf 1210} (2012) 110
\href{http://arxiv.org/abs/arXiv:1207.5018}{\tt [arXiv:1207.5018 [hep-ph]]}.

\bibitem{Butterworth:2015oua}
  J.~Butterworth {\it et al.},
  {\it PDF4LHC recommendations for LHC Run II},
  J.\ Phys.\ G {\bf 43} (2016) 023001
  \href{http://arxiv.org/abs/arXiv:1510.03865}{\tt [arXiv:1510.03865 [hep-ph]]}.

\bibitem{Heinrich:2013qaa}
  G.~Heinrich, A.~Maier, R.~Nisius, J.~Schlenk and J.~Winter,
  {\it NLO QCD corrections to $W^{+} W^{-}b\bar{b}$ production with
    leptonic decays in the light of top quark mass and asymmetry
    measurements},
  JHEP {\bf 1406} (2014) 158
  \href{http://arxiv.org/abs/arXiv:1312.6659}{\tt [arXiv:1312.6659 [hep-ph]]}.
  
\bibitem{Heinrich:2017bqp}
  G.~Heinrich, A.~Maier, R.~Nisius, J.~Schlenk, M.~Schulze, L.~Scyboz and J.~Winter,
  {\it NLO and off-shell effects in top quark mass determinations}, 
  JHEP {\bf 1807} (2018) 129
  \href{http://arxiv.org/abs/arXiv:1709.08615}{\tt [arXiv:1709.08615 [hep-ph]]}.

\bibitem{Ravasio:2018lzi}
  S.~Ferrario Ravasio, T.~Jezo, P.~Nason and C.~Oleari,
  {\it A theoretical study of top-mass measurements at the LHC using
    NLO+PS generators of increasing accuracy},
  Eur.\ Phys.\ J.\ C {\bf 78} (2018) no.6,  458
  \href{http://arxiv.org/abs/arXiv:1801.03944}{\tt [arXiv:1801.03944 [hep-ph]]}.

\bibitem{Beneke:2000hk}
  M.~Beneke {\it et al.},
  {\it Top quark physics},
  1999 CERN Workshop on SM physics (and more) at the LHC,   Geneva,
  Switzerland, 25-26 May 1999
  \href{https://arxiv.org/abs/hep-ph/0003033}{\tt [hep-ph/0003033]}.
  
\bibitem{Liebler:2015ipp}
  S.~Liebler, G.~Moortgat-Pick and A.~S.~Papanastasiou,
  {\it Probing the top-quark width through ratios of resonance
    contributions of $e^+e^-\rightarrow W^+W^-b\bar{b}$},
  JHEP {\bf 1603} (2016) 099
  \href{http://arxiv.org/abs/arXiv:1511.02350}{\tt [arXiv:1511.02350 [hep-ph]]}.
  
\bibitem{Baskakov:2018huw}
  A.~Baskakov, E.~Boos and L.~Dudko,
  {\it Model independent top quark width measurement using a
    combination of resonant and nonresonant cross sections},
  Phys.\ Rev.\ D {\bf 98} (2018) no.11,  116011
  \href{http://arxiv.org/abs/arXiv:1807.11193}{\tt [arXiv:1807.11193 [hep-ph]]}.

  
\end{thebibliography}
\end{document}